\documentclass[twocolumn]{aastex631}

\usepackage{xspace}
\usepackage[version=4]{mhchem} 
\usepackage{hyperref} 
\usepackage{diagbox}

\newcommand{\eureka}{\texttt{Eureka!}\xspace}

\defcitealias{Mor23}{Moran \& Stevenson et al. 2023}
\defcitealias{MayMac23}{May \& MacDonald et al. 2023}
\defcitealias{Liu2025ArXiV}{Liu \& Wang et al. 2025}

\submitjournal{AJ}
\received{January 30, 2025}
\revised{August 19, 2025}
\accepted{August 28, 2025}

\shorttitle{Reanalysis of K2-18 \lowercase{b}'s Near-Infrared JWST Transmission Spectrum}

\shortauthors{Schmidt et al.}

\begin{document}

\title{A Comprehensive Reanalysis of K2-18\,b's JWST NIRISS+NIRSpec Transmission Spectrum}

\author[0000-0001-8510-7365]{Stephen P.\ Schmidt}
\altaffiliation{NSF Graduate Research Fellow}
\affiliation{William H. Miller III Department of Physics and Astronomy, Johns Hopkins University, Baltimore, MD 21218, USA}

\author[0000-0003-4816-3469]{Ryan J. MacDonald}
\altaffiliation{NHFP Sagan Fellow}
\affiliation{Department of Astronomy, University of Michigan, 1085 S. University Ave., Ann Arbor, MI 48109, USA}
\affiliation{School of Physics and Astronomy, University of St Andrews, North Haugh, St Andrews, KY16 9SS, UK}

\author[0000-0002-8163-4608]{Shang-Min Tsai}
\affiliation{Institute of Astronomy \& Astrophysics, Academia Sinica, Taipei 10617, Taiwan}
\affiliation{Department of Earth and Planetary Sciences, University of California, Riverside, CA, USA}

\author[0000-0002-3328-1203]{Michael Radica}
\altaffiliation{NSERC Postdoctoral Fellow}
\affiliation{Department of Astronomy \& Astrophysics, University of Chicago, 5640 South Ellis Avenue, Chicago, IL 60637, USA}
\affiliation{Institut Trottier de Recherche sur les Exoplanètes and Département de Physique, Université de Montréal, 1375 Avenue
Thérèse-Lavoie-Roux, Montréal, QC, H2V 0B3, Canada}

\author[0000-0002-6379-3816]{Le-Chris Wang}
\affiliation{William H. Miller III Department of Physics and Astronomy, Johns Hopkins University, Baltimore, MD 21218, USA}

\author[0000-0003-0973-8426]{Eva-Maria Ahrer}
\affiliation{Max Planck Institute for Astronomy, Heidelberg, 69117, Germany}

\author[0000-0003-4177-2149]{Taylor J. Bell}
\affiliation{AURA for the European Space Agency (ESA), Space Telescope Science Institute, 3700 San Martin Drive, Baltimore, MD 21218, USA}
\affiliation{Bay Area Environmental Research Institute, NASA Ames Research Center, M.S. 245-6, Moffett Field, 94035 CA, USA}
\affiliation{Space Science and Astrobiology Division, NASA Ames Research Center, M.S. 245-6, Moffett Field, 94035 CA, USA}

\author[0000-0003-0652-2902]{Chloe Fisher}
\affiliation{Astrophysics, University of Oxford, Denys Wilkinson Building, Keble Road, Oxford, OX1 3RH, United Kingdom}

\author[0000-0002-5113-8558]{Daniel P. Thorngren}
\affiliation{William H. Miller III Department of Physics and Astronomy, Johns Hopkins University, Baltimore, MD 21218, USA}

\author[0000-0002-0413-3308]{Nicholas Wogan}
\affiliation{NASA Ames Research Center, Moffett Field, CA 94035, US}

\author[0000-0002-2739-1465]{Erin M. May}
\affiliation{Johns Hopkins APL, Laurel, MD 20723, USA}

\author[0000-0001-6096-7772]{Piero Ferrari}
\affiliation{HFML-FELIX, Radboud University, Nijmegen 6525 ED, The Netherlands}

\author[0000-0002-9030-0132]{Katherine A. Bennett}
\affiliation{Morton K. Blaustein Department of Earth \& Planetary Sciences, Johns Hopkins University, Baltimore, MD 21218, USA}

\author[0000-0003-4408-0463]{Zafar Rustamkulov}
\affiliation{Morton K. Blaustein Department of Earth \& Planetary Sciences, Johns Hopkins University, Baltimore, MD 21218, USA}

\author[0000-0003-3204-8183]{Mercedes López-Morales}
\affiliation{Space Telescope Science Institute, 3700 San Martin Drive, Baltimore, MD 21218, USA}

\author[0000-0001-6050-7645]{David K. Sing}
\affiliation{William H. Miller III Department of Physics and Astronomy, Johns Hopkins University, Baltimore, MD 21218, USA}
\affiliation{Morton K. Blaustein Department of Earth \& Planetary Sciences, Johns Hopkins University, Baltimore, MD 21218, USA}

\correspondingauthor{Stephen Schmidt}
\email{sschmi42@jh.edu}

\begin{abstract}
\noindent Sub-Neptunes are the most common type of planet in our galaxy. Interior structure models suggest that the coldest sub-Neptunes could host liquid water oceans underneath their hydrogen envelopes---sometimes called ``hycean'' planets. JWST transmission spectra of the $\sim$ 250\,K sub-Neptune K2-18\,b were recently used to report detections of CH$_4$ and CO$_2$, alongside weaker evidence of (CH$_3$)$_2$S (dimethyl sulfide, or DMS). Atmospheric CO$_2$ was interpreted as evidence for a liquid water ocean, while DMS was highlighted as a potential biomarker. However, these notable claims were derived using a single data reduction and retrieval modeling framework, which did not allow for standard robustness tests. Here we present a comprehensive reanalysis of K2-18\,b's JWST NIRISS SOSS and NIRSpec G395H transmission spectra, including the first analysis of the second-order NIRISS SOSS data. We incorporate multiple well-tested data reduction pipelines and retrieval codes, spanning 60 different data treatments and over 250 atmospheric retrievals. We confirm the detection of CH$_4$ ($\approx 4\sigma$), with a volume mixing ratio range $-2.14 \leq \log_{10} \mathrm{CH_4} \leq -0.53$, but we find no statistically significant or reliable evidence for CO$_2$ or DMS. Finally, we assess the retrieved atmospheric composition using photochemical-climate and interior models, demonstrating that our revised composition of K2-18\,b can be explained by an oxygen-poor mini-Neptune without requiring a liquid water surface or life.

\end{abstract}


\section{Introduction}

James Webb Space Telescope (JWST) transmission spectroscopy has revealed the atmospheric composition of dozens of transiting exoplanets in the three years since its launch. The most amenable planets to atmospheric characterization are hot H$_2$-dominated giant exoplanets. Major JWST results from such hot giant planets include detections of SO$_2$ \citep[e.g.,][]{Rus23, Ald23, Dyr24, Pow24, Sin24, Wel24}, which provides direct evidence for photochemistry in exoplanetary atmospheres \citep[e.g.,][]{Tsa23}, and the direct detection of aerosol species \citep{Grant2023, Dyr24, Ing24}. JWST is also enabling initial forays into population-level trends across properties of giant exoplanet atmospheres \citep[e.g.,][]{Fu25}, which will provide deeper insights as more transmission spectra are analyzed in the coming years.

While JWST transmission spectra have succeeded in probing the giant exoplanet atmospheres, decisive results have been more elusive for the small and cold ends of the exoplanet population.
For super-Earths and terrestrial worlds ($R_{\rm{p}} < 1.7\,R_{\Earth}$), JWST transmission spectra largely rule out thick H$_2$-dominated atmospheres (e.g., \citealt{Lus23}; \citealt{Lim23}; \citealt{Ald24}; \citealt{Rad24}) or are degenerate with unocculted stellar active regions (e.g., \citetalias{Mor23} \citeyear{Mor23}; \citetalias{MayMac23} \citeyear{MayMac23}). However, sub-Neptunes\footnote{We use ``sub-Neptune'' when categorizing planets by size (c.f. super-Earth), and ``mini-Neptune'' when categorizing planets by interior structure (c.f. hycean).}, planets with radii $1.7~\text{R}_\oplus < R < 3.5~\text{R}_\oplus$, likely possess significant atmospheres dominated by relatively light molecules (e.g., H$_2$ or H$_2$O), which aids their atmospheric detectability. While searches for atmospheres on several sub-Neptunes have yielded non-detections \citep[e.g.][]{Wal24} or weak evidence \citep[e.g.,][]{Dam24, Cad24}, an emerging population now has definitive atmospheric detections. To date, JWST has detected atmospheres on five sub-Neptunes: GJ~3470~b \citep{Bea24}, GJ~9827~d \citep{Pia24}, TOI-270~d \citep{Ben24, Hol24}, TOI-421~b \citep{Davenport2025}, and K2-18\,b \citep{Mad23}. Of these planets, the cold ($\sim$ 250\,K) sub-Neptune K2-18\,b has attracted significant observational and theoretical interest in its atmospheric composition and interior structure.

The K2-18 system was identified as a potential transiting exoplanet system in the first campaign of NASA's K2 mission \citep{Mon15}. Further transit observations by Spitzer \citep{Ben17} and radial velocity follow-up \citep{Clo17, Sar18} confirmed K2-18\,b as a habitable zone planet, with a refined radial velocity extraction method further improving its mass precision \citep{Rad22}. The system is now known to host at least two planets \citep{Clo19, Rad22}, though only K2-18\,b transits its early M dwarf host star. The combination of its cool temperature (255\,K, assuming an albedo of 0.3; \citealt{Ben19}), relatively large planetary radius (2.61\,$R_{\Earth}$), and small stellar radius (0.4445\,$R_{\odot}$) has led to K2-18\,b being one of the best targets for the atmospheric characterization of a potentially habitable exoplanet.

Initial transmission spectroscopy observations of K2-18\,b with the Hubble Space Telescope (HST) yielded a detection of a H$_2$-dominated atmosphere, with evidence of either gas phase H$_2$O or CH$_4$. The HST Wide Field Camera 3 (WFC3) transmission spectrum of K2-18\,b was initially interpreted as showing a $> 3\,\sigma$ detection of H$_2$O \citep{Tsi19,Ben19,Mad20}.
However, later work by \citet{Bar21} argued that the low-resolution WFC3 data could alternatively be explained by CH$_4$, due to the similar shapes of H$_2$O and CH$_4$ absorption over the 1.1--1.7\,$\micron$ WFC3 bandpass. These competing interpretations of K2-18\,b's atmospheric composition resulted in an inability to definitively constrain the nature of its atmosphere and interior through HST observations alone.

Atmospheric and interior models of K2-18\,b have ignited a vigorous debate on the nature of this world. Early atmospheric models disfavored a water reservoir below the H$_2$-dominated atmosphere as the source of the apparent H$_2$O vapor \citep{Sch20}, while general circulation models pointed towards a misinterpretation of the HST data as caused by CH$_4$ rather than H$_2$O \citep{Bla21}. Interior structure models of K2-18\,b highlighted that the pressure-temperature profile could allow a habitable liquid water ocean beneath a thin H$_2$ atmosphere \citep{Mad20, Pie20, Nix21} --- sometimes termed a  ``hycean'' planet. Sustaining such a liquid water ocean requires the planet to have a H$_2$-rich atmosphere, an iron-rock core comprising $>$ 10\% of the planet's mass, and a H$_2$O layer with a mass fraction from 10--90\% \citep{Mad21}. Such an ocean can be considered a habitable surface, which renders K2-18\,b an object of potential astrobiological interest in the search for extraterrestrial life. 

Recently, the first JWST transmission spectrum of K2-18\,b revealed multiple prominent atmospheric absorption features \citep{Mad23}. The NIRISS SOSS and NIRSpec G395H observations (spanning 0.8--5.3\,$\micron$) were explained by strong atmospheric CH$_4$ bands (detected at $5\,\sigma$), a peak near 4.3\,$\micron$ attributed to CO$_2$ (at $3\,\sigma$), and ``potential signs'' of (CH$_3$)$_2$S (dimethyl sulfide, or DMS). The JWST detection of CH$_4$, and non-detection of H$_2$O, revealed that the previous inference of H$_2$O from HST was indeed a case of mistaken identity. \citet{Mad23} reported an atmospheric composition of $\sim 1\%$ CH$_4$ and $\sim 1\%$ CO$_2$, which is challenging to explain under standard thermochemistry for a mini-Neptune with a deep atmosphere. They also reported a non-detection of NH$_3$, a gas that would generally be expected in a mini-Neptune at K2-18\,b's temperature, with an upper limit of $\approx$ 30\,ppm. \citet{Mad23} argued that these two lines of evidence point towards a water ocean beneath the H$_2$-dominated outer layer, since an ocean would increase the CO$_2$/CH$_4$ ratio and deplete NH$_3$ \citep{Hu21,Tsai2021}.

Subsequent modeling efforts have argued that the JWST observations of K2-18\,b do not require a habitable liquid water ocean. \citet{Sho24} proposed that the NH$_3$ non-detection can be explained by a deep H$_2$ envelope above a magma ocean, which allows efficient dissolution of nitrogen. \citet{Yang2024} found that the CO$_2$/CH$_4$ ratio of $\sim 1$ from \citet{Mad23} can also be explained by a gas-rich mini-Neptune enriched in water vapor in the deep atmosphere. \citet{Wog24} used photochemical and climate models to investigate three scenarios for K2-18\,b: (i) a lifeless hycean planet, (ii) an `inhabited' hycean planet with a substantial biological CH$_4$ flux into the atmosphere, and (iii) a 100$\times$ solar metallicity mini-Neptune. They ruled out the first scenario (which would have $<$ 1\,ppm of CH$_4$), such that the $\sim 1\%$ CH$_4$ abundance inferred from JWST could be explained by either a methane-producing biosphere or, alternatively, a metal-enriched mini-Neptune. \citet{Wog24} argued that the mini-Neptune scenario is more plausible \emph{a priori}, since this explanation works `out of the box' without requiring one to postulate the existence of life. Most recently, \citet{Cooke2024} investigated the same three scenarios, arguing that the CO$_2$ abundance found by \citet{Mad23} favors the `inhabited' hycean scenario over a mini-Neptune. 

The astrobiological assessment of H$_2$-dominated sub-Neptunes such as K2-18\,b is at an early stage, and hence any claims of life must be tempered with extreme caution. \citet{Mad23} highlighted a low-significance inference of DMS ($\sim 1\,\sigma$ from their best-fitting model) as potential evidence for life, given DMS is significantly produced by marine phytoplankton on Earth \citep{Charlson1987,Barnes2006}. 
Such a DMS flux would have to be equivalent to $\sim$20 times that produced by Earth's marine life to reach detectable levels on a hycean planet with K2-18\,b-like conditions \citep{Tsai2024}.
\citet{Gle24} showed that metabolic reactions consuming CO$_2$ and/or DMS and producing CH$_4$ can potentially produce sufficient free energy to sustain a biosphere \citep[if one assumes a hycean scenario and the atmospheric abundances reported in][]{Mad23} on K2-18\,b.
However, DMS in this metabolic pathway cannot strictly be considered a biosignature due to it being an \emph{energy source} for methane-producing life rather than a \emph{by-product} of life. 
Furthermore, the recent detection of DMS on a comet by \citet{Han24} demonstrates that DMS can also be produced in abiotic environments.
These issues with DMS as potential evidence for a biosphere on K2-18\,b in combination with the lack of detailed cross section data for it complicate any potential investigation of the planet's habitability.
Indeed, the state-of-the-art line data for DMS \citep[HITRAN; ][]{sharpe04} lacks pressure and temperature dependence, potentially biasing retrieval results that include it \citep{Heng2017, WelbanksMadhu2019, Niraula2022}.

The plausibility of the hycean interpretation for K2-18\,b relies critically on the retrieved chemical abundances reported by \citet{Mad23}. However, to date, there has been no independent analysis of the original JWST transmission spectrum of K2-18\,b. The gold standard for interpreting JWST spectra of exoplanet atmospheres is to apply multiple data reduction techniques and retrieval codes to confirm the reproducibility and robustness of the results --- this philosophy has been broadly adopted by the exoplanet community \citep[e.g.,][]{Coulombe2023,Taylor2023,Grant2023,Pow24,Gressier2024,Banerjee2024,Pia24,Wel24}. However, the K2-18\,b analysis by \citet{Mad23} investigated only a single reduction and retrieval model framework. Performing several reductions on the same data set quantifies the impact of choices such as outlier rejection, noise correction, and limb darkening treatments that occur when using independent pipelines. Similarly, performing retrievals with different codes can illustrate how some model-level assumptions (e.g., opacity sources, aerosol parameterization) propagate into atmospheric inferences \citep[e.g.,][]{MaiLine2019, WelbanksMadhu2019, Barstow2020, Niraula2023, Nixon2024}. Atmospheric inferences that withstand these tests are considered robust. On the other hand, severe inconsistencies (e.g., a molecular detection seen only with one reduction), demonstrate that the statistical evidence for such an atmospheric inference must be carefully assessed. Here, we provide the first comprehensive analysis of K2-18\,b's JWST NIRISS SOSS and NIRSpec G395H transmission spectra, using multiple data reduction and retrieval codes, to offer an updated assessment of K2-18\,b's atmospheric composition and possible interior structures.

We note that during the revision of this study, \citet{Madhusudhan2025} presented the MIRI LRS transmission spectrum of K2-18\,b. While \citet{Madhusudhan2025} reported independent evidence for DMS and/or DMDS (dimethyl disulfide) at $\sim 3\sigma$, subsequent studies have demonstrated that the MIRI data is well-fit by a flat line \citep{Taylor2025}, that DMS and DMDS do not provide a uniquely good fit relative to other molecules \citep{Welbanks2025,Pica-Ciamarra2025}, and that DMS or DMDS are not detected when the MIRI data is analyzed in concert with the near-infrared data \citep{Luque2025}. These studies highlight the low information content of the MIRI data relative to the shorter wavelength near-infrared data, which we focus on in this study.

Our study is structured as follows. We describe our new K2-18\,b data reductions in Sections~\ref{sec:niriss} and \ref{sec:nirspec} for NIRISS SOSS and NIRSpec G395H, respectively. Our retrieval analysis and atmospheric composition constraints are presented in Section~\ref{sec:retrievals}. We model plausible atmospheric and interior structures for K2-18\,b in Section~\ref{sec:models}. Finally, we summarize our results and discuss the implications in Section~\ref{sec:disc}.

\section{NIRISS Data Reductions} \label{sec:niriss}

The K2-18 system was observed by JWST's Near Infrared Imager and Slitless Spectrograph \citep[NIRISS;][]{NIRISS1, NIRISS2} instrument using the Single Object Slitless Spectroscopy \citep[SOSS;][]{NIRISS4} mode as part of JWST GO Program 2722 \citep[PI: N. Madhusudhan - ][]{K218JWSTData}.
The details of this observation are described thoroughly in \citet{Mad23}.
We perform two reductions of this data set with two independent pipelines: \texttt{FIREFLy} and \texttt{exoTEDRF}\footnote{We do not perform NIRISS reductions with the \eureka pipeline, as support for this instrument mode is currently pending.}. To allow a transparent comparison between our reduction approaches and \citet{Mad23} (who used the \texttt{JExoRES} pipeline \citealt{Hol23}), we provide an overview of the most important reduction-level configuration choices for NIRISS in Table~\ref{tab:NIRISSChoices}.
Both of our reductions fit the spectroscopic light curves at a two-pixel binning level \citep[the same binning strategy as][]{Mad23} as well as at resolving powers $R\approx 25$ and $R\approx 100$. 
We use two-pixel binning for NIRISS SOSS rather than one-pixel binning as its PSF is two pixels wide.
Additionally, we show the point-by point discrepancies between reductions in Appendix \ref{appendix:comparisons}, as has been done previously for analyses involving multiple reductions where small point-by-point differences could yield different atmospheric interpretations \citep[e.g.,][]{Alderson2024, Alam2024, Louie2025, Alderson2025, Kirk2025}.

\begin{deluxetable*}{llll}
\centering
\tabletypesize{\scriptsize}
\tablewidth{0pt}
\tablecaption{NIRISS Reduction Configuration Comparison \label{tab:NIRISSChoices}}
\tablehead{Step & \texttt{FIREFLy} (this work) & \texttt{ExoTEDRF} (this work) & \citet{Mad23}}
\startdata
\multicolumn{4}{c}{\hspace{10pt} \textbf{Stage 1}}\\
\hline
Saturation & Default & Default & Default\\
Superbias & Default & Default & Default\\
Dark current & Skipped & Default & Skipped\\
Background subtraction& Single-component STScI background& Two-component STScI background& Two-component STScI background\\
\hspace{1pt} (group level) &  model& model& model\\
$1/f$ subtraction& Yes & \texttt{scale-achromatic} & Yes \\
\hspace{1pt} (group level) & & &\\
Linearity & Default & Default & Default\\
Jump & Skipped & \texttt{exoTEDRF}-custom, 10$\sigma$ in time & Default, $5\sigma$\\
Ramp fitting & Default & Default & Default\\
\hline
\multicolumn{4}{c}{\hspace{10pt} \textbf{Stage 2}}\\
\hline
Wavelength calibration & \texttt{pastasoss}\citep{baines2023} & Cross correlation with \texttt{PHOENIX} & ?\\
 & &  stellar model &\\
$1/f$ subtraction& Yes & No & Yes\\
\hspace{1pt} (integration-level) & & &\\
Bad pixel cleaning & $>5\sigma$ outlier pixels & $>5\sigma$ outlier pixels in space and & ?\\
 & &  time &\\
Order tracing & \texttt{pastososs} & \texttt{edgetrigger} & similar to \texttt{edgetrigger}\\
\hline
\multicolumn{4}{c}{\hspace{10pt} \textbf{Stage 3}}\\
\hline
Temporal outliers & $>15\sigma$ & $>5\sigma$ & ?\\
1D Spectrum Extraction & Box extraction, 30 pixel aperture for & Box extraction, 30 pixel & Multi-order \citet{horne1986optimalE} spatial \\
 & order 1, 23 pixel aperture for order 2  & aperture & profile-based extraction, 35 pixel\\
  & & & aperture\\
Exclusions & None & None & Columns with $>20$\% masked flux\\
\hline
\multicolumn{4}{c}{\hspace{10pt} \textbf{Stage 4}}\\
\hline
Light curve model & \texttt{batman} & \texttt{batman} & \texttt{SPOTROD}\\
Spot-crossing event & Integrations trimmed & Gaussian in light curve model & 4-parameter spot model in light\\
\hspace{1pt} treatment & & & curve model\\
Limb darkening & Fixed to Stagger model with offsets& Fit for each bin & Binned to $R \sim 20$,  fit in each bin\\
\hspace{1pt} &  based on white light curve fit & &\\
LD parameterization & \{$u_+, u_-$\} & \{$u_1, u_2$\} & \{$q_1, q_2$\}\\
Systematic trends & linear & linear & linear\\
Code used for white light  & \texttt{emcee} & \texttt{emcee} & \texttt{MultiNest}\\
\hspace{1pt}  curve posterior& & &\\
Fitted WLC $a/R\ast$ & $81.32 \pm 1.74$ & \nodata & $79.9^{+1.4}_{-1.4}$\\
Fitted WLC $b$ & $0.6031\pm0.0237$ & \nodata & \nodata\\
Fitted WLC $i$ & \nodata & \nodata & $89.550^{+0.021}_{-0.020}$\\
Fitted WLC $T_0$ & $60096.72942 \pm 6 \times 10^{-5}$ & 57264.3914 & $60096.729368^{+0.00063}_{-0.00065}$ \\
Binning method & 2-pixel, $R\approx25$, and $R\approx100$ & 2-pixel, $R\approx25$, and $R\approx100$ & 2-pixel level\\
Code used for spectro- & Levenberg-Marquardt; \texttt{lmfit} & MCMC; \texttt{emcee} & Levenberg-Marquardt\\
\hspace{1pt} -photometric fit & & &\\
Error bar inflation & Red noise via residual RMS-bin & Additive error inflation & ?\\
 & size trend & & \\
Code Availability & Not Public & \href{https://github.com/radicamc/exoTEDRF}{\texttt{exoTEDRF} GitHub} & Not Public\\
\hline
\enddata
\tablecomments{``Default'' refers to the default handling of the step in the \texttt{jwst} pipeline. The system parameters ($a/R_\ast$, $i$, and $T_0$) are consistent across each reduction, though they were calculated in \citet{Mad23} based on their reduction up to that point. The listed orbital parameters were used to either confirm consistency with \citet{Mad23} (\texttt{FIREFLy}) or to show negligible differences between resulting spectra with an alternative orbital solution \citep[\texttt{exoTEDRF}, from][]{Rad22}. This is not an exhaustive list of reduction steps. `?' indicates reduction configuration settings not mentioned in \citet{Mad23}.}
\end{deluxetable*}

\subsection{\texttt{FIREFLy}}

We use the \texttt{FIREFLy} pipeline \citep{Rus22, Rus23}, which has been recently updated to better support NIRISS/SOSS observations (\citetalias{Liu2025ArXiV}, L. C. Wang et al. in prep), to re-reduce the K2-18 NIRISS/SOSS observations. 
A more thorough inventory of the steps we use are included in Table \ref{tab:NIRISSChoices}; here we describe the most pertinent choices made.
To correct the 1/$f$ noise at the group level, we first subtract the zodiacal background at each group. 
This is done by scaling the flux jump caused by the reflection of zodiacal light off the pick-off mirror in the STScI SOSS background model to the flux jump present in the observed data and adding a constant offset to match the flux level of the model SOSS background before and after the jump to the observed data. 
We then mask out the bright spectral trace and notable zeroth order contaminants present in the frame and performed 1/$f$ destriping by subtracting the median of each PSF-masked and background-subtracted column. 
In addition, we subtract each image from the temporal median, which we create as a running median of a 7 frame window, to reveal any remaining 1/$f$ noise and subtract it.  
This removes high frequency structure while preserving the slowly changing source flux. 
Because the background levels are real counts and not the detector's bias level, we add the background back after subtracting the 1/$f$ noise in order to ensure that the \texttt{jwst} pipeline's bias correction and up-the-ramp fitting procedures are unaffected. 
We also apply an integration-level 1/$f$ subtraction and clean $>5\sigma$ outlier bad/hot pixels.

In Stage 3, we apply a $>15\sigma$ temporal outlier cut, remove the background for the final time, and perform an additional 1/$f$ noise removal at the integration level. 
Both the background removal and 1/$f$ noise subtraction follow the same algorithm as the group level. 
To extract the white light curves, we sum all the flux in order 1 and the flux from wavelengths within $[0.6, 0.85)$ $\mu$m for order 2. 
We align frames and conduct box extractions on each order with aperture widths that minimize the scatter of the white light curve for each order.

\begin{figure*}[t!]
    \centering
    \includegraphics[width=\linewidth]{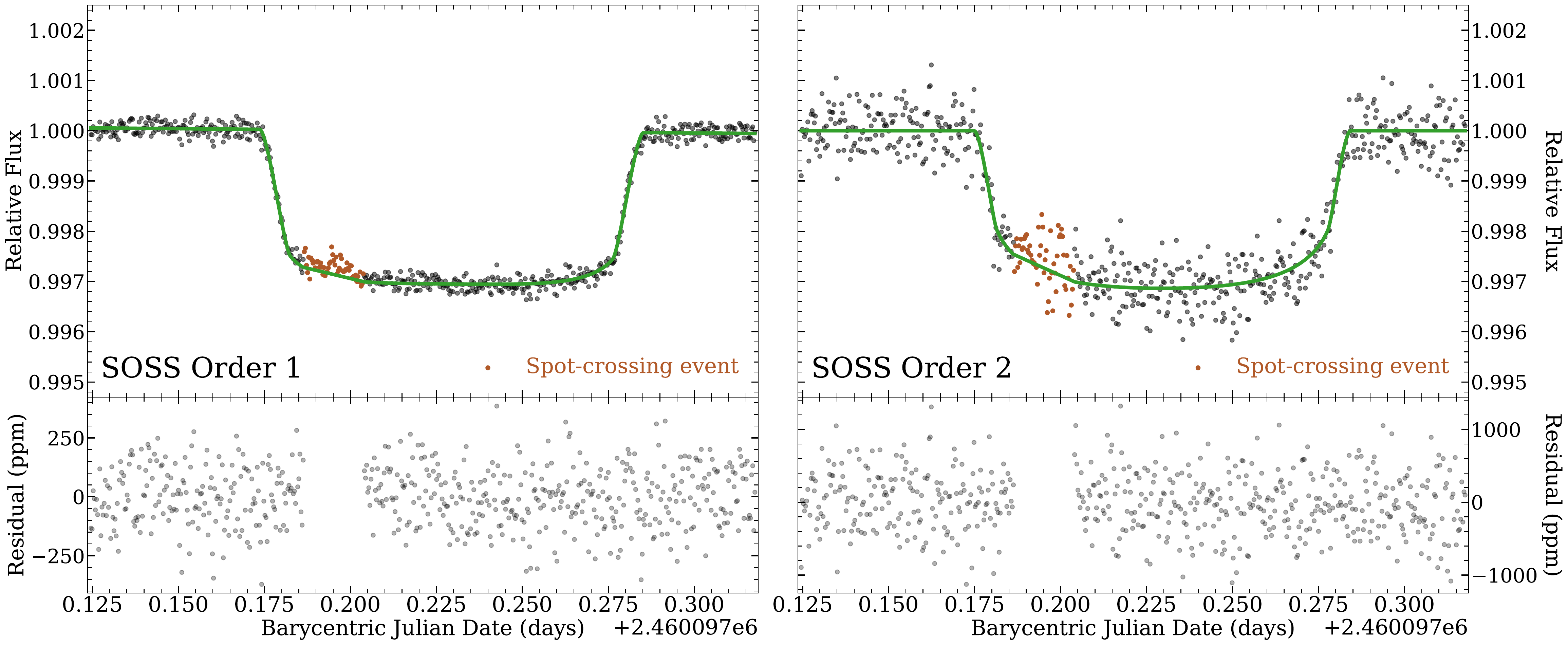}
    \caption{JWST NIRISS SOSS data and white light curve fit for K2-18\,b from the \texttt{FIREFLy} reduction. Left: first order data. Right: second order data. The photometric data points (points, top panels) are compared with the best fit light curve (green lines), resulting in a 124\,ppm residual scatter for order 1 and 414\,ppm for order 2 (bottom panels). Integrations during a spot-crossing event are highlighted (brown points), which are trimmed in the white and spectroscopic light curve fits.
    }
    \label{fig:fireflynirisswlc}
\end{figure*}

Moving on to light curve fitting, we trim out integrations corresponding to a spot-crossing event in the NIRISS light curve. 
Informed by the Bayesian Information Criterion (BIC), we choose the systematics vector to be composed solely of a linear trend for both order 1 and order 2. 
We find that our fitted values are consistent with those reported by \citet{Mad23}; we therefore fix the orbital parameters to theirs for a consistent comparison in both the white light curve and spectrophotometric fits.
For limb darkening, we perform a two-step process of first fitting for spectroscopic limb darkening coefficients and then fixing them to an offset stellar atmosphere model.
We experiment with both the \cite{kipping2013} \{$q_1$, $q_2$\} and the quadratic $u_{+} = u_1 + u_2$ and $u_{-} = u_1 - u_2$ parametrizations, choosing \{$u_+$, $u_-$\} as $q_2$ in particular tends to be uninformative about light curve shape despite both spectra being virtually indistinguishable.
Combining the fitted $u_{+}$ and $u_{-}$ from order 1 and order 2, we find that these fitted limb-darkening coefficients trace well the limb darkening coefficients from the Stagger grid of 3D stellar atmosphere models, with an offset of $-0.14843$ for $u_{+}$ and $-0.07728$ for $u_{-}$.
We show the white light curve fits for the NIRISS SOSS orders 1 and 2 separately in Figure~\ref{fig:fireflynirisswlc}, with the spot-crossing event highlighted.
We account for red noise in our spectrophotometric fit by inflating error bars in quadrature for wavelength bins where the residual RMS-bin size trend is above log linear \citep{Pont2006, Winn2008}. 
This does not impact the error bars for most points, but for those that are, it is typically by a factor of no more than two. 

\subsection{\texttt{exoTEDRF}}

We also reduced the time series observations using the publicly available \texttt{exoTEDRF} pipeline \citep{Fei23, Rad23, Coulombe2023, Cad24, Radica2024b}, starting from the raw, uncalibrated files available on MAST. 
We closely follow the procedure laid out in \citet{Radica2024a}, and we summarize the most pertinent points here. 
In \texttt{stage1} we correct the column-correlated 1/$f$ noise at the group level, using the \texttt{scale-achromatic} method. 
Before ramp fitting, we perform a time-domain outlier rejection \citep{Radica2024a}, using a threshold of 10$\sigma$. 
We perform a piece-wise correction of the SOSS background \citep{Lim23, Fou24}, using the standard SOSS background model provided by STScI\footnote{\href{https://jwst-docs.stsci.edu/jwst-near-infrared-imager-and-slitless-spectrograph/niriss-observing-strategies/niriss-soss-recommended-strategies\#gsc.tab=0}{https://jwst-docs.stsci.edu/jwst-near-infrared-imager-and-slitless-spectrograph/niriss-observing-strategies/niriss-soss-recommended-strategies\#gsc.tab=0}}, and a pre- and post-step scaling factor of 0.95784 and 0.92449 respectively relative to the background model. 
Finally, we extract the stellar spectra using a simple box aperture extraction with a width of 30 pixels, since the order self-contamination is expected to be negligible \citep{Darveau-Bernier2022, Radica2022a}.

For the light curve analysis, we first construct two separate white light curves by summing all the flux in order 1, and wavelengths $\lambda \in (0.6, 0.85]$ for order 2. 
We then jointly fit the white light curves using the flexible \texttt{exoUPRF} library \citep{Radica2024c} following the procedure in \citet{Radica2024a, Rad24}. 
The fitted light curve model is composed of an astrophysical component (i.e., a \texttt{batman} transit model), and a systematics component. 
For the astrophysical component, we share the orbital parameters (i.e., time of mid-transit, orbital inclination and scaled semi-major axis, as well as the orbital eccentricity and argument of periastron) between the two orders, but separately fit the scaled planet radius and two parameters of the quadratic limb darkening law to each. We put Gaussian priors on the orbital parameters based on the findings from \citet{Mad23}.
The systematics component of our light curve fit consists of a linear trend with time, fit independently to each order, as well as a scalar jitter term added in quadrature to the flux errors. 
Instead of commonly-used prescriptions like \texttt{spotrod} \citep{Beky2014, Mad23, Fou24} to model the spot-crossing event, we simply model it with a Gaussian profile, which has been shown to produce equally accurate fits to spot crossing events in transit light curves (Roy et al.~in prep). 
We fit the amplitude of the Gaussian independently to each order, but share the width and position of the Gaussian between the two. 
In all, our white light fit has 19 free parameters, and we use wide, uninformative priors for each parameter. 
For the fits, we use the affine-invariant sampler \texttt{emcee} with 40 chains and using 50000 steps per chain. We discard the first 80\% as burn-in.

For the spectrophotometric fits, we fix the orbital parameters to the best-fitting values from the white light curve fits.
We test fixing the orbital parameters to the updated solution presented in \citet{Rad22}, but find negligible changes in the resulting transmission spectra. 
We also fix the location and width of the Gaussian spot model to our best-fitting white light curve values, but fit the amplitude for each wavelength bin. 
Finally, we freely fit the quadratic limb darkening parameters, systematic slope, and additive jitter term to each bin, resulting in a total of eight free parameters.

\vspace{0.5cm}

\section{NIRSpec Data Reductions} \label{sec:nirspec}
The K2-18 system was also observed by JWST's Near Infrared Spectrograph \citep[NIRSpec;][]{NIRSpec1, NIRSpec2} instrument using the G395H grating as part of JWST GO Program 2722 \citep[PI: N. Madhusudhan - ][]{K218JWSTData}.
As with NIRISS, the details of these observations are discussed in \citet{Mad23}.
We perform a total of four NIRSpec reductions: one with the \texttt{FIREFLy} pipeline, one with the \texttt{ExoTEDRF} pipeline, and two with the \texttt{Eureka!} pipeline. 
We bin the data in four ways: resolving powers $R\approx100$, $R\approx200$, and $R\approx300$, and at native pixel resolution.
We summarize the most relevant choices made in each reduction, as well as a comparison to the reduction choices made in \citet{Mad23}'s \texttt{JExoRES} NIRSpec reduction, in Table \ref{tab:NIRSpecChoices}. Similarly with NIRISS, we show the point-by point discrepancies between reductions in Appendix \ref{appendix:comparisons}.
Combining these reductions with our NIRISS reductions, we compare the low-resolution data-variants from each reduction code in Figure~\ref{fig:reductions} ($R \approx 25$ for NIRISS; $R \approx 100$ for NIRSpec) and directly compare our pixel-level data, binned to the same resolution as \citet{Mad23}, in Figure~\ref{fig:madhucomparison}.

\begin{deluxetable*}{llllll}[ht!]
\centering
\tabletypesize{\scriptsize}
\tablewidth{\textwidth}
\tablecaption{NIRSpec Reduction Configuration Comparison \label{tab:NIRSpecChoices}}
\tablehead{
Step & \texttt{FIREFLy} (this work) & \texttt{ExoTEDRF} (this work) & \texttt{Eureka!} A (this work) & \texttt{Eureka!} B (this work) & \citet{Mad23}
}
\startdata
\multicolumn{6}{c}{\textbf{Stage 1}}\\
\hline
Saturation & Default & Default & Default & Default & Default \\
Superbias & Default & Default & Default & \texttt{Eureka!} custom & Default \\
Dark current & Skipped & Default & Default & Default & Default \\
Background subtraction & \texttt{FIREFLy} custom & \texttt{exoTEDRF} custom & \texttt{Eureka!} custom & \texttt{Eureka!} custom & \texttt{JExoRES} custom \\
\hspace{1pt} (group level) & & & & & \\
1/$f$ subtraction & \texttt{FIREFLy} custom & \texttt{exoTEDRF} custom & \texttt{Eureka!} custom & \texttt{Eureka!} custom & \texttt{JExoRES} custom \\
\hspace{1pt} (group level) & & & & &\\
Linearity & Default & Default & Default & Default & Default \\
Jump & Skipped & \texttt{exoTEDRF} custom, 12$\sigma$ & Skipped & Default, 15$\sigma$ & Default, 5$\sigma$ \\
 & & in time & & & \\
Ramp fitting & Default & Default & Default & Default & Default \\
\hline
\multicolumn{6}{c}{\textbf{Stage 2}}\\
\hline
Wavelength calibration & Default & Cross correlation with & Default & Default & Default \\
& & \texttt{PHOENIX} model & & & \\
$1/f$ correction & Yes & Yes & No & No & No \\
\hspace{1pt} (integration level) & & & & & \\
\hline
\multicolumn{6}{c}{\textbf{Stage 3}}\\
\hline
Bad pixel mask \& & \texttt{lacosmic} \& known & Outliers $ >10\sigma$ in time & Outliers $> 3\sigma$ in time  & Outliers $>4\sigma$ in time & Data Quality flags only \\
\hspace{1pt} temporal outliers & bad pixels & and $>10\sigma$ in space & and $>5 \times$ median in space & and $>2.5\sigma$ in space & \\
1D Spectrum extraction & Box extraction with & Box extraction with & \citet{horne1986optimalE} spatial & \citet{horne1986optimalE} spatial & \citet{horne1986optimalE} spatial \\
 & 5.2 pixel (for NRS1) \& & \texttt{edgetrigger}-defined & profile-based extraction, & profile-based extraction, & profile-based extraction \\
 & 2.41 pixel (for NRS2) & centroids & 3 pixel half width & 4 pixel half width & with 3 principal \\
 & aperture & & & & component PSF \\
Exclusions & None & $\geq 5\sigma$ outliers & $\geq 4\sigma$ outliers & $\geq 4\sigma$ outliers & Columns with $> 20$\% \\
 & & & & & masked flux \\
\hline
\multicolumn{6}{c}{\textbf{Stage 4}}\\
\hline
Trim & First 575 pixels in & None & None  & First 25 integrations & First 5 minutes \\
 & NRS1; first 8 \& last 18 & & & & \\
 & pixels in  NRS2 & & & & \\
Light curve model & \texttt{batman} & \texttt{batman} & \texttt{batman} & \texttt{batman} & \texttt{SPOTROD} \\
Starspot treatment & None & None & None & None & 4-parameter spot model \\
  & & & & & \\
Limb darkening & Fixed to fitted white & Fit for each bin & \texttt{MPS-ATLAS} model set 2, & \texttt{PHOENIX}~model, & Binned to $R \sim 20$, then \\
 & light curve values for & &calculated with \texttt{exotic-ld}  &  calculated with~\texttt{exotic-ld} & fit in each bin and all \\
 & the respective detector & & & & pixels within the bin \\
LD parameterization & \{$u_+, u_-$\} & \{$u_1, u_2$\} & \{$u_1, u_2$\} & Quadratic & \{$q_1, q_2$\} \\
Systematic trends & Linear & Linear & Linear & Linear in time and linear & Quadratic and linear \\
& & & & correlation with position & \\
& & & & \& spatial PSF width & \\
Code used for white & \texttt{emcee} & \texttt{emcee} & \texttt{emcee} & \texttt{dynesty} & \texttt{MultiNest} \\
\hspace{1pt} light curve posterior & & & & & \\
Fitted WLC $a/R\ast$ & $80.27 \pm 0.98$ & \nodata & $80.68\pm0.10$ & $81.56 \pm 0.85$ & $80.92^{+0.78}_{-0.72}$\\
Fitted WLC $b$ & $0.621 \pm 0.013$ & \nodata & \nodata & \nodata & \nodata\\
Fitted WLC $i$ ($^{\circ}$) & \nodata & \nodata & $89.564\pm0.002$ & 89.577 $\pm$ 0.013 & $89.567^{+0.012}_{-0.011}$\\
Fitted WLC $T_0$ &  $59964.96946 \pm 0.00004$ & 57264.3914 & $59964.96947\pm0.00006$ &  $59964.969451 \pm 0.000034$ & $59964.969453^{+0.00035}_{-0.00034}$\\
Binning method & Pixel-level, $R\approx100$, & Pixel-level, $R\approx100$, & Pixel-level, $R\approx100$, & Pixel-level, $R\approx100$, & Pixel-level \\
 & $R\approx200$, and $R\approx300$ & $R\approx200$, and $R\approx300$ & $R\approx200$, and $R\approx300$ & $R\approx200$, and $R\approx300$ & \\
Code used for spectro- & Levenberg-Marquardt & MCMC; \texttt{emcee} & MCMC; \texttt{emcee} & Nested Sampling; \texttt{dynesty} & Levenberg-Marquardt \\
\hspace{1pt} -photometric fit & algorithm; \texttt{lmfit} & & & & algorithm \\
Error bar inflation & Red noise via residual  & Jitter term & White noise term & White noise term & ? \\
 &  RMS-bin size trend & & & & \\
Code Availability & Not Public & \href{https://github.com/radicamc/exoTEDRF}{\texttt{exoTEDRF} GitHub} & \href{https://github.com/kevin218/Eureka}{\eureka GitHub} & \href{https://github.com/kevin218/Eureka}{\eureka GitHub} & Not Public \\
\enddata
\tablecomments{``Default'' refers to the default handling of the step in the \texttt{jwst} pipeline. The system parameters ($a/R_\ast$, $i$, and $T_0$) are the same for each reduction, and are calculated in \citet{Mad23}. The listed orbital parameters were used to either confirm consistency with \citet{Mad23} (\texttt{FIREFLy}, \texttt{Eureka!} B) or to show negligible differences between resulting spectra with an alternative orbital solution \citep[\texttt{exoTEDRF}, from][]{Rad22}. Note that this is not an exhaustive list of reduction steps. `?' indicates reduction configuration settings not mentioned in \citet{Mad23}.}
\end{deluxetable*}

\subsection{\texttt{FIREFLy}}

\begin{figure*}[t!]
    \centering
    \includegraphics[width=\linewidth]{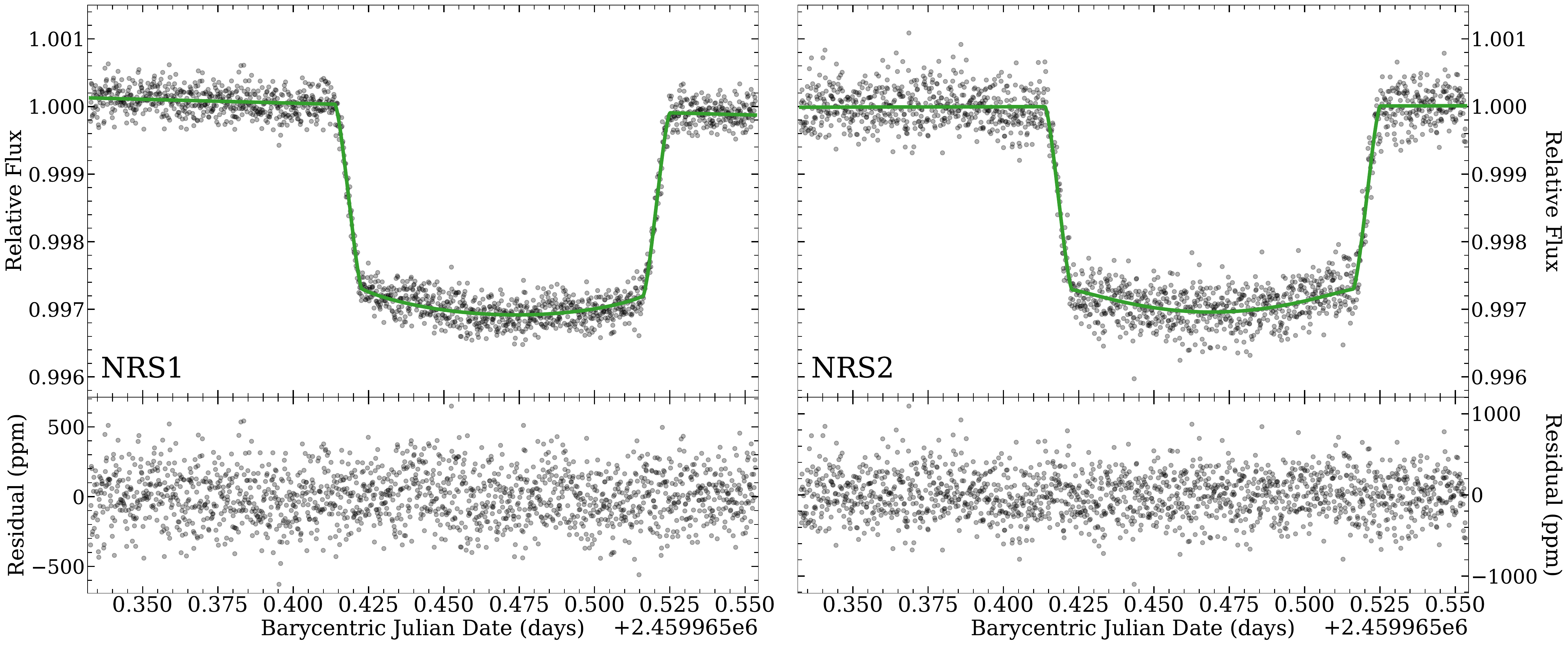}
    \caption{JWST NIRSpec G395H data and white light curve fit for K2-18\,b from the \texttt{FIREFLy} reduction.  Left: NRS1 detector data. Right: NRS2 detector data.  The photometric data points (points, top panels) are compared with the best fit light curve (green lines), resulting in a 180\,ppm residual scatter for NRS1 and 269\,ppm for NRS2 (bottom panels).
    }
    \label{fig:fireflynirspecwlc}
\end{figure*}

We also used the \texttt{FIREFLy} pipeline to conduct a full time-series analysis of the K2-18 NIRSpec/G395H data.
We again list a more thorough inventory of the steps in Table \ref{tab:NIRSpecChoices}, describing here the most pertinent choices made.
We skip the jump step in Stage 1 to avoid cosmic ray false positives due to a less well-constrained ramp linear slope, as typically occurs with data sets with fewer than 25 groups per integration.
In Stage 3, we clean bad pixels by flagging ones with sharp variance spikes of over 100$\sigma$ using \texttt{lacosmic} \citep{lacosmic} and other known bad pixels in NIRSpec G395H as the first part of the stellar extraction. 
Our box extraction's aperture full-widths were selected as they minimize the scatter of the out-of-transit white light curve.

In the light curve fitting step, we trim the first 575 pixel columns of NRS1, first 8 pixel columns of NRS2, and last 18 pixel columns of NRS2. 
We find that our fitted weighted average orbital parameters are consistent with those reported by \citet{Mad23}; we therefore fix the orbital parameters to theirs for our white light curve fit.
For consistency with \citet{Mad23}, we experiment with including both a linear and quadratic term in time, but we find that only a linear term is favored; we therefore choose to only include a linear term for our systematics model. 
We show the results of our white light curve fit in Figure~\ref{fig:fireflynirspecwlc}.

\begin{figure*}[t!]
    \centering
    \vspace{0.1cm}
    \includegraphics[width=\linewidth]{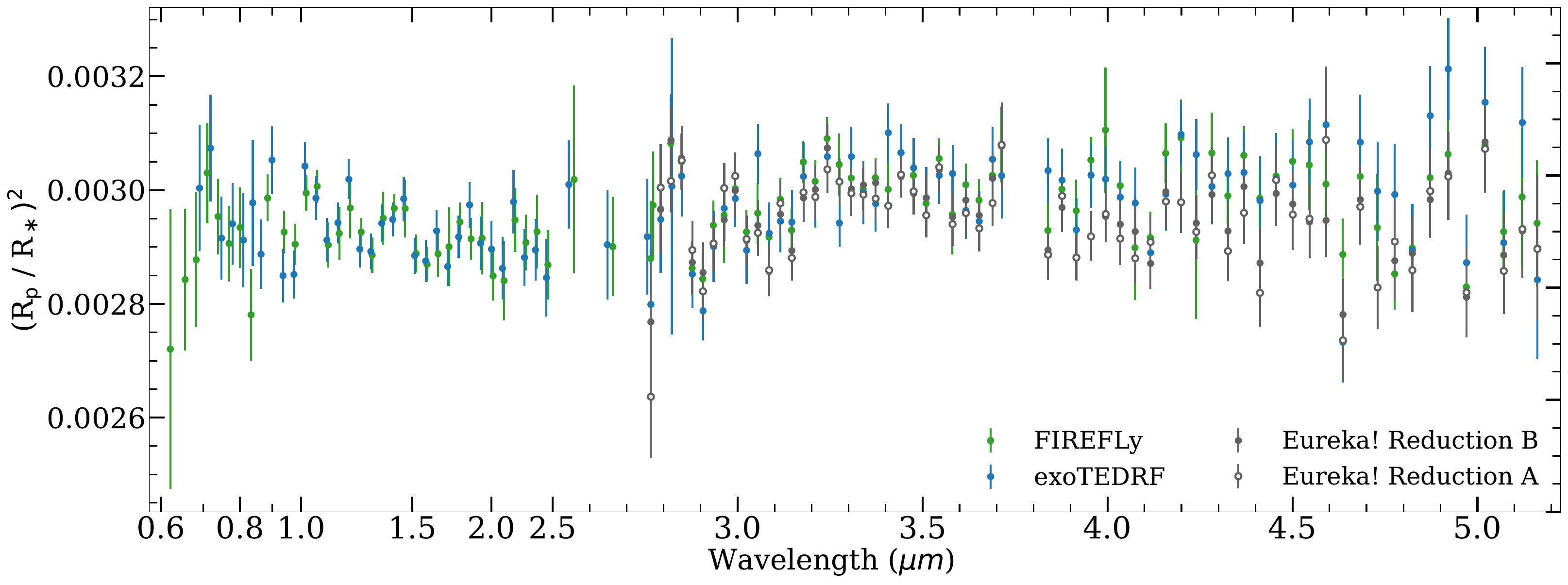}
    \caption{
    Four independently-reduced JWST transmission spectra of K2-18\,b. NIRISS SOSS and NIRSpec G395H data sets reduced using the \texttt{FIREFLy} pipeline (green error bars; \citealt[][Liu \& Wang et al.~in prep, Wang et al. in prep]{Rus22, Rus23}) are compared with the same data reduced with the \texttt{exoTEDRF} pipeline (blue error bars; \citealt{Radica2024c, Rad23, Fei23}) and two NIRSpec G395H reductions using the \texttt{Eureka!} pipeline (error bars with grey (B)/white (A) central points; \citealt{Eureka!}). NIRISS data are shown at $R\approx25$ and NIRSpec data are shown at $R\approx100$. The four reductions demonstrate broad agreement.}
    \label{fig:reductions}
\end{figure*}

For our spectrophotometric fit, we fix the limb darkening parameters and linear systematic term to the fitted white light curve values in addition to the already fixed orbital parameters. 
We elect to fix the spectrophotometric limb darkening values to the white light curve values as limb darkening is not expected to be strongly wavelength dependent at the redder wavelengths of NIRSpec (\citetalias{MayMac23} \citeyear{MayMac23}; \citetalias{Mor23} \citeyear{Mor23}). 
We account for red noise in our spectrophotometric fit by inflating error bars in quadrature for wavelength bins where the residual RMS-bin size trend is above log linear.  
For most points this does not result in error bar inflation, and for points that are inflated, it is typically by a factor of no more than two. 
To demonstrate the fidelity of our fitting procedure given that \texttt{FIREFLy} is proprietary software, we show all of the spectroscopic light curve fits and residuals for each bin of our $R\approx100$ reduction in Appendix~\ref{appendix:firefly_spectroresiduals}. 

\subsection{\texttt{exoTEDRF}} 

To reduce the NIRSpec/G395H observations, we also use the \texttt{exoTEDRF} pipeline \citep{Rad23, Fei23, Radica2024b}, which has recently been updated to support NIRSpec observations \citep[][Ahrer et al.~submitted]{Radica2024b}. We start from the raw, uncalibrated data files and follow the procedure outlined in Ahrer et al.~(submitted). For completeness, we summarize the key points here. We perform standard saturation and superbias corrections on the data frames \citep[e.g.,][]{Ald23} before correcting the background and 1/$f$ noise at the group-level, using the \texttt{median} method and a trace mask width of 16 pixels. As with SOSS, we perform a time-domain outlier flagging \citep{Radica2024a} with a rejection threshold of 12$\sigma$. We then repeat the background and 1/$f$ correction at the integration-level, after ramp fitting, in order to remove any remaining traces of the background flux. We then interpolate all pixels which have a non-zero data quality flag, or are flagged as 10$\sigma$ temporal outliers using a median of the surrounding pixels in space and time, respectively. For the spectral extraction, we use the \texttt{edgetrigger} algorithm \citep{Radica2022a} to determine the centroids of the NRS1 and NRS2 spectral traces, and then perform a box aperture extraction with a width of 8 pixels.

We follow the same light curve fitting procedure as with the SOSS light curves --- jointly fitting the NRS1 and NRS2 white light curves, sharing the planet's orbital parameters between the two and independently fitting for chromatic parameters like the scaled planet radius, limb darkening and systematics. As with SOSS, we fix the orbital parameters to the values used by \citet{Mad23} (there is, once again, no difference in the resulting spectra using the \citet{Rad22} orbital solution). There is no spot crossing in the NIRSpec light curves, and so we do not include a Gaussian spot model. However, we keep the linear slope and error inflation term added in quadrature to the flux errors that were included in the SOSS fits. In all, our white light curve fits have 15 free parameters. We again use \texttt{emcee} for the fits, using the same amount of chains, steps, and burn in as above.

\subsection{\texttt{Eureka!} Reductions}

We conduct two independent reductions of the NIRSpec data using the open-source python package \eureka \citep{Eureka!} to test the extent to which the reduction choices within the same pipeline can yield differing final results. The \eureka control and parameter files we used for each reduction, allowing replication of our results, are available for download on Zenodo: \href{https://doi.org/10.5281/zenodo.14735688}{doi:10.5281/zenodo.14735688}.

\begin{figure*}[ht!]
    \centering
    \vspace{0.2cm}
    \includegraphics[width=\linewidth]{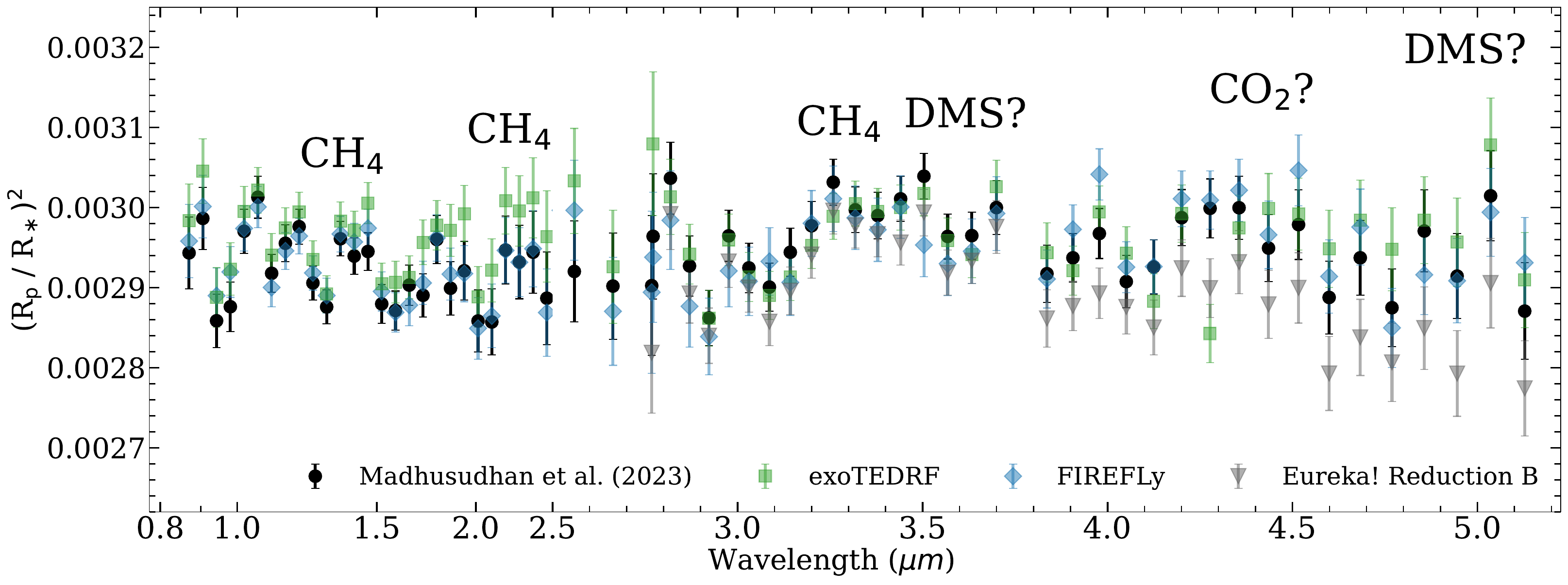}
    \caption{Comparison between our transmission spectra and the reduction from \citet{Mad23}. Three of our reductions, which were performed at the pixel level and then binned to $R\approx55$, are shown: \texttt{exoTEDRF} (green squares), \texttt{FIREFLy} (blue diamonds), and the NIRSpec-only \eureka reduction B (gray downward-pointing triangles). We show only \eureka reduction B, since reduction A was performed at $R = 100$. We compare these binned pixel-level data to the low-resolution data from \citet{Mad23}, which were obtained using the \texttt{JExoRES} pipeline (black error bars; \citealt{Hol23}). The \citet{Mad23} data were binned directly from their pixel-level data, so we bin our pixel-level data using the same binning approach. These $R \approx 55$ datasets are purely for visual comparison with \citet{Mad23}, and are not used for retrievals later in our study. Wavelength locations of CH$_4$, CO$_2$, and DMS absorption featured are annotated.
    }
    \label{fig:madhucomparison}
\end{figure*}

For our first \eureka reduction, which we henceforth refer to as reduction ``A'', we start with the uncalibrated files and run Stage 1 and Stage 2 which are wrapped around the \texttt{jwst} pipeline. We use the default steps in both stages with the exception of the jump step and the photom step, which we skip. We also use \texttt{Eureka!}'s group-level background subtraction and a custom bias scale factor (using a smoothing filter with a window length of 30 integrations and group 1) in Stage 1\footnote{See also \citetalias{Mor23} \citeyear{Mor23}}.

In Stage 3 we extract the stellar time-series spectra. First, we correct the curved trace of the spectra and perform a background subtraction where we used the median of the entire frame, excluding the area within 5 pixels of the center of the trace. Prior to the calculation of the median background, we perform two iterations of outlier masking using thresholds of 3$\sigma$ along the time and 5 times the median along the spatial axis. We use an aperture of 7 pixels, corresponding to a 3-pixel half width, to extract the time-series spectra using the optimal extraction method \citep{horne1986optimalE}.

We generate pixel-level and binned light curves in Stage 4 of \eureka following the same binning scheme as \texttt{FIREFLy}. We use a 3$\sigma$-clipping compared to a 10-pixel-rolling-median to reduce the number of outliers. We fit the light curves by fixing the system parameters to those reported in \citet{Mad23} to allow for direct comparison, and we used the MCMC python package \texttt{emcee}. We opt to use a quadratic limb-darkening law, where we fix one parameter ($u_1$) to limb darkening parameters generated using the \texttt{exotic-ld} Python package \citep{grant2024exoticLDJoss} and the \texttt{MPS-ATLAS} (set 2) stellar atmosphere model grid \citep{Kostogryz2023mpsatlas2}. We use a simple linear trend in time to account for a systematic trend in our light curves.  

Our second \eureka reduction, which we will henceforth refer to as reduction ``B'', used \eureka version 0.11.dev446+gf5d684ee.d20240712; we started with the \textunderscore uncal files produced with SDP\textunderscore VER 2023\textunderscore 3b, and we used version 1.15.1 of the \texttt{jwst} pipeline with CRDS version 11.17.26 and CRDS context \texttt{jwst\textunderscore 1252}. Our reduction procedure largely followed the \eureka reduction described in \citet{schlawin2024gj1214}; we summarize the important differences between this \eureka reduction and that of \citet{schlawin2024gj1214} below.

We increased the Stage 1 jump step rejection threshold to 15 (instead of 6). In Stage 3, we 4$\sigma$-clipped background pixels along the time axis (instead of 10$\sigma$) and subtracted the mean background flux per column and per integration computed using only the pixels $\geq$6 px away from the center of the spectral trace (instead of 7 px). Our optimal spectral extraction \citep{horne1986optimalE} used a spatial profile with a half-width of 4 px (instead of 5 px). In Stage 4, we manually masked several pixel columns (18 in NRS1 and 9 in NRS2) that exhibited unusually high noise levels compared to their neighbors, likely due to unmasked bad pixels in an earlier stage. We then spectrally binned the data into multiple different spectral resolutions and masked any integration that was a $\geq$4$\sigma$ outlier (instead of 3.5$\sigma$) with respect to a 20-integration wide boxcar filter, likely due to a missed cosmic ray.

When fitting the light curves, our astrophysical model consisted of a \texttt{batman} transit model with quadratic limb-darkening coefficients fixed to a \texttt{PHOENIX} limb-darkening model computed by \texttt{exotic-ld} \citep{husser2013phoenix,grant2024exoticLDJoss}. We first verified that our broadband light curves were consistent with the orbital parameters reported by \citet{Mad23}, with our joint fit of the NRS1 and NRS2 broadband lightcurves finding $T_0 = 59964.969451 \pm 0.000034$ BMJD\textunderscore TDB, $i = 89.577^{\circ} \pm 0.013^{\circ}$, and $a/R_* = 81.56 \pm 0.85$ (we had fixed the orbital period to that of \citealt{Benneke2019}). We then fixed our orbital parameters to those of \citet{Mad23}. Our systematic model consisted of a linear trend in time and a linear correlation with the position and the width of the spatial PSF. We also fitted for a white noise multiplier to account for any excess white noise in the data beyond the estimated photon noise. We trimmed the first 25 integrations to remove any initial detector settling. To estimate the best-fit model values and their corresponding uncertainties, we used the \texttt{dynesty} nested sampling algorithm \citep{speagle2020dynesty} with 121 live points, `multi' bounds, the `rwalk' sampling algorithm, and a convergence criterion of $\Delta\log\mathcal{Z}\leq0.01$, where $\mathcal{Z}$ is the Bayesian evidence.

\section{Atmospheric Retrieval Analysis of K2-18 \texorpdfstring{\lowercase{b's}}{b's} Transmission Spectrum} \label{sec:retrievals}

Here we detail our atmospheric interpretation of K2-18\,b's JWST transmission spectrum. Our central goal is to quantify the sensitivity of atmospheric inferences to both data-level and model-level choices, establishing which atmospheric properties are robust and which are not. To this end, we conduct an extensive atmospheric retrieval analysis across the 60 possible combinations of re-analyzed NIRISS and NIRSpec data produced in this study. We use two independent retrieval codes to verify the robustness of our results. In total, our analysis comprises over 250 retrievals. We describe our retrieval configurations in Section~\ref{sec:retrieval_config} and our results in Section~\ref{sec:retrieval_results}.

\subsection{Retrieval Configuration} \label{sec:retrieval_config}

We employ two open source retrieval codes, \texttt{POSEIDON} and \texttt{BeAR}, to explore the range of atmospheric properties consistent with the revised K2-18\,b's transmission spectra derived above. We additionally show retrieval results using the data from \citet{Mad23} in Appendix~\ref{appendix:Madhu_reproduction}. Our retrieval configurations are constructed to encompass the parameter space explored by \citet{Mad23}, while also considering additional factors (such as planetary mass uncertainties and including the 2$^{\rm{nd}}$ order NIRISS SOSS data).
Our model settings, retrieval configurations\footnote{In this case, the ``model'' describes the structure of the planet's atmosphere in the context of an observed transmission spectrum, with different potential ``assumptions/settings'' made for it. These include which potential aspects of the model to consider, the preferred parameterization of each component of the model, which parameters are fixed as well as the values they are fixed to, and the priors on each free parameter in the model. On the other hand, the ``retrieval framework'' describes how the model is fit to the data. This includes the way that the planet's atmosphere and transmission through it are numerically modeled, the data used in these numerical models, and the code used to explore the parameter space. The combination of these is what we describe as the overall ``retrieval configuration.''}, and a comparison with \citet{Mad23} are summarized in Table~\ref{tab:retrieval_configurations}.

\begin{deluxetable*}{lccc}
\centering
\tablewidth{0pt}
\tablecaption{Atmospheric Retrieval Configuration Comparison}
\tablehead{
Model Setting & \colhead{\texttt{POSEIDON} (this work)} & \colhead{\texttt{BeAR} (this work)} & \colhead{\texttt{AURA} \citep{Mad23}}
}
\startdata
    \multicolumn{4}{c}{\textbf{System Properties}} \\
    Stellar Radius & 0.4445\,$R_{\odot}$ & 0.41\,$R_{\odot}$ & 0.4445\,$R_{\odot}$ \\
    \hspace{1pt} (citation) & \citet{Ben19} & \citet{Ben17} & \citet{Ben19} \\
    Planetary Radius & Fit (from $\log_{\rm{10}} P_{\rm{ref}}$) & Fit (free parameter) & Fit (from $\log_{\rm{10}} P_{\rm{ref}}$) \\
    Planetary Mass & Fit (free parameter) & Fit (from $\log_{\rm{10}} g_{\rm{p}}$) & 8.63\,$M_{\Earth}$ \\
    Planetary Gravity & Fit (from $M_{\rm{p}}$) & Fit (free parameter) & 12.4\,ms$^{-2}$ \\
    \hline
    \multicolumn{4}{c}{\textbf{Atmospheric Model}}\\
    Pressure Grid & $10^{-8}$--10\,bar & $10^{-8}$--10\,bar & ? \\
    Number of Layers & 100 & 200 & ? \\
    Background Gas & H$_2$ + He & H$_2$ + He & H$_2$ + He \\
    Molecules Included & H$_2$O, CH$_4$, NH$_3$, HCN, CO, CO$_2$ & H$_2$O, CH$_4$, NH$_3$, HCN, CO, CO$_2$ & H$_2$O, CH$_4$, NH$_3$, HCN, CO, CO$_2$ \\
    & DMS, CS$_2$, CH$_3$Cl, OCS, N$_2$O & DMS, CS$_2$, CH$_3$Cl, OCS & DMS, CS$_2$, CH$_3$Cl, OCS, N$_2$O \\
    He/H$_2$ Ratio & 0.17 & 0.17 & ? \\
    P-T Profile Treatment & Free Profile & Isotherm & Free Profile \\
    \hspace{1pt} (citation) & \citep{Madhusudhan2009} & --- & \citep{Madhusudhan2009} \\
    Cloud Treatment & Patchy Clouds + Haze & Opaque Cloud & Patchy Clouds + Haze \\
    \hspace{1pt} (citation) & \citep{MacDonald2017} & --- & \citep{MacDonald2017} \\
    Hydrostatic Boundary Condition & $\log_{\rm{10}} P_{\rm{ref}}$ & $R_{\mathrm{p, \, ref}}$ & $\log_{\rm{10}} P_{\rm{ref}}$ \\
    \hline
    \multicolumn{4}{c}{\textbf{Spectral Model}}\\
    Model Wavelength Grid & 0.58--5.3\,$\micron$ & 0.5996--5.176\,$\micron$ & ? \\
    Native Opacity Resolution & 0.01\,cm$^{-1}$ & 0.01\,cm$^{-1}$ & ? \\
    Opacity Sampling Resolution & $R = $ 100,000 & $R = $ 60,000 & ? \\
    CH$_4$ Line List & MM & YT10to10 & Ames-2016 \\
    \hspace{1pt} (citation) & \citep{Yurchenko2024} & \citep{yurchenko14} & \citep{Huang2013,Huang2017} \\
    CO$_2$ Line List & UCL-4000 & UCL-4000 & HITEMP \\
    \hspace{1pt} (citation) & \citep{yurchenko20} & \citep{yurchenko20} & \citep{Hargreaves2020} \\
    DMS Cross Section & HITRAN & HITRAN & HITRAN \\
    \hspace{1pt} (citation) & \citet{sharpe04} & \citet{sharpe04} & \citet{sharpe04} \\
    \hline
    \multicolumn{4}{c}{\textbf{Stellar Contamination Model}}\\
    Stellar Heterogeneity Treatment & One-het & Two-het & None$^{\dagger}$ \\
    Stellar Heterogeneity Parameters & $T_{\rm{phot}}$, $T_{\rm{het}}$, $f_{\rm{het}}$ & $T_{\rm{phot}}$, $\Delta T_{\rm{spot}}$, $\Delta T_{\rm{fac}}$, $f_{\rm{spot}}$, $f_{\rm{fac}}$ & ---$^{\dagger}$ \\
    Stellar Model Grid & \texttt{PHOENIX}  & blackbody & ---$^{\dagger}$ \\
    Stellar Metallicity ([Fe/H]) & -0.08 & --- & ---$^{\dagger}$ \\
    Stellar Surface Gravity ($\log_{\rm{10}} g_{\rm{*}}$) & 4.93 (cgs) & --- & ---$^{\dagger}$ \\
    \hspace{1pt} (citation) & \citep{Tabernero2024} & --- & ---$^{\dagger}$ \\
    \hline
    \multicolumn{4}{c}{\textbf{Retrieval Priors and Settings}}\\
    Planetary Mass ($M_{\rm{p}}$) & $\mathcal{N}(8.63, 1.35^2)\,M_{\Earth}$ & --- & 8.63\,$M_{\Earth}$ (fixed) \\
    Planetary Gravity ($\log_{\rm{10}} g_{\rm{p}}$) & --- & $\mathcal{N}(3.19, 0.17^2)$ (cgs) & 3.09 (fixed; cgs) \\
    Atmospheric Temperature ($T_{\mathrm{ref}}$) & $\mathcal{U}(0, 800)$\,K & $\mathcal{U}(50, 1000)$\,K & $\mathcal{U}(0, 800)$\,K \\
    P-T Profile Curvature 1 ($\alpha_{1}$) & $\mathcal{U}(0.02, 2.00)$\,K$^{-\frac{1}{2}}$ & --- & $\mathcal{U}(0.02, 2.00)$\,K$^{-\frac{1}{2}}$ \\
    P-T Profile Curvature 2 ($\alpha_{2}$) & $\mathcal{U}(0.02, 2.00)$\,K$^{-\frac{1}{2}}$ & --- & $\mathcal{U}(0.02, 2.00)$\,K$^{-\frac{1}{2}}$ \\
    P-T Profile Region 1 ($\log_{10} (P_{1}$ / bar)) & $\mathcal{U}(-8, 1)$ & --- & $\mathcal{U}(-6, 0)$ \\
    P-T Profile Region 2 ($\log_{10} (P_{2}$ / bar)) & $\mathcal{U}(-8, 1)$ & --- & $\mathcal{U}(-6, 0)$ \\
    P-T Profile Region 3 ($\log_{10} (P_{3}$ / bar)) & $\mathcal{U}(-2, 1)$ & --- & $\mathcal{U}(-2, 0)$ \\
    Molecule Volume Mixing Ratio ($\log_{\rm{10}} X_i$) & $\mathcal{U}(-12, -0.3)$ & $\mathcal{U}(-12, -0.522)$ & $\mathcal{U}(-12, -0.3)$ \\
    Reference Radius ($R_{\mathrm{p, \, ref}}$) & 2.61\,$R_{\Earth}$ (fixed) & $\mathcal{U}(2.0, 2.8)\,R_{\Earth}$ & 2.61\,$R_{\Earth}$ (fixed) \\
    Reference Pressure ($\log_{\rm{10}} (P_{\rm{ref}}$ / bar)) & $\mathcal{U}(-6, 0)$ & 1.0 (fixed) & $\mathcal{U}(-6, 0)$ \\
    Cloud Pressure ($\log_{\rm{10}} (P_{\rm{cloud}}$ / bar)) & $\mathcal{U}(-8, 1)$ & $\mathcal{U}(-7, 0)$ & $\mathcal{U}(-6, 1)$ \\
    Cloud Fraction ($\bar{\phi}$) & $\mathcal{U}(0, 1)$ & --- & $\mathcal{U}(0, 1)$ \\
    Haze Rayleigh Enhancement ($\log_{10} a$) & $\mathcal{U}(-4, 10)$ & --- & $\mathcal{U}(-4, 10)$ \\
    Haze Slope ($\gamma$) & $\mathcal{U}(-20, 2)$ & --- & $\mathcal{U}(-20, 2)$ \\
    Data Offset 1 ($\delta_{\mathrm{rel, \, 1}}$) & $\mathcal{U}(-100, 100)$\,ppm & $\mathcal{U}(-100, 100)$\,ppm & $\mathcal{U}(-100, 100)$\,ppm \\
    Data Offset 2 ($\delta_{\mathrm{rel, \, 2}}$) & $\mathcal{U}(-100, 100)$\,ppm & $\mathcal{U}(-100, 100)$\,ppm & $\mathcal{U}(-100, 100)$\,ppm \\
    $T_{\rm{phot}}$ & $\mathcal{N}(3563, 25^{2})$\,K & $\mathcal{N}(3457, 39^{2})$\,K & ---$^{\dagger}$ \\
    $T_{\rm{het}}$ & $\mathcal{U}(2300, 4275.6)$\,K & --- & ---$^{\dagger}$ \\
    $f_{\rm{het}}$ & $\mathcal{U}(0, 0.5)$ & --- & ---$^{\dagger}$ \\
    $\Delta T_{\rm{spot}}$ & --- & $\mathcal{U}(-1500, 0)$\,K & --- \\
    $\Delta T_{\rm{fac}}$ & --- & $\mathcal{U}(0, 1000)$\,K & --- \\
    $f_{\rm{spot}}$ & --- & $\mathcal{U}(0, 0.5)$ & --- \\
    $f_{\rm{fac}}$ & --- & $\mathcal{U}(0, 0.5)$ & --- \\
    \texttt{MultiNest} Live Points & 1000 & 1000 & ? \\
    Retrieval Code Availability & \href{https://github.com/MartianColonist/POSEIDON}{\texttt{POSEIDON} GitHub} & \href{https://github.com/NewStrangeWorlds/BeAR}{\texttt{BeAR} GitHub} & Not Public \\
    \hline
\enddata
\tablecomments{`?' indicates model settings not specified in \citet{Mad23}. Line list references for other molecules included in each retrieval code are provided in Sections~\ref{sec:POSEIDON} and \ref{sec:BeAR}. $^{\dagger}$~All the results presented in \citet{Mad23} did not include stellar contamination, but the study stated they ran additional (not shown) retrievals considering a one-heterogeneity model. Gaussian priors are summarized as $\mathcal{N}(\mu, \sigma^2)$, where $\mu$ and $\sigma$ are the mean and standard deviation, respectively. $T_{\mathrm{ref}}$ refers to the top-of-atmosphere temperature for \texttt{POSEIDON} and \citet{Mad23}, while for \texttt{BeAR} it represents the isothermal temperature.}
\label{tab:retrieval_configurations}
\end{deluxetable*}

\subsubsection{\texttt{POSEIDON}} \label{sec:POSEIDON}

We first retrieved K2-18\,b's transmission spectrum using the retrieval code \texttt{POSEIDON}\footnote{\url{https://github.com/MartianColonist/POSEIDON}} \citep{MacDonald2017,MacDonald2023}. \texttt{POSEIDON} is widely applied to interpret JWST transmission spectra, including hot Jupiters \citep[e.g.,][]{Grant2023,Fournier-Tondreau2024b}, sub-Neptunes \citep{Pia24}, super-Earths (e.g., \citetalias{Mor23} \citeyear{Mor23}; \citetalias{MayMac23} \citeyear{MayMac23}), and terrestrial worlds \citep[e.g.,][]{Lim23,Cad24}. Here we use \texttt{POSEIDON} to conduct a systematic retrieval survey of the sensitivity of K2-18\,b's atmospheric inferences to 60 different data variants (reduction codes and/or data spectral resolution). Our \texttt{POSEIDON} retrieval configuration is constructed to be as close as possible to that described in \citet{Mad23}, allowing a fair comparison, but we emphasize and justify below several areas where we adopted different model choices.

Our atmosphere model for K2-18\,b assumes a H$_2$-He background gas (with a solar ratio of He/H$_2$ = 0.17; \citealt{asplund09}), commensurate with its planetary radius and mass ($\approx 2.6\,R_{\Earth}$ and $\approx 8.6\,M_{\Earth}$, respectively). We construct atmospheres ranging from $10^{-8}$--10\,bar with 100 layers spaced uniformly in log-pressure under hydrostatic equilibrium. We adopt $10^{-8}$\,bar as the lowest atmosphere pressure (vs. the $10^{-6}$\,bar limit implied by \citealt{Mad23}) to reduce the risk of strong bands saturating the atmosphere. We solve hydrostatic equilibrium using a boundary condition where the freely retrieved reference pressure corresponds to the radial distance where $r = 2.61\,R_{\Earth}$ (the radius from \citealt{Ben19}). The atmospheric pressure-temperature (P-T) profile is freely retrieved following the prescription in \citet{Madhusudhan2009}. We consider the same potential molecular species as \citet{Mad23}, which includes the major O-, C-, and N-bearing species expected at K2-18\,b's equilibrium temperature (255\,K; \citealt{Ben19}): H$_2$O, CH$_4$, NH$_3$, HCN, CO, and CO$_2$; as well as several disequilibrium species proposed as biosignatures in cool sub-Neptunes \citep{Mad21}: C$_2$H$_6$S (DMS), CS$_2$, CH$_3$Cl, OCS, and N$_2$O. We fit the volume mixing ratios of these gases independently, with their number densities then determined via the ideal gas law. Given the large uncertainties on K2-18\,b's mass ($8.63 \pm 1.35\,M_{\Earth}$; \citealt{Clo19}), we prescribe the planet's mass as a free parameter using a Gaussian prior (\citealt{Mad23} fixed K2-18\,b's mass), allowing mass uncertainties to propagate into the retrieved atmospheric composition \citep{Batalha2019a}. Finally, we consider aerosols following the inhomogeneous cloud and haze parameterization of \citet{MacDonald2017}.

We calculate model spectra by solving the equation of radiative transfer in a cylindrical coordinate system for 100 incident stellar rays attenuated by atmospheric opacity along the line of sight. The \texttt{POSEIDON} opacity database recently underwent a major update in \texttt{POSEIDON} v1.2 \citep{Mullens2024}, including a state-of-the-art new ExoMol CH$_4$ line list \citep{Yurchenko2024}. Our \texttt{POSEIDON} opacities are pre-computed at a high resolution ($\Delta\nu$ = 0.01\,cm$^{-1}$, equivalent to $R = \lambda/\Delta\lambda = 10^6$ at 1\,$\micron$) for a grid of temperatures and pressures using the open source Python package \texttt{Cthulhu}\footnote{\url{https://github.com/MartianColonist/Cthulhu}} \citep{Agrawal2024}. Our K2-18\,b retrievals use opacities derived from the following line lists/measured cross sections: H$_2$O (POKAZATEL; \citealt{polyansky18}), CH$_4$ (MM; \citealt{Yurchenko2024}), NH$_3$ (CoYuTe; \citealt{coles19}), HCN (Harris; \citealt{barber14}, CO (Li2015; \citealt{li15}), CO$_2$ (UCL-4000; \citealt{yurchenko20}), C$_2$H$_6$S (DMS) (HITRAN measured xsec; \citealt{sharpe04}) CS$_2$ (HITRAN-2020; \citealt{Gordon2022}), CH$_3$Cl (HITRAN-2020; \citealt{Gordon2022}), OCS (OYT8; \citealt{Owens2024}), and N$_2$O (HITEMP; \citealt{hargreaves19}). We further include continuum collision-induced absorption from H$_2$-H$_2$ and H$_2$-He pairs \citep{Karman2019} and Rayleigh scattering for all gases \citep{MacDonald2022}. We calculate model transmission spectra using opacity sampling on a wavelength grid from 0.58--5.3\,$\micron$ at a spectral resolution of $R =$ 100,000. We adopt this exceptionally high spectral resolution for all our \texttt{POSEIDON} retrievals for three reasons: (i) to ensure negligible opacity sampling errors when interpreting pixel-level data; (ii) to consistently sample cross sections for every data combination (even at lower resolutions); and (iii) to reduce the risk of opacity accuracy differences between our work and \citet{Mad23} (since the latter did not state their assumed sampling resolution). While our $R =$ 100,000 models impose a greater computational burden than is strictly necessary for retrievals of JWST data\footnote{Through the JWST Transiting Exoplanet Early Release Science Program, we found a useful heuristic that $R_{\rm{model}} =$ 20,000 suffices for $R_{\rm{data}}$ = 100 and $R_{\rm{model}} =$ 60,000 suffices for $R_{\rm{data}} \sim$ 3,000 (pixel-level).}, this choice ensures high accuracy in our retrieved atmospheric properties and reduces the risk of an insufficiently high opacity resolution inconsistent with the methodology of \citet{Mad23}.

Our \texttt{POSEIDON} retrievals additionally account for stellar contamination from unocculted active regions, commonly known as the transit light source effect \citep{Rackham2018}. We employ a one-heterogeneity model \citep[e.g.,][]{Rathcke2021,Lim23,Fou24}, defined by the stellar photosphere temperature, the heterogeneity temperature (corresponding to spots/faculae if the heterogeneity is cooler/warmer than the photosphere), and the heterogeneity covering fraction. The stellar contamination contribution is calculated by interpolating \texttt{PHOENIX} models \citep{husser2013phoenix} using the \texttt{PyMSG} package \citep{Townsend2023}. For fixed stellar properties, we adopt the stellar radius from \citet{Ben19} ($R_{*} = 0.4445\,R_{\odot}$) and the stellar surface gravity, metallicity, and effective temperature from \citet{Tabernero2024} ([Fe/H] = $-0.08$, $\log_{10} g_{*}$ = 4.93, $T_{\rm{eff}}$ = 3563\,K). These updated K2-18 stellar properties were published after \citet{Mad23}, so we adopt them for a more accurate characterization of the transit light source effect. 

Our reference \texttt{POSEIDON} retrieval model has 28 free parameters, with priors as given in Table~\ref{tab:retrieval_configurations}. We generally adopt the same priors as \citet{Mad23} for consistency, with a few exceptions: (i) we adjust the pressure parameters describing the P-T profile and cloud-top pressure to span our wider atmospheric pressure range; (ii) we apply a Gaussian prior on the planetary mass, using the standard deviation from \citet{Clo19}; and (iii) we use newly updated constraints on the host star's effective temperature from \citet{Tabernero2024} to set the Gaussian prior on $T_{\rm{phot}}$. We allow for two free offsets, one between the NIRISS SOSS and NIRSpec G395H NRS1 data and one between the NIRISS and NIRSpec G395H NRS2 data, as both \citet{Mad23} and our Appendix~\ref{appendix:Madhu_reproduction} demonstrate the need for at least one free offset. Alongside the reference model, we additionally ran nested retrievals, with CH$_4$, CO$_2$, and DMS removed in turn, to calculate the Bayes factors (and hence the equivalent detection significances) for these molecules. Therefore, we ran 4 retrievals for each combination of NIRISS SOSS (6 variants: 2 reduction pipelines each with 3 resolutions) and NIRSpec G395H data (10 variants: 2 reduction pipelines each with 4 resolutions, and 2 reduction pipelines with 1 resolution), for a total of 240 retrievals spanning the 60 different data combinations. We use \texttt{PyMultiNest} \citep{Feroz2009,Buchner2014} with 1,000 live points to explore the parameter space. Each retrieval took 2--5 days on 24 CPUs --- with longer run-times for pixel-level data variants --- for a total compute time of $\gtrsim$50 CPU years.

\subsubsection{BEAR} \label{sec:BeAR}

To validate our \texttt{POSEIDON} results, we also perform a second set of retrievals using the open-source Bern Atmospheric Retrieval code (\texttt{BeAR}\footnote{\url{https://newstrangeworlds.github.io/BeAR}}). \texttt{BeAR} is an updated version of the retrieval code formerly known as \texttt{Helios-r2}, which was originally developed for retrievals on Brown Dwarf spectra \citep{kitzmann20}. \texttt{BeAR} has been adapted for use on exoplanet spectra \citep{Fisher2024} and has the capabilities to analyze transmission, emission, and phase curve spectra. \texttt{BeAR} uses opacity sampling of line-by-line cross sections to calculate model spectra, coupled to \texttt{PyMultiNest} \citep{Feroz2009,Buchner2014} for statistical exploration.

Our \texttt{BeAR} retrievals use a similar retrieval configuration to that described above, but here we highlight differences in the model setup. First, we assume for simplicity an isothermal temperature throughout the atmosphere (with a uniform prior of 50--1000\,K). The atmosphere is divided into 200 equal layers in log-pressure, ranging from 10--$10^{-8}$\,bar. Second, we fit for the 10\,bar planet radius, rather a reference pressure, as our hydrostatic boundary condition (uniform from 2.0--2.8\,$R_\Earth$). We retrieve the planetary log-surface gravity as a proxy for the mass uncertainties (Gaussian with mean 3.19 and standard deviation 0.17, both in cgs units). Third, we restrict our aerosol model to an opaque cloud deck (with a log-uniform pressure prior of 1.0--$10^{-7}$ bar). Fourth, we sample the opacities at a slightly lower resolution of $R =$ 60,000, due to GPU memory limitations. We include the following molecules in our \texttt{BeAR} retrievals, with their associated line lists: H$_2$O \citep{polyansky18}, CH$_4$ \citep{yurchenko14}, NH$_3$ \citep{coles19}, HCN \citep{harris06,barber14}, CO \citep{li15}, CO$_2$ \citep{yurchenko20}, DMS \citep{sharpe04}, CS$_2$ \citep{sharpe04}, CH$_3$Cl \citep{owens18}, and OCS \citep{wilzewski16}. We also include collision-induced absorption from H$_2$-H$_2$ \citep{abel11} and H$_2$-He \citep{abel12}, alongside Rayleigh scattering from H$_2$ \citep{vardya62}. Our retrievals assume free chemistry, meaning the molecular abundances (aside from H$_2$ and He) are retrieved independently, making no assumptions about chemical processes. The abundances are drawn from log-uniform prior distributions, ranging from volume mixing ratios of $10^{-12}$ to 0.3. The remaining atmosphere is filled with H$_2$ and He, assuming a solar He/H$_2$ ratio of 0.17 \citep{asplund09}. Finally, we include the effects of stellar contamination using a two-heterogeneity model with five free parameters: the stellar photosphere temperature (following a Gaussian prior with a mean of 3457\,K and a standard deviation of 39\,K --- corresponding to the effective temperature values from \citealt{Clo17}); the faculae fraction (uniform from 0--0.5) and temperature difference relative to the photosphere (uniform from 0--1000\,K); and equivalent parameters for spots (with a temperature difference uniform from -1500--0\,K). The stellar contamination is calculated assuming blackbodies for the photosphere and active regions. Alongside the atmospheric and stellar parameters described above, our \texttt{BeAR} retrievals account for possible offsets between different filters and instruments \citep[e.g.,][]{yip21,Ald23} by fitting two offsets --- one between NIRISS SOSS and NIRSpec G395H, and one between the NRS1 and NRS2 filters of NIRSpec G395H (both uniform from $-$100--100\,ppm). Table~\ref{tab:retrieval_configurations} provides a full comparison between our \texttt{BeAR} and \texttt{POSEIDON} retrieval setups, alongside that used in \citet{Mad23}.

We applied our \texttt{BeAR} retrievals to a subset of our re-analyzed NIRISS SOSS and NIRSpec G395H data. Specifically, we focus on low-resolution data combinations using a single reduction code ($R\approx100$ \texttt{FIREFly} and \texttt{exoTEDRF} data for both NIRISS and NIRSpec) and the highest-resolution pixel-level data produced by a common reduction code. Therefore, we considered 4 data combinations with \texttt{BeAR}. For each data combination, we ran 4 retrievals, a reference model and 3 nested models, as above, to calculate detection significances of CH$_4$, CO$_2$ and DMS.

\subsection{Retrieval Results} \label{sec:retrieval_results}

\subsubsection{Molecular Detection Significances} \label{sec:det_significances}

\begin{figure}[ht!]
    \centering
    \includegraphics[width=\columnwidth]{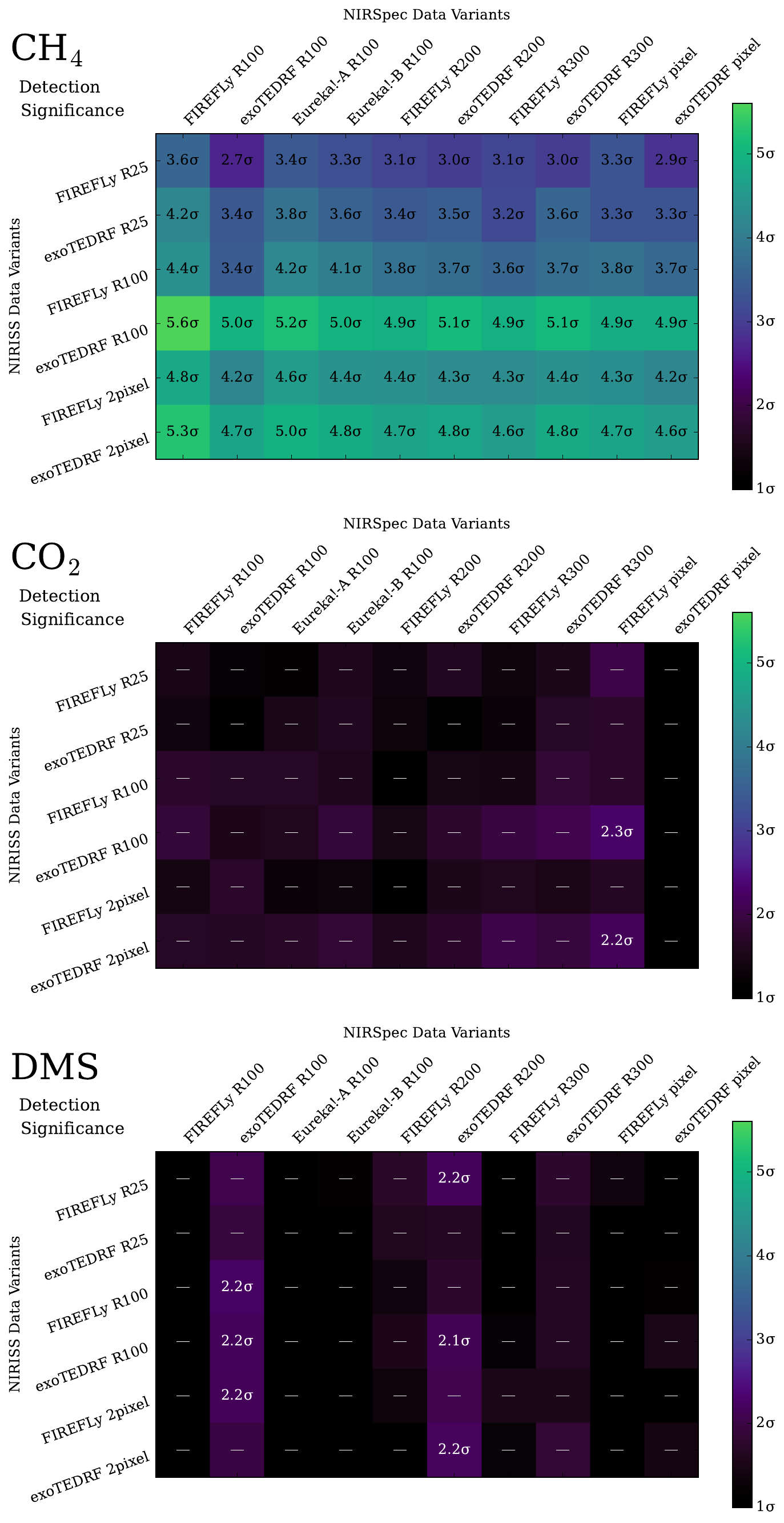}
    \caption{Detection significances of CH$_4$, CO$_2$, and DMS from \texttt{POSEIDON} retrievals of K2-18\,b's JWST NIRISS SOSS + NIRSpec G395H transmission spectrum. Each grid cell represents a Bayesian model comparison between two retrievals, one with and one without a molecule, for a specific NIRISS SOSS and NIRSpec G395H data variant pair. The data variants are ordered with increasing spectral resolution to the right (for NIRSpec) and down (for NIRISS). The cells are colored and annotated according to the detection significance equivalent to the calculated Bayes factor \citep[e.g.,][]{Trotta2008,Benneke2013}. Bayes factors $<3$ (equivalent to $< 2.1\sigma$), corresponding to no evidence on the Jeffreys' scale, are donated by `---'. Our retrievals demonstrate strong evidence for CH$_4$, but find no reliable statistical evidence for CO$_2$ or DMS in K2-18\,b's atmosphere.
    }
    \label{fig:retrieval_detection_significances}
\end{figure}

Our atmospheric retrieval analysis confirms a strong detection of CH$_4$ in K2-18\,b's atmosphere. Figure~\ref{fig:retrieval_detection_significances} shows a grid of 180 Bayesian model comparisons from \texttt{POSEIDON}, colored by the detection significance corresponding to each Bayes factor, spanning every data combination produced in this study. We see that CH$_4$ is detected at $> 2.7\,\sigma$ for all data combinations, with most combinations favoring CH$_4$ at $> 4\,\sigma$. Considering our CH$_4$ significances as independent draws from the distribution of possible data-level choices, we find a detection significance of $4.2^{+0.7}_{-0.9}\,\sigma$. While slightly lower than the $5\,\sigma$ detection reported in \citet{Mad23}, we consider these results a strong confirmation that CH$_4$ is robustly present in K2-18\,b's atmosphere.

Figure~\ref{fig:retrieval_detection_significances} additionally contains a wealth of information pertaining to the sensitivity of CH$_4$ detections to both data resolution and reduction code. First, we see that increasing the NIRISS SOSS data resolution from $R\approx25$ to $R\approx100$ leads to a clear increase in the CH$_4$ detection significances from $\sim 3.3\,\sigma$ to $\sim 4.5\,\sigma$. This behavior is not surprising, since $R\approx25$ NIRISS data makes it harder for retrievals to uniquely resolve CH$_4$ bands from other molecules with similar absorption features (especially H$_2$O), resulting in lower Bayes factors and hence lower significances. We generally find higher CH$_4$ significances for the NIRISS data with \texttt{exoTEDRF} (mean significances of $3.5\,\sigma$, $5.0\,\sigma$, and $4.8\,\sigma$ for the $R\approx25$, $R\approx100$, and pixel-level data, respectively) compared with \texttt{FIREFLy} ($3.1\,\sigma$, $3.8\,\sigma$, and $4.4\,\sigma$, for the same data resolutions). We note that the pixel-level \texttt{exoTEDRF} results are in good agreement with the $R\approx100$ \texttt{exoTEDRF} results, while the $R\approx100$ \texttt{FIREFLy} results lead to notably lower detection significances. A potential reason for the discrepancies at $R\approx100$, but not at pixel-level, could be differences in how each reduction code handles systematics and limb darkening, which manifests more at lower resolutions than at pixel-level.

We do not find any reliable or significant evidence for CO$_2$ or DMS in the observed transmission spectrum of K2-18\,b. Figure~\ref{fig:retrieval_detection_significances} demonstrates that almost every combination of NIRISS and NIRSpec data results in no evidence for CO$_2$ or DMS (Bayes factors $<$ 3, denoted by `---' in Figure~\ref{fig:retrieval_detection_significances}, are ``not worth mentioning'' according to the Jeffreys' scale of Bayesian model comparison --- see \citealt{Trotta2008}). Even choosing the data combination with the maximum Bayes factors results in 4.2 for CO$_2$ ($2.3\,\sigma$) and 3.9 for DMS ($2.2\,\sigma$), which are considered as ``weak evidence at best'' on the Jeffreys' scale. However, one can see in Figure~\ref{fig:retrieval_detection_significances} that it is only the coupling between the \texttt{FIREFLy} pixel-level NIRSpec data and the \texttt{exoTEDRF} $R\approx100$ or 2-pixel NIRISS data that result in any marginal evidence for CO$_2$. Similarly, even the marginal evidence for DMS is only present for $R\approx100$ or $R\approx200$ NIRSpec data from \texttt{exoTEDRF} and disappears for higher data resolutions. Therefore, any marginal evidence for CO$_2$ or DMS depends on either the choice of a specific data reduction code or insufficient data resolution. In Appendix~\ref{appendix:Madhu_reproduction}, we demonstrate that if we run retrievals on the data from \citet{Mad23} then we recover a maximum significance of $2.5\,\sigma$ for CO$_2$ --- this shows that the claimed CO$_2$ detection arises from the specific data treatments in that study. Our results demonstrate that, contrary to the findings of \citet{Mad23}, there is no robust detection of CO$_2$ from K2-18\,b's initial NIRISS SOSS and NIRSpec G395H transmission spectra, nor are there even potential signs of DMS. Figure~\ref{fig:retrieval_detection_significances} offers a clear visual representation of the difference between a robust detection and non-detections.

Our \texttt{BeAR} retrievals similarly find strong detections of CH$_4$ and non-detections of CO$_2$ and DMS. For the \texttt{FIREFly} reductions, for CH$_4$ we obtain detection significances of $3.5\,\sigma$ and $4.5\,\sigma$ for the pixel-level and $R \approx 100$ retrievals, respectively. For CO$_2$ and DMS, the Bayes factor is $<3$ for both resolutions, corresponding to no evidence. The results are extremely similar for the \texttt{exoTEDRF} reductions, for which we obtain $3.8\,\sigma$ and $4.4\,\sigma$ for CH$_4$ for the pixel-level and $R \approx 100$ retrievals, respectively, and Bayes factors $< 3$ for CO$_2$ and DMS. These results are in good agreement with our retrievals using \texttt{POSEIDON} (Figure~\ref{fig:retrieval_detection_significances}), indicating that our conclusions are robust to the choice of retrieval code.

Having established that only CH$_4$ is evidenced by K2-18\,b's JWST NIRISS SOSS and NIRSpec G395H transmission spectrum, we now turn to present abundance constraints consistent with these data.

\subsubsection{Molecular Abundance Constraints} \label{sec:molecule_abundances}

\begin{figure*}[ht!]
    \centering
    \includegraphics[width=0.85\textwidth, trim = 0.0cm 0.5cm 0.0cm 0.5cm]{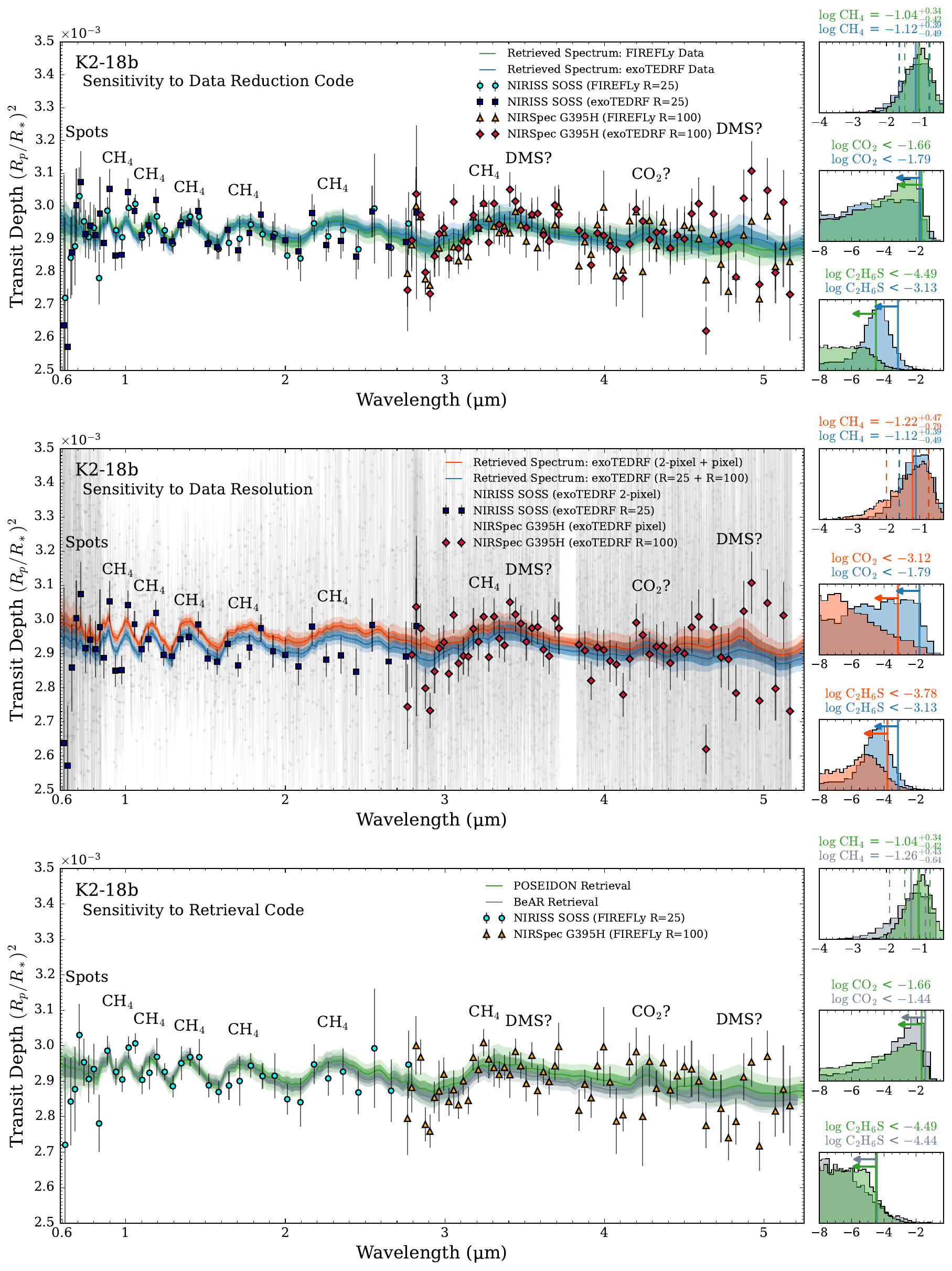}
    \caption{Sensitivity of K2-18\,b's atmospheric composition to different data treatments and retrieval configurations. Top: comparison between \texttt{POSEIDON} retrievals of data reduced by the \texttt{FIREFLy} and \texttt{exoTEDRF} pipelines. Middle: comparison between \texttt{POSEIDON} retrievals of low-resolution (NIRISS SOSS: $R\approx 25$, NIRSpec G395H: $R\approx 100$) and high-resolution (pixel-level) \texttt{exoTEDRF} data. Bottom: comparison between the retrieval codes \texttt{POSEIDON} and \texttt{BeAR} on \texttt{FIREFLy} data. The retrieved model transmission spectra, plotted at $R\approx100$ for clarity, show the median (solid lines) and $\pm$ 1$\sigma$ and $\pm$ 2$\sigma$ credible regions (contours) corresponding to the dataset(s) shown in each panel. The median retrieved offsets have been applied to the NIRSpec G395H NRS1 and NRS2 datasets. Right: the retrieved CH$_4$, CO$_2$, and DMS abundances corresponding to each retrieval. Our retrievals find broadly consistent CH$_4$ abundances, indicating a robust detection (the median and $\pm$ 1$\sigma$ abundance constraints are represented by solid and dashed lines, respectively). However, any suggestions of CO$_2$ and DMS depend on the specific data reduction and data resolution, indicating unreliable inferences (hence the 95\% upper limits, represented by arrows). 
    }
    \label{fig:retrieval_sensitivity}
\end{figure*}

We first consider the sensitivity of the atmospheric abundances of CH$_4$, CO$_2$, and DMS to the choice of data reduction code, data resolution, and retrieval code. Figure~\ref{fig:retrieval_sensitivity} provides a pairwise retrieval comparison of how these three factors can affect the inferred atmospheric composition of K2-18\,b. First, we highlight a comparison between \texttt{POSEIDON} retrievals of the lowest-resolution \texttt{FIREFLy} and \texttt{exoTEDRF} data ($R\approx25$ NIRISS; $R\approx100$ NIRSpec). We selected these combinations to allow a clear demonstration of why the \texttt{exoTEDRF} NIRSpec $R\approx100$ data leads to weak evidence of DMS (per Figure~\ref{fig:retrieval_detection_significances}) but \texttt{FIREFLy} does not. We see in Figure~\ref{fig:retrieval_sensitivity} (top panel) that several \texttt{exoTEDRF} data points near $3.5\,\micron$ and $4.9\,\micron$ have greater transit depths than \texttt{FIREFLy}, which coincide with opacity maxima for DMS. These transit depth differences result in a peak in the DMS mixing ratio posterior near $\log_{10} X_{\rm{DMS}} \sim -4$ for \texttt{exoTEDRF}, but an upper limit for \texttt{FIREFLy}. The non-robustness of this DMS inference can be seen in the second pairwise comparison in Figure~\ref{fig:retrieval_sensitivity} (middle panel), where we show that our retrieval of the highest resolution \texttt{exoTEDRF} data (2-pixel for NIRISS; pixel-level for NIRSpec) no longer favors DMS. We note that the higher model transit depths for the pixel-level retrieval arise from a higher mean transit depth in the \texttt{exoTEDRF} NIRISS data compared to the $R\approx25$ data variant (resulting in smaller offsets between the NIRISS and NIRSpec data for the pixel-level retrieval). Finally, we demonstrate the excellent agreement between \texttt{POSEIDON} and \texttt{BeAR} in Figure~\ref{fig:retrieval_sensitivity} (lower panel). This shows that our main results (CH$_4$ abundance constraints, upper limits on the CO$_2$ and DMS abundances) are robust to the different atmospheric modeling and retrieval configuration choices made by \texttt{POSEIDON} and \texttt{BeAR} (Table~\ref{tab:retrieval_configurations}). We additionally provide the full corner plots from our \texttt{POSEIDON} and \texttt{BeAR} retrieval grids as supplementary material on Zenodo: \href{https://doi.org/10.5281/zenodo.14735688}{doi:10.5281/zenodo.14735688}.

\begin{figure}[ht!]
    \centering
    \includegraphics[width=\columnwidth]{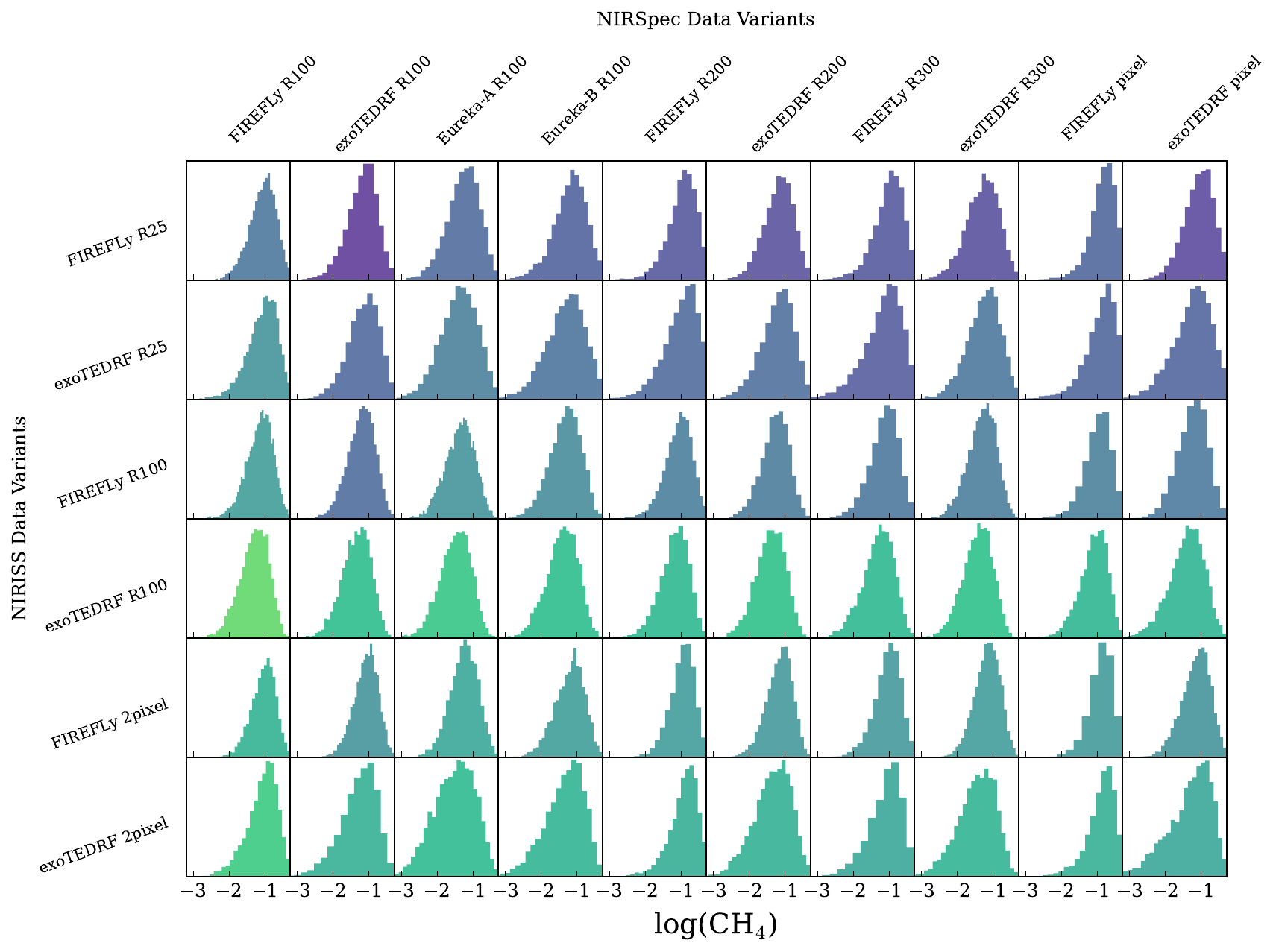}
    \includegraphics[width=\columnwidth]{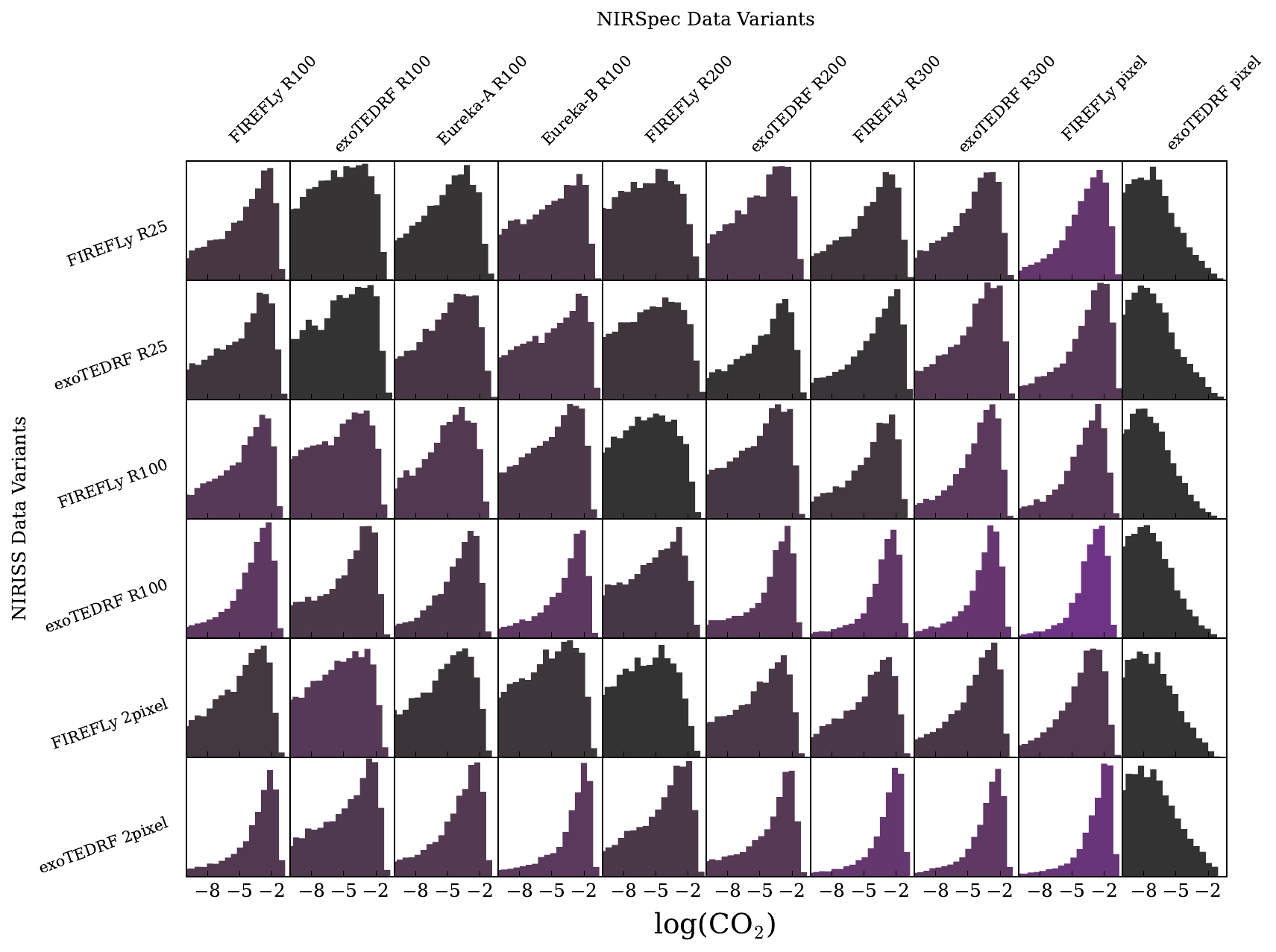}
    \includegraphics[width=\columnwidth]{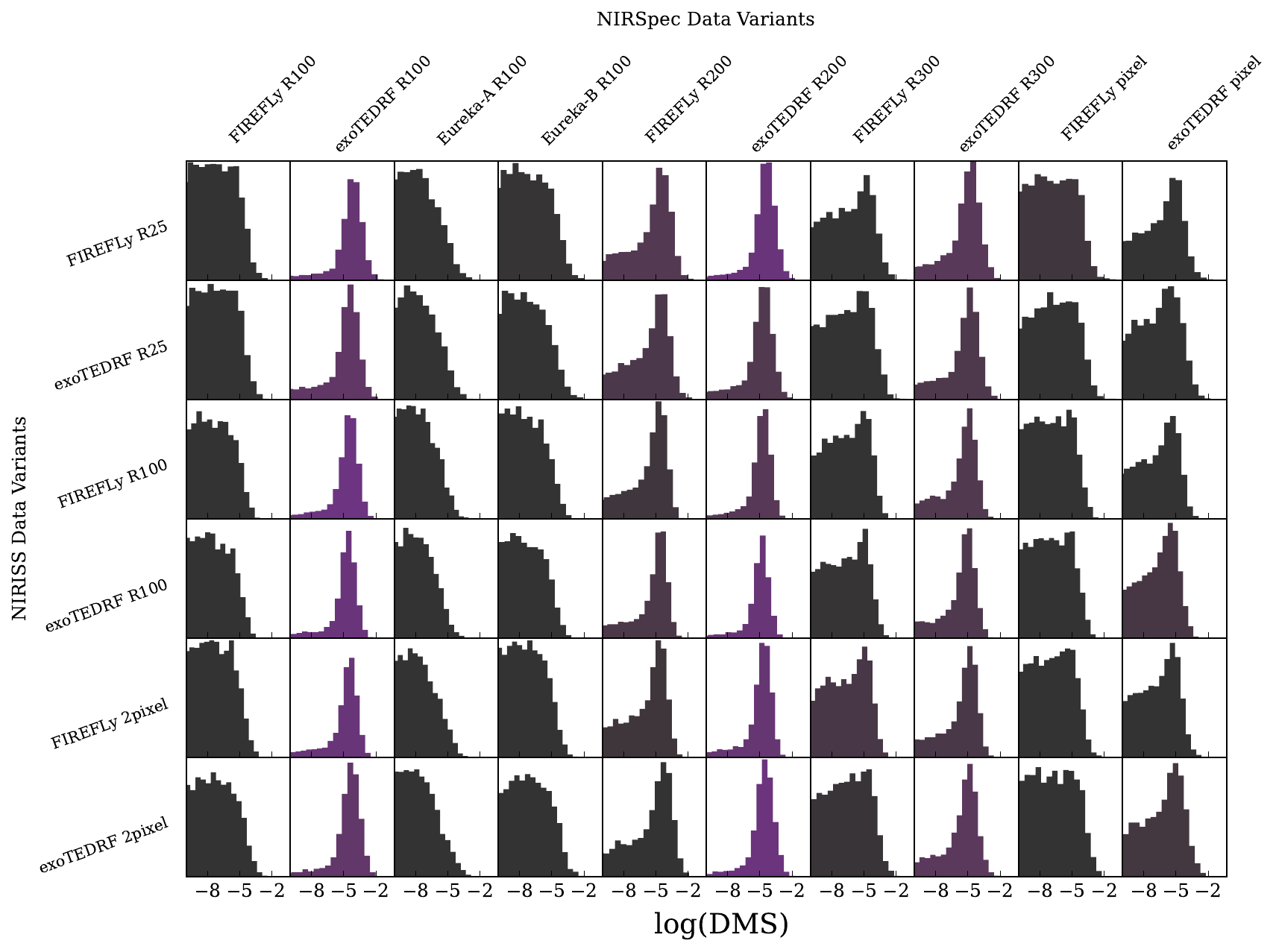}
    \caption{Grid of histograms showing the posteriors for CH$_4$, CO$_2$, and DMS for each of our retrievals. The arrangement and color map are the same as in Figure \ref{fig:retrieval_detection_significances}, with the color of the histogram representing the detection significance. While the CH$_4$ posteriors exhibit little variance across reduction choices, there are major differences for CO$_2$ and DMS that are indicative of a non-detection. 
    }
    \label{fig:retrieval_CH4_CO2_DMS_abundances}
\end{figure}

CH$_4$ is the only molecule with an abundance lower limit across all data combinations (Figure~\ref{fig:retrieval_CH4_CO2_DMS_abundances}), which indicates a robust detection. To account for the spread in abundances across different data treatments, we consider a conservative range spanning the lowest $-1\,\sigma$ and highest $+1\,\sigma$ across all the CH$_4$ posteriors --- similar to the approach by \citet{Pow24} for SO$_2$ in WASP-39b. We find a $1\,\sigma$ range of $\log_{10} \rm{CH_4} =$ -2.14 to -0.53 and a $2\,\sigma$ range of $\log_{10} \rm{CH_4} =$ -2.86 to -0.35, with the most probable CH$_4$ abundance near $\sim$ 10\%. Our retrieved CH$_4$ is consistent with the value reported in \citet{Mad23} using a single data reduction code ($\log_{10} \rm{CH_4} = -1.74^{+0.59}_{-0.69}$; see also Appendix~\ref{appendix:Madhu_reproduction}). The high abundance of CH$_4$ in K2-18\,b's atmosphere is consistent with a metal-enriched mini-Neptune (as discussed in Section~\ref{sec:atmointerp}) and is similar to the deep abundance of CH$_4$ on Neptune ($4 \pm 1$\%; \citealt{Karkoschka2011}). This high CH$_4$ abundance also raises the background mean molecular weight of K2-18\,b's atmosphere to $\mu_{\rm{atm}} = $ 2.46--7.64\,amu.

Abundance inferences for CO$_2$ and DMS depend critically on the data combination adopted, with some combinations showing suggestive peaks but most showing broad posteriors with no evidence supporting their presence (Figure~\ref{fig:retrieval_CH4_CO2_DMS_abundances}). We therefore report $2\,\sigma$ upper limits on the abundances of CO$_2$ and DMS in K2-18\,b's atmosphere, since neither of these molecules are robustly detected. Considering $2\,\sigma$ extrema upper limits, as for CH$_4$ above, we find $\log_{10}$ CO$_2 < -1.02$ and $\log_{10}$ DMS $< -2.51$. However, while these limits are pulled to higher abundances by the few data combinations exhibiting posterior peaks (Figure~\ref{fig:retrieval_CH4_CO2_DMS_abundances}). Therefore, we consider the median $2\,\sigma$ upper limits to provide a more representative statistic for the abundances of molecules that are not robustly detected: $\log_{10}$ CO$_2 < -1.58$ and $\log_{10}$ DMS $< -3.58$. We stress that these upper limits do not imply that either molecule is absent at \emph{any} atmospheric abundance, as abundances lower than our 95\% limits are still compatible with the present JWST transmission spectrum. Indeed, some mini-Neptune scenarios predict CO$_2$ abundances of $\log_{10} \rm{CO_2} \sim -3$ (Section~\ref{sec:atmointerp}), consistent with our upper limit.

We additionally report upper limits on several gases important for understanding the atmospheres and interiors of sub-Neptunes. Water and ammonia are depleted from the gas phase (95\% upper limits of $\log_{10} \rm{H_2 O} < -2.44$ and $\log_{10} \rm{NH_3} < -4.70$), which can be explained by cloud condensation \citep{Hu21,Mad23} and dissolution into a deep magma ocean \citep{Sho24}, respectively. Similarly, we find an upper limit of $\log_{10} \rm{CO} < -2.04$. We show in Section~\ref{sec:models} that these non-detections do not require a surface liquid ocean on K2-18\,b. Alongside our non-detection of DMS, we find upper limits on four other gases proposed as biosignatures \citep[e.g.,][]{Segura2005,Seager2013,Schwieterman2018,Mad21}: $\log_{10}$ CS$_2 < -2.63$, $\log_{10} $CH$_3$Cl $< -1.81$, $\log_{10}$ OCS $< -2.77$, and $\log_{10}$ N$_2$O $< -3.56$. Therefore, there is no evidence of biosignatures in the JWST NIRISS SOSS + NIRSpec G395H transmission spectrum of K2-18\,b.

\subsubsection{Atmospheric Temperature} \label{sec:temperature}

K2-18\,b's transmission spectrum is consistent with an isothermal upper-atmosphere temperature structure. While our \texttt{POSEIDON} retrievals allow for temperature gradients, our top-of-atmosphere temperature ($T_{10\,\rm{nbar}} =$ 98--339\,K) is consistent with the temperature at significantly higher pressures (e.g. $T_{10\,\rm{mbar}}$ = 120--358\,K). Our temperature constraints are consistent with \citet{Mad23}, who reported $T_{10\,\rm{mbar}} = 242^{+57}_{-79}$\,K. Since low-resolution transmission spectra only probe atmospheric pressures from $\sim 10^{-5}$--$10^{-1}$\,bar, we note that the lack of any detected temperature gradient does not preclude steep adiabatic temperature gradients at deeper pressures.

\subsubsection{Aerosols} \label{sec:aerosols}

We do not detect any evidence of aerosols in K2-18\,b's atmosphere. Our retrievals offer no constraints (either upper or lower limits) on the haze parameters or the terminator cloud fraction. \citet{Mad23} similarly reported a non-detection of aerosols. Since scattering hazes can play an important role in moderating K2-18\,b's temperature under a hycean interpretation \citep{Mad23,Jordan2025}, our non-detection of a haze --- based on data that include the bluer 0.6--0.8\,$\micron$ 2$^{\rm{nd}}$ order NIRISS SOSS data that are more sensitive to a haze slope --- provides important context for the viability of a liquid water ocean. However, given the present signal-to-noise ratio at the shortest wavelengths (e.g. Figure~\ref{fig:retrieval_sensitivity}), additional observations with NIRISS SOSS would be required before hazes can be definitively ruled out.

\subsubsection{Unocculted Starspots} \label{sec:starspots}

\begin{figure}[t!]
    \centering
    \vspace{0.4cm}
    \includegraphics[width=\columnwidth]{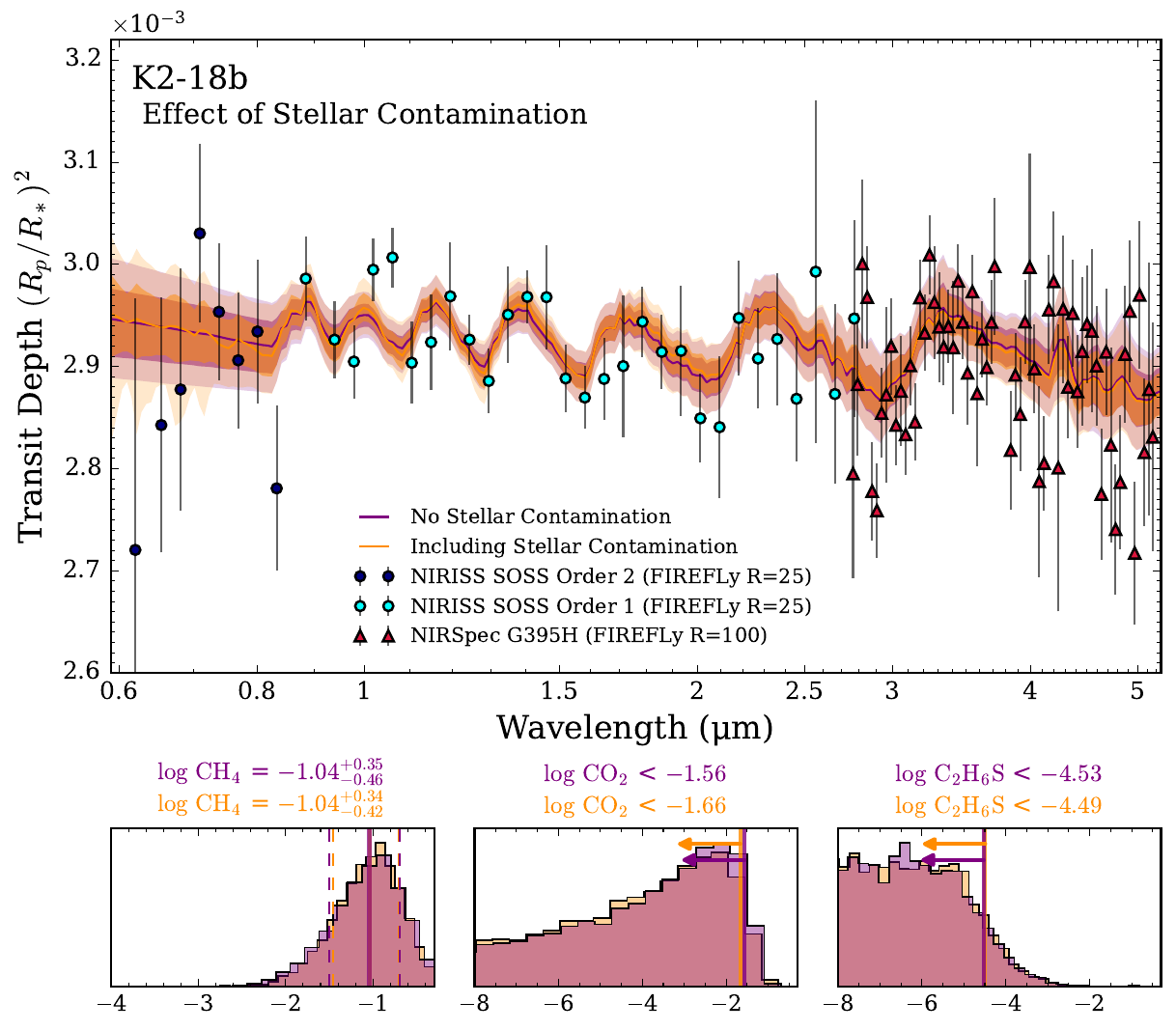}
    \caption{Influence of stellar contamination on K2-18\,b retrievals. Top: two \texttt{POSEIDON} retrieval models for the low-resolution \texttt{FIREFLy} data variant, one including a planetary atmosphere and stellar contamination (orange) and an alternative atmosphere-only model (purple). Bottom: corresponding posterior distributions for the CH$_4$, CO$_2$, and DMS abundances. The inferred atmospheric composition is insensitive to whether or not stellar contamination is included.}
    \label{fig:retrieval_stellar_contam}
\end{figure}

By including stellar contamination in our retrievals, we obtain constraints of unocculted active regions on K2-18. We find a small coverage fraction (2--10\%) of cool spots ($\Delta T \approx 500 \pm 350$\,K cooler than the photosphere). While this may appear contrary to the finding of no stellar contamination reported by \citet{Mad23}, we note that our inclusion of the 2$^{\rm{nd}}$ order NIRISS data provides coverage of shorter wavelengths where starspot contamination is more prominent \citep[e.g.,][]{Lim23, Fou24, Cad24, Rad24}. We investigated the impact of stellar contamination by running several additional \texttt{POSEIDON} retrievals with stellar contamination excluded, with one example shown in Figure~\ref{fig:retrieval_stellar_contam}. We find that this potential low-level stellar contamination does not notably affect the inferred atmospheric composition of K2-18\,b. However, given the stochastic nature of the transit light source effect, further repeat observations will be necessary to confirm this. Therefore, while it is likely that future transit observations of K2-18\,b will be able to be stacked to increase the transmission spectrum precision, it would be prudent to continue observations with NIRISS SOSS (as opposed to only observing at longer wavelengths) to further constrain the possibility of stellar heterogeneities.

With our inferences from K2-18\,b's JWST transmission spectrum in hand, we next turn to an assessment of viable atmospheric and interior structures. 

\section{\texorpdfstring{Atmosphere and Interior Modeling: \\ The Nature of K2-18}{Atmosphere and Interior Modeling: The Nature of K2-18} \texorpdfstring{\lowercase{b}}{b}} \label{sec:models}

K2-18\,b's mass and radius allow a wide range of planetary structures, with the most widely considered scenarios being a mini-Neptune or a hycean world. Here we conduct a careful assessment of the range of planetary structures and climates compatible with our retrieved composition  (Section~\ref{sec:atmointerp}). We also calculate interior models to assess which planetary structures are compatible with the observed mass and radius (Section~\ref{sec:interior}). We proceed to elucidate viable natures for K2-18\,b.

\subsection{Atmosphere Modeling} \label{sec:atmointerp}

We first quantify the expected compositional and climate properties for K2-18\,b across various proposed scenarios and compare them to our retrieved atmospheric abundance constraints. Our analysis builds upon a similar narrative to \citet{Wog24}, but here we use our revised retrieved abundances to evaluate if a hydrogen-rich mini-Neptune or a hycean planet with a liquid water surface is favored. We consider four scenarios: (i) a mini-Neptune with water clouds; (ii) an oxygen-poor mini-Neptune; (iii) a hycean planet; and (iv) a supercritical ocean. 

\subsubsection{Mini-Neptune with Water Clouds}\label{ref:sec_cloudy_Nep}

First, we consider K2-18\,b as a mini-Neptune with a thick hydrogen envelope. The key distinction from previous studies \citep{Tsai2021,Wog24,Cooke2024} is that here we compute temperature profiles consistent with the disequilibrium composition predicted by photochemistry and vertical mixing. We simulate mini-Neptune atmospheres with two independent sets of coupled photochemical-climate models: (1) \texttt{Photochem} \citep[v0.6.2,][]{Wog24code,Wog24} \& \texttt{PICASO} \citep[v3.3.0,][]{Batalha2024,Mukherjee2024}; and (2) \texttt{VULCAN} \citep{Tsai2017,Tsai2021b} \& \texttt{HELIOS} \citep{Malik2017,Malik2019}. For each simulation, we iterate between the photochemical and climate models, passing the chemical composition and temperature profiles between the codes until a photochemical-climate steady-state is achieved. 

\begin{figure}[ht!]
  \centering
  \includegraphics[width=\linewidth]{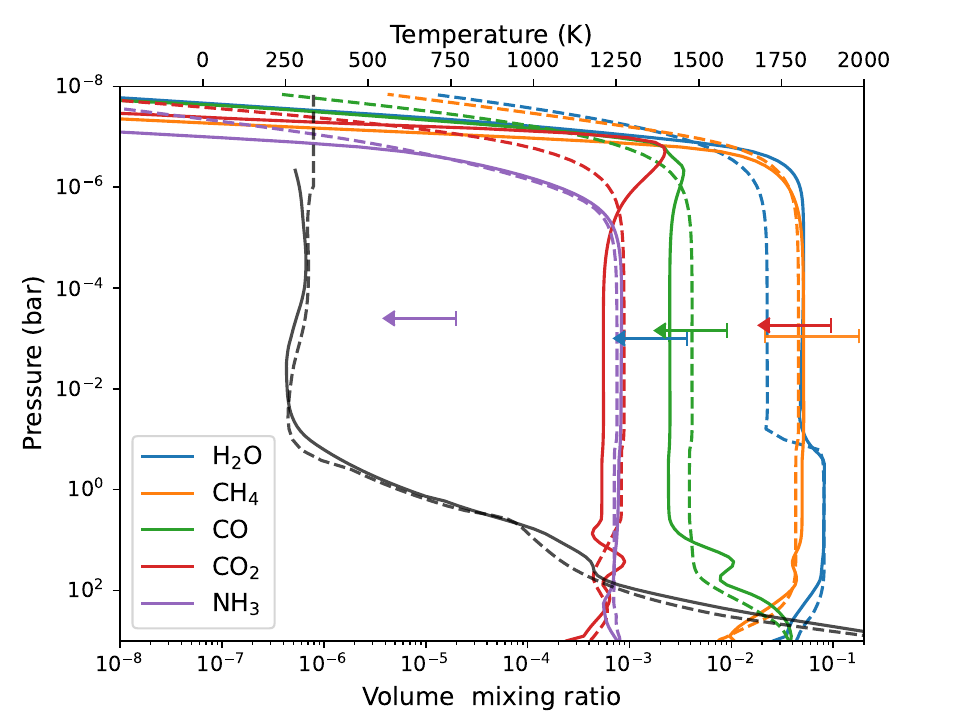}
  \includegraphics[width=\linewidth]{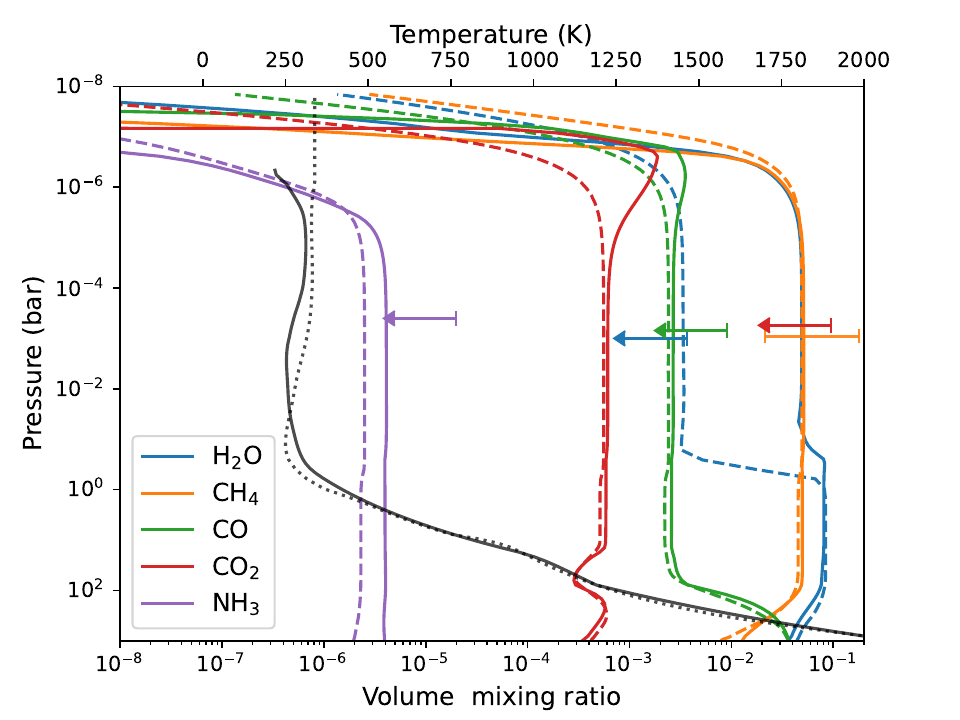}
  \caption{Self-consistent mini-Neptune models of K2-18\,b's atmosphere. Upper panel: 100$\times$ solar metallicity case. Lower panel: the same metallicity, but with 5000 $\times$ nitrogen depletion. The temperature (black lines) and composition profiles (colored lines) from \texttt{VULCAN}-\texttt{HELIOS} (solid lines) and \texttt{Photochem}-\texttt{PICASO} (dashed lines) with the 1\,$\sigma$ CH$_4$ range (orange bar) and 95\% upper limits for other molecules (arrows).}
  \label{fig:Nep_clouds}
\end{figure}

Figure~\ref{fig:Nep_clouds} (upper panel) illustrates the temperature and chemical composition computed by our self-consistent models, assuming 100$\times$ solar metallicity, an intrinsic temperature of 60\,K, and a uniform eddy diffusion coefficient (K$_{zz}$) of $10^{7}$ cm$^2$/s (similar to \citealt{Hu21,Tsai2021}). The simulated \ce{CH4} matches the retrieved abundance, while the modeled \ce{CO} and \ce{CO2} remain compatible with their respective 2-$\sigma$ upper limits of log$_{10}$ \ce{CO} = $-$2.45 and log$_{10}$ \ce{CO2} = 
 $-$1.70. However, the modeled \ce{H2O} and \ce{NH3} exceed our inferred upper limits. \cite{Cooke2024} highlighted that K2-18\,b's temperature as a mini-Neptune with 100$\times$ solar metallicity equilibrium composition is too high to condense enough water to meet the \ce{H2O} upper limit, unless the stellar irradiation is artificially reduced by $40\%$. Our self-consistent calculations yield similar results, with the temperature remaining too warm and the \ce{H2O} mixing ratio only reduced by a factor of a few due to condensation. The \texttt{Photochem}-\texttt{PICASO} model produces a slightly cooler temperature than \texttt{VULCAN}-\texttt{HELIOS} and, hence, a marginally lower water abundance. Without ad hoc irradiation reduction, the predicted \ce{H2O} abundance exceeds the 2\,$\sigma$ upper limit by more than an order of magnitude for both simulations.

Secondly, the lack of detected \ce{NH3} is one of the main arguments against a mini-Neptune scenario, as \ce{NH3} is hypothesized to be present at detectable abundances ($\log_{10} \mathrm{NH_3} \approx -3$, see Figure~\ref{fig:Nep_clouds}) in a thick H$_2$-dominated atmosphere \citep{Yu2021,Hu21,Tsai2021}. A potential explanation for lower \ce{NH3} abundances, suggested by \cite{Sho24}, is that a reducing magma layer at the bottom of the gaseous envelope can efficiently dissolve nitrogen into the magma and hence deplete \ce{NH3} from the gas phase.

We explore the extent of nitrogen depletion required to explain our non-detection of \ce{NH3} in the lower panel of Figure~\ref{fig:Nep_clouds}. This second mini-Neptune model also has a 100$\times$ solar metallicity atmosphere, but with the total nitrogen depleted by a factor of 5000 (i.e., N/H = 0.02$\times$ solar) to account for nitrogen dissolution into a magma ocean. Such a low N/H ratio is close to, but within the lower bound, of expected nitrogen abundances under low magma oxygen fugacity \citep{Sho24,Rig24}. This nitrogen-depletion factor lowers the atmospheric \ce{NH3} abundance at the pressure probed by transmission spectroscopy to $<$ 10\,ppm, compatible with our retrieved 95\% upper limit. Furthermore, a depleted \ce{NH3} abundance moves the stratosphere to a slightly higher pressure, which allows additional \ce{H2O} condensation (especially for the \texttt{Photochem}-\texttt{PICASO} model). The condensed water would form clouds around 0.1\,bar, which is consistent with our retrieval constraints without damping the \ce{CH4} features over the NIRISS spectral range.

Should K2-18\,b's atmosphere possess a scattering haze (not considered in these models), the tropopause could be cooler than shown in Figure~\ref{fig:Nep_clouds} with a corresponding lower \ce{H2O} abundance in the upper atmosphere \citep{Pie20}. However, a fine balance is needed to avoid producing a sufficiently strong haze that would dampen the shortest wavelength CH$_4$ bands to a degree not seen in the JWST NIRISS transmission spectrum \citep{Lec24}. These considerations motivate additional modeling efforts under the mini-Neptune scenario to explore the sensitivity of the H$_2$O and \ce{CH4} abundance to aerosols. Nevertheless, these results demonstrate that a mini-Neptune scenario for K2-18\,b, including nitrogen depletion from a basal magma ocean, is a plausible interpretation of our retrieved atmospheric abundances -- though an additional albedo source is required to cold-trap water.

\subsubsection{Mini-Neptune with high C/O ratio}

An alternative way to explain the low water abundance on K2-18\,b is an oxygen-poor atmosphere with a high carbon-to-oxygen (C/O) ratio. K2-18\,b's current orbit sits between the soot \citep{Li2021} and water ice lines. If the planet migrated early to its present location within the ice line, it is plausible to form a carbon-rich and oxygen-poor envelope \citep{Bergin2023,Yang2024}. We now consider the necessary C/O ratio to explain our non-detection of \ce{H2O} under the mini-Neptune scenario without an additional albedo source.

We first consider the simplest calculation of thermochemical equilibrium to illustrate the impact of the C/O ratio on the atmospheric \ce{CH4} and \ce{H2O} abundances. We perform these equilibrium calculations using Fastchem \citep{Stock2022,Kitzmann2024} over the 1--100\,bar pressure range in the deep atmosphere for a range of temperatures. Since this pressure range encloses possible quench points, these deep equilibrium abundances should broadly reflect the abundances from $\sim 10^{-4}$--$10^{-1}$\,bar probed by the JWST spectrum --- as confirmed by the rather uniform vertical distributions of \ce{CH4} and \ce{H2O} within this region in the absence of condensation (e.g. Figure~\ref{fig:Nep_clouds}; see also \citealt{Tsai2021,Yu2021,Cooke2024}).

\begin{figure}[t!]
    \centering
    \includegraphics[width=\linewidth]{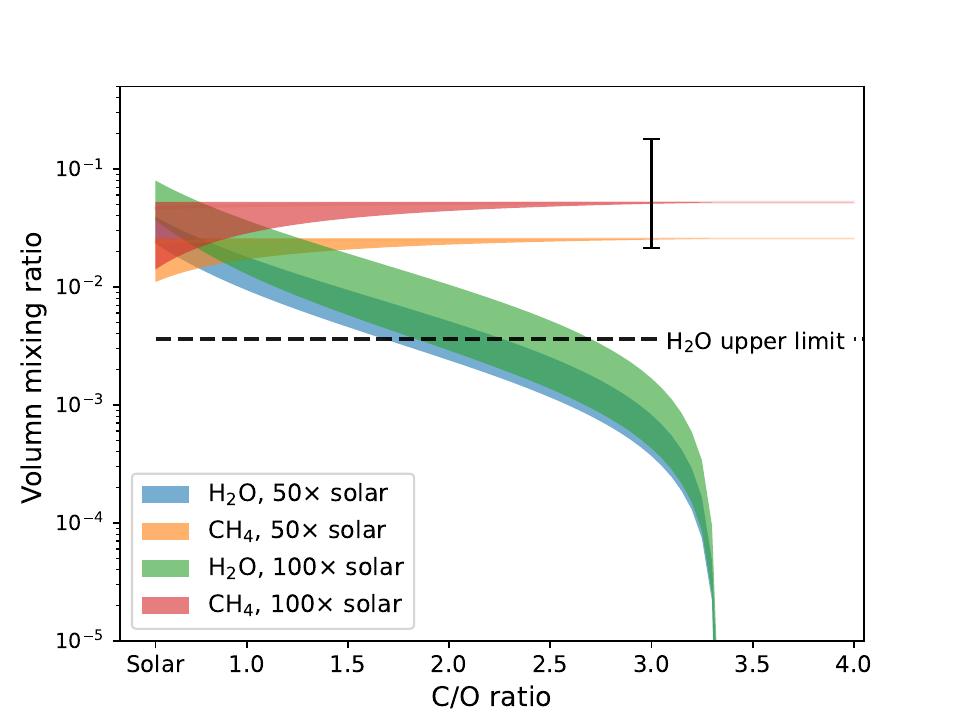}
    \caption{The equilibrium abundances of \ce{H2O} and \ce{CH4} as a function of C/O ratio. The shaded regions represent the range across pressures from 100--1\,bar and temperatures from 800--1000\,K, which cover the expected quench levels. The dashed line is the 2-$\sigma$ upper limit of \ce{H2O} from our retrieval analysis, and the open circle with an error bar represented our retrieved 1\,$\sigma$ \ce{CH4} abundance.}
    \label{fig:eq_CtoO}
\end{figure}

Figure~\ref{fig:eq_CtoO} shows the equilibrium \ce{H2O} and \ce{CH4} abundances as C/O varies. For simplicity, we fixed carbon and only changed the oxygen abundance. From the reasoning above, the equilibrium \ce{H2O} and \ce{CH4} abundances shown by the shaded regions in Figure~\ref{fig:eq_CtoO} can be considered as the observable abundances. We see that \ce{H2O} decreases with increasing C/O --- well-known from previous studies \citep[e.g.][]{Madhusudhan2012,Molliere2015}, as oxygen is preferentially sequestered within silicates --- while \ce{CH4} remains nearly constant. We require C/O $\gtrsim$ 3 under equilibrium chemistry to lower the \ce{H2O} abundance below the 95\% upper limit from our retrievals. This simple analysis suggests that an oxygen-depleted (high C/O ratio) mini-Neptune scenario, without effective water condensation, can be consistent with the observed \ce{H2O} and \ce{CH4} constraints. Having demonstrated \ce{H2O} depletion under equilibrium conditions, we next turn to self-consistent photochemical-climate models for a high C/O ratio.
 
\begin{figure}[t!]
    \centering
    \includegraphics[width=\linewidth]{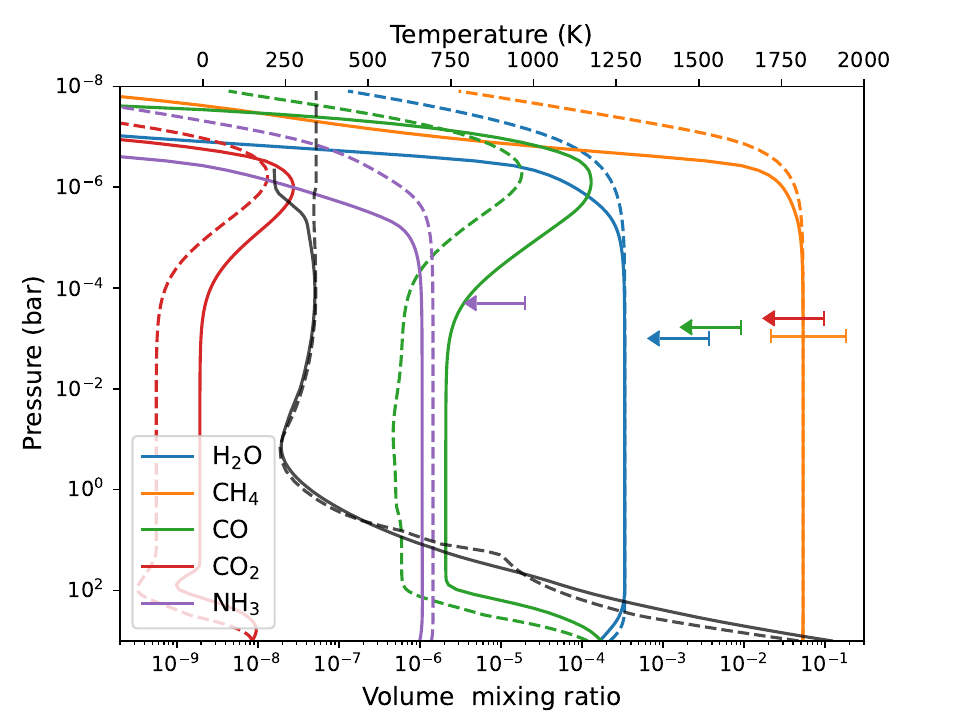}
    \caption{Self-consistent mini-Neptune models for K2-18\,b with an oxygen-depleted atmosphere. The atmospheric conditions and models are the same as the lower panel in Figure~\ref{fig:Nep_clouds}, but with C/O = 3.25.}
    \label{fig:deq_CtoO3}
\end{figure}

Figure~\ref{fig:deq_CtoO3} shows our self-consistent photochemical climate models for K2-18\,b under a high C/O ratio mini-Neptune scenario. The atmospheric properties are the same as Section~\ref{ref:sec_cloudy_Nep} (100$\times$ solar metallicity and 5000$\times$ N depletion), but with C/O = 3.25. We see that this oxygen-poor scenario naturally explains the retrieved \ce{CH4} abundance, alongside the non-detections of \ce{H2O} and \ce{CO2}, without invoking water condensation.
Additionally, \ce{CO2} is negligible due to the scarcity of oxygen. Future observations with tighter constraints on \ce{CO2} could help differentiate this oxygen-depleted scenario from the cold-trapped water scenario discussed in Section~\ref{ref:sec_cloudy_Nep}.

\subsubsection{Hycean Scenario}
Previous arguments for hycean atmospheres have emphasized the presence of \ce{CO2} \citep{Hu21,Mad23}, since \ce{CO2} is expected to be the dominant carbon-bearing molecule. Its abundance, possibly ranging from $\sim$ 1 ppm to 1\%, would be controlled by the solubility balance with the ocean \citep{Hu21,Kite2018}. Conversely, \ce{CH4} is not photochemically stable in a small atmosphere. The high amount of \ce{CH4} ($\sim 10$\% from our retrievals) 
requires an additional supply, as demonstrated by \cite{Wog24}.\footnote{See also \cite{Cooke2024} for a thorough analysis of its dependence on various parameters.}

Given our non-detection of \ce{CO2} from K2-18\,b's JWST transmission spectrum, while \ce{CH4} is robustly detected, we now consider whether a hycean scenario remains viable for K2-18\,b. Since a hycean world possesses a thin H$_2$ atmosphere to sustain a liquid water ocean, \ce{CH4} is expected to be photochemically converted to \ce{CO2} in the upper atmosphere, as discussed in \citet{Tsai2021} and \citet{Wog24}. To explore whether our \ce{CH4} and \ce{CO2} constraints fit the hycean scenario, we conducted a test to demonstrate how \ce{CH4} undergoes such conversion. We adopted a representative temperature profile for a hycean atmosphere from \cite{Tsai2024} (assuming a surface albedo of 0.3) and fixed the surface \ce{CH4} abundance at 1$\%$, without any initial \ce{CO2}, representing a surface source continuously replenishing \ce{CH4}.

\begin{figure}[t!]
    \centering
    \includegraphics[width=\linewidth]{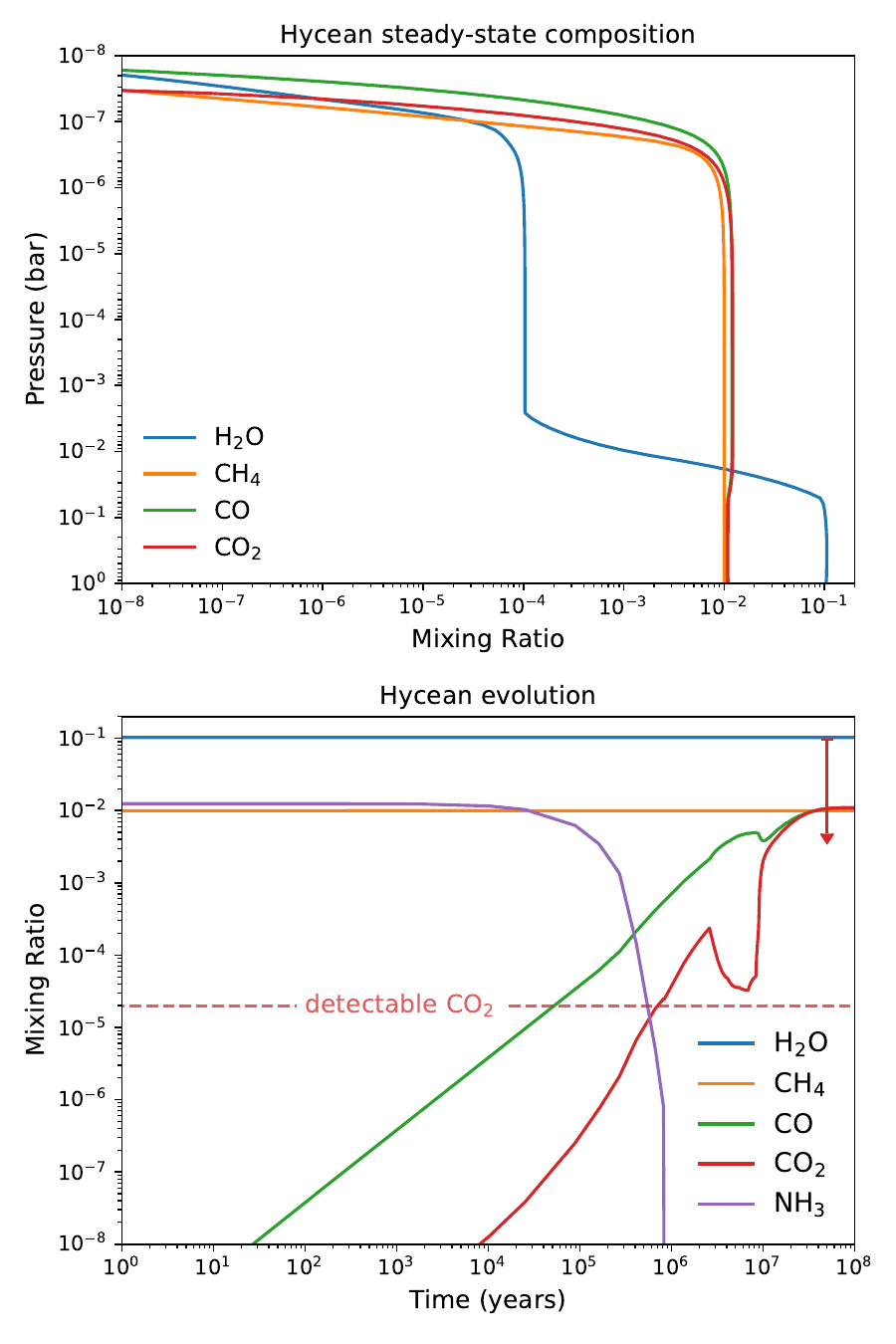}
    \caption{Atmospheric composition over time for a hycean scenario for K2-18\,b. Upper panel: the steady-state composition profiles computed by VULCAN in our hycean scenario. Lower panel: the composition evolution at 0.1 bar. The red arrow corresponds to the 2 $\sigma$ upper limit of \ce{CO2} from our retrievals, while the dashed line indicates the \ce{CO2} abundance required to produce a potentially detectable ($\sim$100 ppm) transmission feature at 4.2 $\mu$m.}
    \label{fig:mix_evo}
\end{figure}

The resulting steady-state composition and the evolution are shown in the upper panel of Figure~\ref{fig:mix_evo}. Compared to the mini-Neptune scenario (Figures \ref{fig:Nep_clouds} and \ref{fig:deq_CtoO3}), \ce{CO2} and CO exhibit significantly higher abundances. 

 The evolution in the lower panel of Figure~\ref{fig:mix_evo} illustrates
 the photochemical conversion of \ce{CH4} into \ce{CO2}, occurring on a geologically short timescale of $\sim$ 10--100 Myr. While the exact \ce{CO2} abundance depends on the precise levels of \ce{H2O}, \ce{CH4}, and the surface temperature, the main takeaway of our test is that \ce{CO2} and \ce{CO} are expected to have comparable abundances to \ce{CH4} within a \ce{CH4}-rich ($\sim$ 1$\%$) and thin atmosphere. Therefore, regardless of the source of \ce{CH4}, our well-constrained measured \ce{CH4} abundance, which lies above the 95\% upper limits for both \ce{CO2} and \ce{CO}, poses a challenge to reconcile the absence of \ce{CO2}, as \ce{CH4} should naturally convert to \ce{CO2} in a hycean atmosphere.

Should K2-18\,b indeed have a water-rich interior covered by a hydrogen-helium envelope, as allowed by the observed mass and radius, the water will most likely exist in a supercritical state. This water phase arises from the greenhouse effect of \ce{H2} collision-induced absorption and the suppression of moist convection \citep{Pie23,Inn23}. Under such a super-runaway scenario, supercritical \ce{H2O} becomes miscible with \ce{H2} \citep{Soubiran2015, Gupta2025}. This was recently proposed to be the case for TOI-270 d \citep{Ben24} as well as a possible explanation for GJ 1214 b \citep{Nixon2024b}. Without an appropriate climate model to track the transition from the supercritical ocean to the subcritical upper atmosphere, it is unclear whether the temperature in the upper atmosphere allows a condensing layer to cold-trap water enough to match our retrieved abundances.

Under a supercritical ocean scenario, a wide range of \ce{CH4}/\ce{CO2} ratios are allowed due to the unknown hydrogen molarity of K2-18\,b \citep{Luu24}. Consequently, it is not currently possible to conduct a detailed comparison between our retrieved \ce{CH4} abundance and the \ce{CO2} upper limit. Future observations that provide a tighter \ce{CO2} abundance constraint would offer useful insights into the potential thermal properties of a supercritical ocean.

\subsection{Interior Structure Modeling} \label{sec:interior}
To aid in evaluating the possibility of a liquid water ocean on K2-18\,b, we also construct several interior structure models of the planet which match its observed mass and radius ($M_{\rm{p}} = 8.63 \pm 1.35 M_{\oplus}$, $R_{\rm{p}} = 2.61 \pm 0.087 R_\oplus$).  We use the mass and radius presented by \citet{Ben19} as the basis of our models\footnote{If the planet were smaller, as reported in earlier work, the required water fraction for a thin H/He atmosphere drops to a more plausible range.}. We prefer this set of planet parameters due to the use of a Gaia parallax in their calculation. These interior models solve the equations of hydrostatic equilibrium, mass conservation, and the material equations of state \citep[see][]{Thorngren2016}.  This is similar to the analysis of \citet{Mad20}, but includes methane in the atmosphere to comport with the observed spectra, setting its mass fraction prior as normally distributed with mean 0.3 and standard deviation 0.05.  We used the \citet{Cha21} equation of state for H/He, \citet{Maz19} for water, \citet{Nettelmann2016} and \citet{Bethkenhagen17} for methane, ANEOS \citep{Thompson1990} for rock, and SESAME \citep{Lyon1992} for iron.  The methane and H/He are mixed in the atmosphere via the additive volumes approximation, with the adiabats calculated through thermal integration \citep[see][Eqs. 1.07-1.18]{ThorngrenPhDT}.  We thermally evolve the planet using atmosphere models from \citet{For07} to regulate heat flow from the interior.

\begin{figure}
    \centering
    \includegraphics[width=\columnwidth]{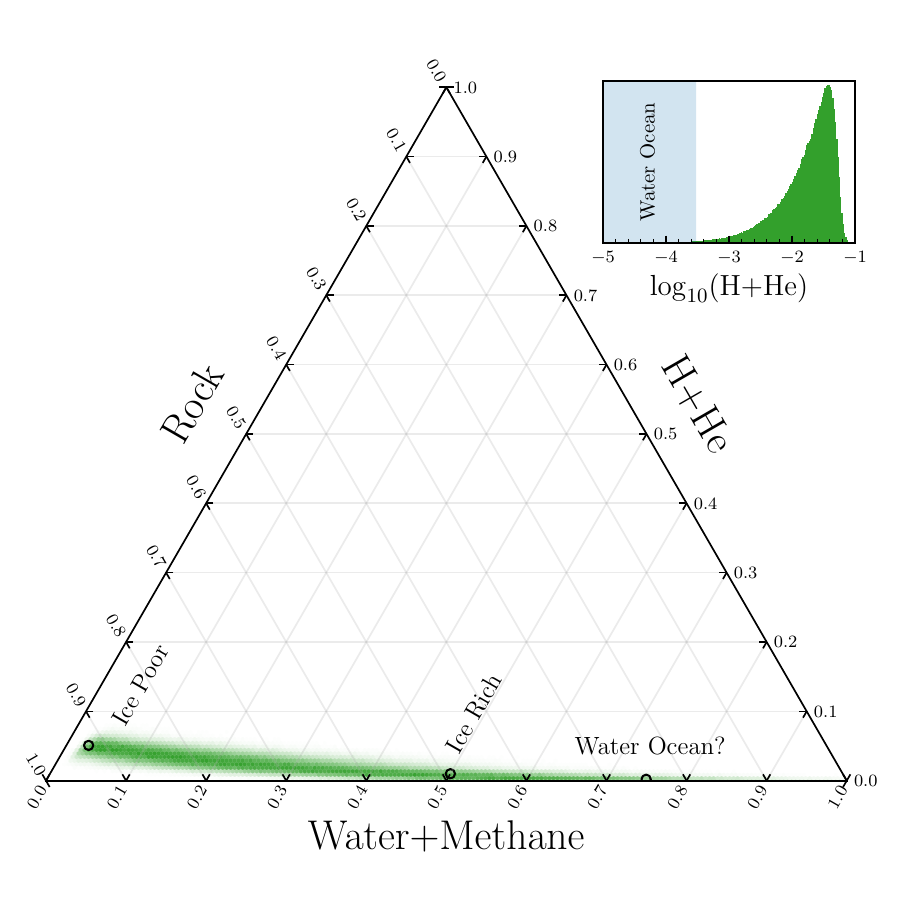}
    \caption{Potential interior compositions for K2-18\,b given our retrieval analysis. The green shaded region on the ternary diagram shows an MCMC posterior of compositions consistent with K2-18\,b's mass, radius, host star flux, atmosphere metallicity ($\sim30$\% CH$_4$), and system age.  We group methane with water as volatile species for analysis in terms of ice-to-rock ratios, but is physically located in the H/He layer in our model.  The three cases discussed in Sec. \ref{sec:interior} are circled and labeled.  The inset shows the posterior histogram of the log(H/He) mass fraction, with the amount permitting a liquid water ocean (shaded region) containing only 0.58\% of our posterior.  Above that the water layer is a supercritical-fluid.}
    \label{fig:ternary}
\end{figure}

We used MCMC to find compositions of the planet which match the planet's observed properties, especially radius and the atmospheric methane abundance.  Figure \ref{fig:ternary} shows the results as a ternary diagram of the mass fractions of rock (and iron, assuming an Earth-like ratio), nebular ices (here, water and methane), and H/He (at a solar ratio) for uniformly-distributed composition priors. Our models find H/He makes up less than 10\% of the planet’s mass, with the exact limit varying with the water abundance. Any of the models along the shaded region in Figure~\ref{fig:ternary} fits the observed data, but we identify a few specific cases of interest (denoted with circles) for further discussion.

First, we consider the ice-poor case, which represents a typical Neptune-like planet that could form within the ice line, experiencing limited volatile accretion. In this case, approximately 6\% of the mass must be H/He to match the observed radius. Such a planet features a deep envelope on top of a solid or molten rock surface. A more ice-rich planet could result from forming outside the ice line, with rock-to-water ratios around 50-50 as seen in the solar system icy moons \citep{Showman1999, Kuskov2005}. Stratified into separate layers, the planet would have a H/He envelope of about 1\% of the planet's mass. This results in the surface of the water layer occurring at about 2.21\,R$_\oplus$, a pressure of 100\,kbar and a temperature of 3200\,K; a supercritical fluid rather than a liquid.

To support a liquid water ocean, the surface must not exceed the critical point of water.  Our model atmospheres pass almost directly through the water critical point at 647\,K and 220\,bar.  Thus the H/He atmosphere must not pressurize the water beyond this, which corresponds to an H/He layer mass fraction of no more than 0.03\%. Our posterior probability of this composition is just 0.59\% (Fig. \ref{fig:ternary}, inset). Furthermore, accounting for the possibility of a runaway greenhouse effect could push this pressure limit down to tens of bars \citep{Inn23}.

In order to match the observed radius, such a low H/He fraction must be compensated for by a high water fraction of at least 70\%. This is improbable from a formation perspective -- \citep{Lodders2003} find the solar water-to-rock ratio to be 1.17:1 (54\%), similar to the solar system icy moons \citep{Kuskov2005}. \citet{Marcus2010} argue that this initial water-to-rock ratio provides an upper limit for the final planet composition, as giant impacts cannot enhance the water fraction, only deplete it. Including other ices (methane and ammonia) gives a max ice-to-rock ratio of 2.41:1 (70\%), but this would require formation outside the \emph{methane} ice line and subsequent migration, and methane's low solubility in water suggests it would likely contribute to a thick atmosphere rather than permit a water ocean.  Thus, while we cannot absolutely rule out a water ocean, such a structure is very improbable for K2-18\,b due to its large radius.

\section{Summary \& Discussion} \label{sec:disc}
The nature of K2-18\,b is of broad interest for sub-Neptune studies and for exoplanet habitability. We have conducted a comprehensive reanalysis of the JWST transmission spectrum of K2-18\,b from NIRISS SOSS and NIRSpec G395H. Our approach considers multiple well-tested data reduction and atmospheric retrieval codes to ensure robust and reliable results. Thus, our analysis assessed what can be confidently inferred about K2-18\,b's atmosphere and interior from the initial JWST observations. Our main results are as follows:

\begin{itemize}
    \item We confirm the detection of CH$_4$ in K2-18\,b's atmosphere ($\approx 4\,\sigma$), with a volume mixing ratio of $-2.14 \leq \log_{10} \mathrm{CH_4} \leq -0.53$ ($1-30$\%).
    \item We do not detect CO$_2$ or DMS, contrary to the findings of \citet{Mad23}, placing 95\% confidence upper limits of $\log_{10} \mathrm{CO_2} < -1.02$ and $\log_{10} \mathrm{DMS} < -2.51$.
    \item Any inferences of CO$_2$ and DMS from these JWST data are unreliable and low-significance, exhibiting high sensitivity to choices made during the reduction process and retrieval model assumptions (Figure~\ref{fig:retrieval_detection_significances}).
    \item K2-18\,b's retrieved atmospheric composition can be explained by a 100$\times$ solar mini-Neptune with an oxygen-poor and nitrogen depleted (e.g. from a magma ocean) composition.
    \item A hycean scenario for K2-18\,b is only viable with an anonymously high water-to-rock ratio, a low H/He fraction, atmospheric hazes, and positing a substantial CH$_4$ flux. We regard the necessary fine-tuning as an unlikely product of nature.
\end{itemize}

Based on our reanalysis, we summarize the observational constraints, challenges, and viability for each interpretation of K2-18\,b in Table~\ref{tab:interpretation}. Our findings raise many implications about K2-18\,b, and the sub-Neptune population more broadly, as we now discuss.

\begin{deluxetable*}{l|cc}
\tablecaption{Mini-Neptune vs Hycean interpretations for K2-18\,b}
\label{tab:interpretation}
\tablehead{
    \diagbox[width=10em]{Constraint}{Scenario}&
  mini-Neptune & Hycean}
\startdata
Climate  & Compatible & Requires high-albedo aerosols \\
Interior & Compatible & Requires high water-to-rock ratio \\
\ce{H2O} & Requires high-albedo aerosols or oxygen depletion & Requires high-albedo aerosols \\
\ce{CH4} & Compatible & Requires external source \\
\ce{CO2} & Compatible & Compatible, but not compatible with the coexistence with \ce{CH4}\\
\ce{NH3} & Requires high depletion from a basal magma ocean & Compatible through dissolution into the water\\
\enddata
\end{deluxetable*}

\subsection{Plausibility of Hycean Scenarios \& Biosphere Viability for K2-18 b}
Our revised atmospheric composition of K2-18\,b, accounting for data-level uncertainties, can be naturally explained by a mini-Neptune without requiring a liquid water ocean or life. Our non-detection of CO$_2$ lowers the potential CO$_2$/CH$_4$ ratio from $\sim 1$ to $<$ 0.3 (at 95\% confidence). This lower ratio complicates a hycean interpretation for K2-18\,b, as a high CO$_2$/CH$_4$ is one of the primary lines of evidence for a liquid water ocean underlying a thin H$_2$-dominated envelope \citep{Hu21,Tsai2021}. However, absence of evidence is not evidence of absence, and we stress that even mini-Neptune scenarios will contain some level of CO$_2$ (e.g., Figure~\ref{fig:Nep_clouds} predicts $\log_{10} \mathrm{CO_2} \sim -3$). Consequently, should future, more sensitive, JWST observations result in a CO$_2$ detection with an abundance lower than our upper limit ($\log_{10} \mathrm{CO_2} < -1.7$), this would be an additional line of evidence \emph{against} a hycean interpretation (see also Section~\ref{sec:future_obs}). Our photochemical modeling offers an additional line of evidence against inhabited hycean scenarios. We show in Figure~\ref{fig:mix_evo} that, even if one assumes a hycean scenario with a large unknown CH$_4$ flux (e.g. from methane-producing life), over time the atmospheric CO$_2$ and CO abundances would build to levels comparable to or above our upper limits. Consequently, our non-detections of CO$_2$ and CO are not consistent with the observed CH$_4$ abundance under an inhabited hycean scenario. Our results therefore render the mini-Neptune scenario the most likely interpretation for K2-18\,b given current observational constraints.

Our revised atmospheric composition also motivates a careful reevaluation of the viability of methanogenesis on K2-18\,b. Our non-detections of CO$_2$ and DMS prompt a re-assessment of whether a sufficient free energy source exists for methane-producing life under the metabolic pathway studied by \citet{Gle24}, rendering it less plausible that K2-18\,b sustains a biosphere. Further, our models in Figure~\ref{fig:mix_evo} demonstrate that a hycean scenario with the CH$_4$ abundance we retrieve would necessarily have a high $\sim$ 1\% CO$_2$ abundance that is in slight tension with our non-detection of CO$_2$ (though it is consistent with our 2-$\sigma$ upper limit). Updated metabolic calculations that consider the minimum viable CO$_2$ abundance to sustain methanogenesis will be required to assess whether a biosphere is ruled out by our revised atmospheric composition.

\subsection{\texorpdfstring{On Establishing Robust and Reliable \\ Molecular Detections from Atmospheric Retrievals}{On Establishing Robust and Reliable Molecular Detections from Atmospheric Retrievals}}

Our revised atmospheric composition for K2-18\,b demonstrates the critical importance of using multiple data reductions and multiple retrieval codes to establish which molecular detections are robust. In particular, our non-detection of CO$_2$ stands in stark contrast to the finding of a robust detection reported by \citet{Mad23}. However, this is not the first time that reported molecular detections in K2-18\,b's atmosphere have proven unreliable. Pre-JWST, multiple studies reported detections of H$_2$O vapor from HST WFC3 transmission spectroscopy at compelling statistical significances: 3.6\,$\sigma$ \citep{Tsi19}, 3.9\,$\sigma$ \citep{Ben19}, and 3.3\,$\sigma$ \citep{Mad20}. We now know from the JWST spectrum of K2-18\,b that the spectral feature seen by WFC3 was, in fact, CH$_4$ masquerading as H$_2$O (likely due to WFC3 edge effects near 1.7\,$\micron$). 
Here we discuss here how future retrieval analyses can safeguard against overstating the statistical evidence of a molecular detection.

First, our grid of retrievals demonstrates that detection significances naturally possess their own uncertainty due to data-level assumptions. Figure~\ref{fig:retrieval_detection_significances} clearly illustrates a spread of CH$_4$ significances (clustered around $4\,\sigma$) as a function of the data reduction code and the degree of spectral binning pre-retrieval. We propose that visualizations such as Figure~\ref{fig:retrieval_detection_significances} are a valuable tool to intuitively establish if a molecular detection is robust (as with CH$_4$) or whether there is no reliable statistical evidence (as with CO$_2$ and DMS). An additional value of Figure~\ref{fig:retrieval_detection_significances} is the clear representation of lost information when retrieving NIRISS spectra at $R = 25$ rather than $R = 100$, resulting in a significantly lower CH$_4$ detection significance, which suggests $R = 100$ is a safe minimum resolution for inferring atmospheric properties from JWST transmission spectra. Our first takeaway is thus that the term `robust' must be reserved for molecular detections that hold across multiple data reduction and retrieval codes.

Second, we note that the retrieval literature tends to present overconfident detection significances and inconsistent terminology from Bayesian model comparisons. When establishing if a molecule is present in an atmosphere, the common practice is to conduct a Bayesian model comparison between a retrieval model containing the molecule in question and one with it removed, calculate the Bayes factor, and then convert the Bayes factor into an `equivalent detection significance' (using the inverse of Equation~27 in \citealt{Trotta2008}). This approach was first introduced in the exoplanet retrieval context by \citet{Benneke2013} and has been widely used in retrieval codes since \citep[e.g.,][]{Waldmann2015,MacDonald2017,Wel21}. This procedure often provides useful insights, provided that: (i) both the Bayes factor and equivalent significance are provided, and (ii) well-defined terminology is adopted (e.g. the Jeffreys' scale; see \citealt{Trotta2008}). Not providing Bayes factors can produce overconfidence in a result. For example, a Bayes factor of 3 corresponds to an equivalent significance, according to the approach from \citet{Benneke2013}, of $2.1\,\sigma$, when this model is actually only favored at a 3:1 odds ratio. Similarly, a claim of a $3\,\sigma$ `detection' from a Bayesian model comparison is only a Bayes factor of~21. Recently, \citet{Kipping2025} highlighted this point and suggested that a more intuitive detection significance is given by equating the false positive detection probability with the $p$ value (i.e. $p = 1/(1 + \mathcal{B}$) and deriving $p = \mathrm{erfc}[N\sigma_{\rm{intuitive}}/\sqrt{2}]$. This results in values of $N\sigma_{\rm{intuitive}}$ that are approximately 1 lower than the relation from \citet{Benneke2013} (see also \citealt{Trotta2008}). Using this logic, the `$3\,\sigma$ detection' of CO$_2$ in K2-18\,b's atmosphere reported by \citet{Mad23} (without quoted Bayes factors) should be better thought of as a $2\,\sigma$ result, for which the term `detection' is too strong. It is for this reason that we report all Bayes factors $<$ 3 (equivalent to $N\sigma_{\rm{intuitive}} \lesssim 1$) as non-detections in Figure~\ref{fig:retrieval_detection_significances}.

\begin{deluxetable*}{lccccccccc}[ht!]
\tablecaption{Evidence for Molecular Signatures in K2-18\,b's JWST Transmission Spectrum}
\tablewidth{\textwidth}
\tablehead{
& \multicolumn{4}{|c}{This Work} & \multicolumn{5}{|c}{\citet{Mad23}} \\
\hline
Molecule & \multicolumn{1}{|c}{$\mathcal{B}_{\mathrm{Ref}, \, i}$} & $N\sigma_{\rm{classic}}$ & $N\sigma_{\rm{intuitive}}$ & \multicolumn{1}{c|}{Jeffreys' Scale} & $\mathcal{B}_{\mathrm{Ref}, \, i}$ & $N\sigma_{\rm{classic}}$ & $N\sigma_{\rm{intuitive}}$ & Their Classification & Jeffreys' Scale
}
\startdata
    CH$_4$ & 12--$10^6$ & 2.7--5.6$\,\sigma$ & 1.8--4.9$\,\sigma$ & moderate-- & 11,000--44,000$^{\dagger}$ & 4.7--5.0$\,\sigma$ &  3.9--4.2$\,\sigma$ & robust detection & strong evidence \\
    & & & & strong evidence & & & & & \\
    CO$_2$ & 0.6--4.2 & $< 2.3\,\sigma$ & $< 1.3\,\sigma$ & no evidence & 18--40$^{\dagger}$ & 2.9--3.2$\,\sigma$ & 1.9--2.3$\,\sigma$ & robust detection & moderate evidence \\
    DMS & 0.6--3.9 & $< 2.2\,\sigma$ & $< 1.3\,\sigma$ & no evidence & $\sim$1--5$^{\dagger}$ & $< 2.4\,\sigma$ & $< 1.4\,\sigma$ & potential signs & no evidence-- \\   
    & & & & & & & & & weak evidence \\
\enddata
\tablecomments{$\mathcal{B}_{\mathrm{Ref}, \, i}$ is the Bayes factor between models with and without molecule `i'. $N\sigma_{\rm{classic}}$ uses the relation from \citet{Benneke2013}, while $N\sigma_{\rm{intuitive}}$ is derived from equating $p = 1/(1 + \mathcal{B}_{\mathrm{Ref}, \, i}$) \citep{Kipping2025}. The Jeffreys' scale uses the following terms: `no evidence' ($\mathcal{B}_{\mathrm{Ref}, \, i} < 3$), `weak evidence' ($3 < \mathcal{B}_{\mathrm{Ref}, \, i} < 12$), `moderate evidence' ($12 < \mathcal{B}_{\mathrm{Ref}, \, i} < 150$), and `strong evidence' ($\mathcal{B}_{\mathrm{Ref}, \, i} > 150$). 
$^{\dagger}$Bayes factors were not provided by \citet{Mad23}, so we compute them from their quoted detection significances. The range of Bayes factors and significances quoted for \citet{Mad23} correspond to their no offset, one offset, and two offsets cases. 
Our lowest Bayes factor for CH$_4$ corresponds to $R = 25$ NIRISS data, which  hinders the CH$_4$ detectability (see Section~\ref{sec:det_significances} and Figure~\ref{fig:retrieval_detection_significances}), but for $R >= 100$ data the minimum CH$_4$ Bayes factor is 86 ($N\sigma_{\rm{intuitive}} = 2.5\,\sigma$). Since the formula mapping Bayes factors to $N\sigma_{\rm{classic}}$ breaks down for low Bayes factors ($\lesssim 3$), we quote only the highest significance for CO$_2$ and DMS.
}
\label{tab:molecular_detections_classification}
\end{deluxetable*}

In Table~\ref{tab:molecular_detections_classification}, we summarize and compare the statistical evidence for CH$_4$, CO$_2$, and DMS in K2-18\,b's atmosphere from our study with \citet{Mad23}. We present the Bayes factors, the `classical' detection significance (following the scheme from \citealt{Benneke2013}) and the `intuitive' detection significance \citep{Kipping2025}. First, we consider the evidence for CH$_4$ in K2-18\,b's atmosphere. Figure~\ref{fig:retrieval_detection_significances} shows that 33/40 Bayesian model comparisons with $R_{\rm{data}} \geq 100$ have $N\sigma_{\rm{classic}} > 4.0\,\sigma$ --- or $N\sigma_{\rm{intuitive}} > 3.0\,\sigma$ --- which corresponds to `strong evidence' on the Jeffreys' scale (only the \texttt{FIREFLy} $R = 100$ NIRISS data results in slightly lower significances). At this level of model preference, the term `detection' is clearly warranted for CH$_4$ in both our study and \citet{Mad23}. Second, we see that the statistics reported for CO$_2$ by \citet{Mad23} should be downgraded from `robust detection' to `moderate evidence' according to the Jeffreys' scale, while our retrieval model comparisons are classified as `no evidence' for CO$_2$ (see Figure~\ref{fig:retrieval_detection_significances}). These results highlight the reliable and reproducible CH$_4$ inference, but show that the JWST data from \citet{Mad23} do not provide sufficient evidence supporting a detection of CO$_2$ (see also Appendix~\ref{appendix:Madhu_reproduction}). Finally, we see that our retrievals yield `no evidence' for DMS, while the results from \citet{Mad23} correspond to `no evidence' or `weak evidence' at most when using the Jeffreys' scale. We therefore advocate for reporting more conservative statistical measures for detection significances from Bayesian model comparisons (e.g. $N\sigma_{\rm{intuitive}}$, see \citealt{Kipping2025}), alongside exploring the sensitivity of detections to specific data regions \citep[e.g.][]{Welbanks2023}, to avoid overstating the statistical evidence of molecules from JWST spectra of exoplanet atmospheres.

\subsection{Compatibility Between Near-infrared and Mid-infrared Transmission Spectra of K2-18\,b} \label{sec:nir_vs_mir_compatability}
\begin{figure*}[ht!]
    \centering
    \includegraphics[width=0.85\textwidth, trim = 0.0cm 0.5cm 0.0cm 0.5cm]{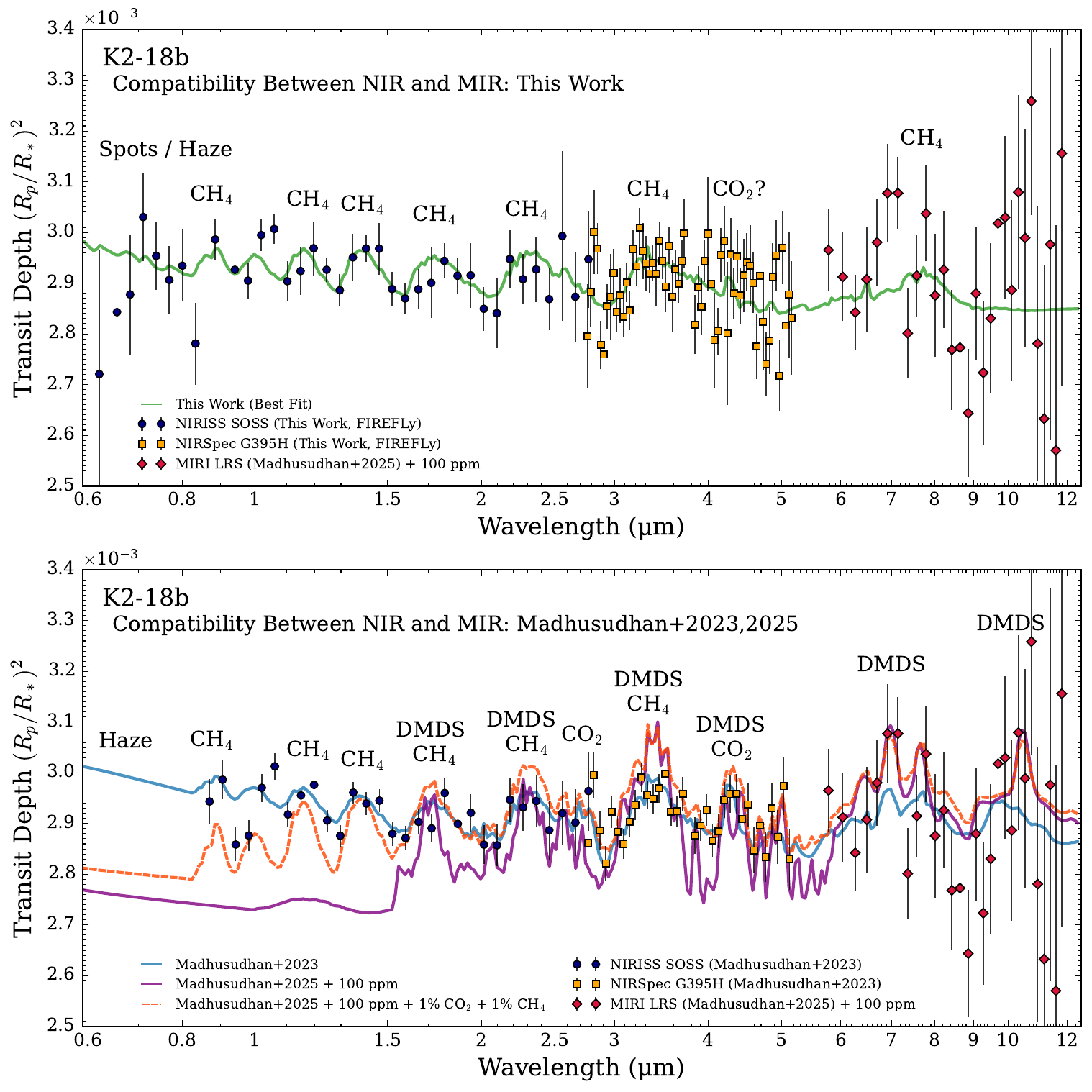}
    \caption{Extrapolated best-fit models compared to additional JWST transmission spectra of K2-18\,b. Top: best-fitting \texttt{POSEIDON} model to the low-resolution \texttt{FIREFLy} near-infrared data (NIRISS SOSS: $R\approx 25$, NIRSpec G395H: $R\approx 100$) extrapolated into the mid-infrared for comparison with the MIRI LRS data from \citet{Madhusudhan2025} ($R\approx 40$). Bottom: reproduction of the median model spectra from \citet{Mad23} (near-infrared only) and \citet{Madhusudhan2025} (mid-infrared only) with \texttt{POSEIDON}, each extrapolated over the wavelength range of the other datasets. All models are shown binned to $R = 100$ for clarity and a 100\,ppm offset is applied to the MIRI LRS data, and the models corresponding to \citet{Madhusudhan2025}, to account for differences in the baseline transit depth. The model from this study, exhibiting no absorption features from DMS or DMDS, provides an adequate fit to the MIRI data. However, the median atmospheric properties from \citet{Madhusudhan2025} predict DMDS absorption features with amplitudes inconsistent with the near-infrared data. 
    }
    \label{fig:nir_vs_mir_best-fit_models}
\end{figure*}

Here, we briefly examine the implications of our non-detections of DMS and CO$_2$ with the mid-infrared MIRI LRS spectrum of K2-18\,b recently presented by \citet{Madhusudhan2025}. In Figure~\ref{fig:nir_vs_mir_best-fit_models} (upper panel), we show our best-fitting transmission spectrum to our \texttt{FIREFLy} low-resolution data (NIRISS SOSS at $R\approx 25$, NIRSpec G395H at $R\approx 100$) extrapolated to the mid-infrared. We see that this extrapolated model --- which contains only a minor CH$_4$ feature and no DMS absorption in the mid-infrared --- provides a reasonable fit to the MIRI LRS \texttt{JExoRES} data from \citet{Madhusudhan2025}, consistent with there being no significant spectral features in the mid-infrared \citep{Taylor2025,Welbanks2025}. However, the same is not true when extrapolating the transmission spectrum corresponding to the median retrieved atmospheric properties reported by \citet{Madhusudhan2025} backward to the near-infrared. We show in Figure~\ref{fig:nir_vs_mir_best-fit_models} (lower panel) that the DMDS features reported by \citet{Madhusudhan2025} would produce significantly stronger absorption near 2.3\,$\micron$ and 3.3\,$\micron$ than is seen in the near-infrared data (or in the corresponding median retrieved models from \citealt{Mad23}). This mismatch in scale heights is primarily driven by the MIRI LRS data favoring a temperature of $\approx$ 350\,K, while the near-infrared data favors cooler temperatures of $\approx$ 240\,K. We note that \citet{Luque2025} similarly found that simultaneous retrievals of the near-infrared and mid-infrared JWST data offer no significant evidence for DMS or DMDS in K2-18\,b's atmosphere.

\subsection{Additional JWST Observations of K2-18 b} \label{sec:future_obs}
JWST will continue to observe spectra of K2-18\,b, offering deeper insights into its atmospheric composition and internal structure.
In particular, Program GO-2372 (PI: Renyu Hu) is obtaining new near-infared observations of K2-18\,b's transmission spectrum. We briefly discuss how our results relate to potential outcomes from these upcoming observations.

Program GO-2372 is in the process of observing 4 additional transits with NIRSpec G395H (with 2 pending at the time of writing) and 2 transits with NIRSpec G235H. These observations will significantly improve the observational precision over the 1.7--5.2$\,\micron$ wavelength range, offering additional sensitivity to the allowed abundances of molecules including CH$_4$, CO$_2$, DMS, CO, H$_2$O, NH$_3$, and HCN. The improved precision of these observations may allow the detection of CO$_2$, which would be expected for a mini-Neptune with a slightly lower abundance than our present CO$_2$ upper limit (see Figure~\ref{fig:Nep_clouds}). Therefore, a detection of CO$_2$ in new multi-transit JWST data would not be not at odds with the non-detection of CO$_2$ from the single NIRISS SOSS and NIRSpec G395H transits we report here.  

\subsection{K2-18\,b is Unlikely to Harbor a Water Ocean}
Our finding that K2-18\,b is unlikely to be a hycean planet adds to the current dearth of observational evidence for the existence of hycean planets. A similar example is provided by JWST transmission spectra of the sub-Neptune TOI-270\,d, which \citet{Hol24} identified as a hycean candidate using only NIRSpec data, but \citet{Ben24}, using both NIRISS and NIRSpec data, found is better explained via a miscible supercritical envelope. Future JWST observations of sub-Neptunes will have a valuable role to play in assessing the existence of hycean worlds. Our results provide a cautionary tale for the value of exploring a spread of data-level and model-level approaches when retrieving sub-Neptune transmission spectra, with the consequence that there is currently no spectroscopic evidence for the existence of hycean planets in nature.

\section*{Acknowledgments}

We thank the anonymous referee for their helpful comments. We also thank Adina Feinstein, Duncan Christie, Emma Esparza-Borges, Evert Nasedkin, and Jake Taylor for helpful comments. We thank Hamish Innes for an illuminating discussion on the supercritical ocean scenario.
This work is based in part on observations made with the NASA/ESA/CSA JWST. The data were obtained from the Mikulski Archive for Space Telescopes at the Space Telescope Science Institute, which is operated by the Association of Universities for Research in Astronomy, Inc., under NASA contract NAS 5-03127 for JWST. The specific observations analyzed can be accessed via \dataset[DOI:10.17909/3ds1-8z15]{https://doi.org/10.17909/3ds1-8z15} on MAST \citep{K218JWSTData}.
S.P.S. is supported by the National Science Foundation Graduate Research Fellowship Program under Grant No. DGE2139757. Any opinions, findings, and conclusions or recommendations expressed in this material are those of the author and do not necessarily reflect the views of the National Science Foundation.
R.J.M. is supported by NASA through the NASA Hubble Fellowship grant HST-HF2-51513.001, awarded by the Space Telescope Science Institute, which is operated by the Association of Universities for Research in Astronomy, Inc., for NASA, under contract NAS 5-26555.
S.-M.T. acknowledges support from NASA Exobiology grant No. 80NSSC20K1437 and the University of California, Riverside.
M.R.\ acknowledges support from the Natural Sciences and Engineering Research Council of Canada (NSERC).
T.J.B.\ acknowledges funding support from the NASA Next Generation Space Telescope Flight Investigations program (now JWST) via WBS 411672.07.04.01.02.
This research was supported in part through computational resources and services provided by Advanced Research Computing at the University of Michigan, Ann Arbor.
This research has made use of the NASA Exoplanet Archive, which is operated by the California Institute of Technology, under contract with the National Aeronautics and Space Administration under the Exoplanet Exploration Program. 
This research has made use of NASA’s Astrophysics Data System.

\vspace{0.2cm}

\facilities{JWST, MAST, ADS}

\vspace{0.2cm}

\noindent \software{ \vspace{0.1cm}
\\ \eureka \citep{Eureka!},
\\ \texttt{exoTEDRF}  \citep{Fei23, Rad23, Radica2024b},
\\ \texttt{FIREFLy} \citep{Rus22, Rus23},
\\ \texttt{BeAR} \citep{kitzmann20},
\\ \texttt{POSEIDON} \citep{MacDonald2017, MacDonald2023},
\\ \texttt{Photochem} \citep{Wog24code,Wog24},
\\ \texttt{PICASO} \citep{Batalha2019b,Mukherjee2023},
\\ \texttt{VULCAN} \citep{Tsai2017,Tsai2021b},
\\ \texttt{HELIOS} \citep{Malik2017,Malik2019},
\\ IPython \citep{ipython},
\\ \texttt{astropy} \citep{AstropyI, AstropyII, AstropyIII},
\\ \texttt{batman} \citep{Kreidberg2015},
\\ \texttt{dynesty} \citep{speagle2020dynesty},
\\ \texttt{emcee} \citep{Foremak-Mackey2013}
\\ \texttt{ExoTiC-LD} \citep{grant2024exoticLDJoss},
\\ \texttt{lacosmic} \citep{lacosmic},
\\ \texttt{lmfit} \citep{lmfit},
\\ \texttt{matplotlib} \citep{hunter2007matplotlib},
\\ \texttt{numpy} \citep{harris2020array},
\\ \texttt{PyMultiNest} \citep{Feroz2009,Buchner2014},
\\ \texttt{scipy} \citep{jones2001scipy, 2020SciPy-NMeth}
}

\bibliography{K2-18b}{}
\bibliographystyle{aasjournal}

\newpage
\appendix

\section{Reproduction of the Retrieval Results from Madhusudhan et al. (2023)} \label{appendix:Madhu_reproduction}

Here we show that our atmospheric retrieval framework can reproduce consistent results to those presented in \citet{Mad23}, if we use the same full-resolution data presented in that study. This demonstrates that the non-detections of CO$_2$ and DMS presented above are primarily driven by differences in the data reduction and/or light curve fitting approaches from our respective studies, rather than differences in the atmospheric modelling and/or retrieval approaches.

We used \texttt{POSEIDON} to replicate the results of \citet{Mad23}. For this purpose, we used the exact same atmospheric model configuration as the `canonical' model described in \citet{Mad23}. To summarise, this retrieval model fits for the log$_{10}$ volume mixing ratios of \ce{H2O, CH4, NH3, HCN, CO, CO2, DMS, CS2, CH3Cl, OCS, and N2O}, the 6-parameter P-T profile from \citet{Madhusudhan2009}, the 4-parameter inhomogeneous cloud and haze parameterisation from \citet{MacDonald2017}, and the reference pressure corresponding to $2.61\,R_{\Earth}$. We consider three retrieval models: (i) no offset between the NIRISS and NIRSpec G395H data; (ii) one offset between NIRISS and NIRSpec G395H; and (iii) two offsets, one offset between NIRISS and the NIRSpec G395H NRS1 detector and one offset between NIRISS and the NIRSpec G395H NRS2 detector. Therefore, the three models have 22, 23, and 24 free parameters, respectively. We use identical priors for these parameters as in Table~4 from \citet{Mad23}. For fixed system properties (e.g. planetary mass, stellar radius), we used the values from \citet{Ben19} for consistency with \citet{Mad23}.

Several important retrieval model settings were not mentioned by \citet{Mad23}. Therefore, we adopted conservative choices to ensure the reliability of our retrieval results. Our opacities are sampled on a model wavelength grid with a resolution of $R = \lambda/\Delta\lambda = $ 100,000, ensuring negligible errors from opacity sampling. We used 100 atmospheric layers from 10$^{-6}$--10\,bar, spaced uniformly in log-pressure. We note that the minimum atmospheric pressure used here differs from the value of 10$^{-8}$\,bar we adopt for our main retrieval results, but we chose 10$^{-6}$\,bar here to match the value implied in Table 4 of \citet{Mad23}. The line lists in \texttt{POSEIDON} (described in the main text and Table~\ref{tab:retrieval_configurations}) do differ from \citet{Mad23} for several important molecules (in particular CH$_4$ and CO$_2$), but we stress that \texttt{POSEIDON}~v1.2 is a state-of-the-art opacity database with the latest ExoMol line lists (as of 2024). Finally, we used 1,000 \texttt{MultiNest} live points to ensure fine sampling of the parameter space.

Table~\ref{tab:Madhu_reproducton_Bayesian_model_comparison} summarises the retrieval model statistics from our analysis using the data from \citet{Mad23}. We find that only CH$_4$ is robustly detected, with a detection significance $>$ 4.5$\sigma$ across the different offset treatments. None of our retrievals provide notable evidence for CO$_2$ or DMS, with peak Bayes factors of $\approx$ 6 and $\approx$ 3, respectively (compared to $\mathcal{B}_{\mathrm{Ref}, \, i} >$ 12 for `moderate evidence' and $\mathcal{B}_{\mathrm{Ref}, \, i} >$ 150 for `strong evidence' on the Jeffreys' scale, see \citealt{Trotta2008}). Our retrievals favour a single offset between the NIRISS SOSS and NIRSpec G395H data ($44^{+13}_{-12}$\,ppm), consistent with \citet{Mad23}. The statistically preferred `one offset' model has a Bayes factor of $\approx$ 5 for CO$_2$ (2.4$\sigma$), corresponding to `weak evidence' at best, while there is no evidence for DMS (Bayes factor $\approx$ 1).

\begin{figure*}[ht!]
    \centering
    \includegraphics[width=\linewidth]{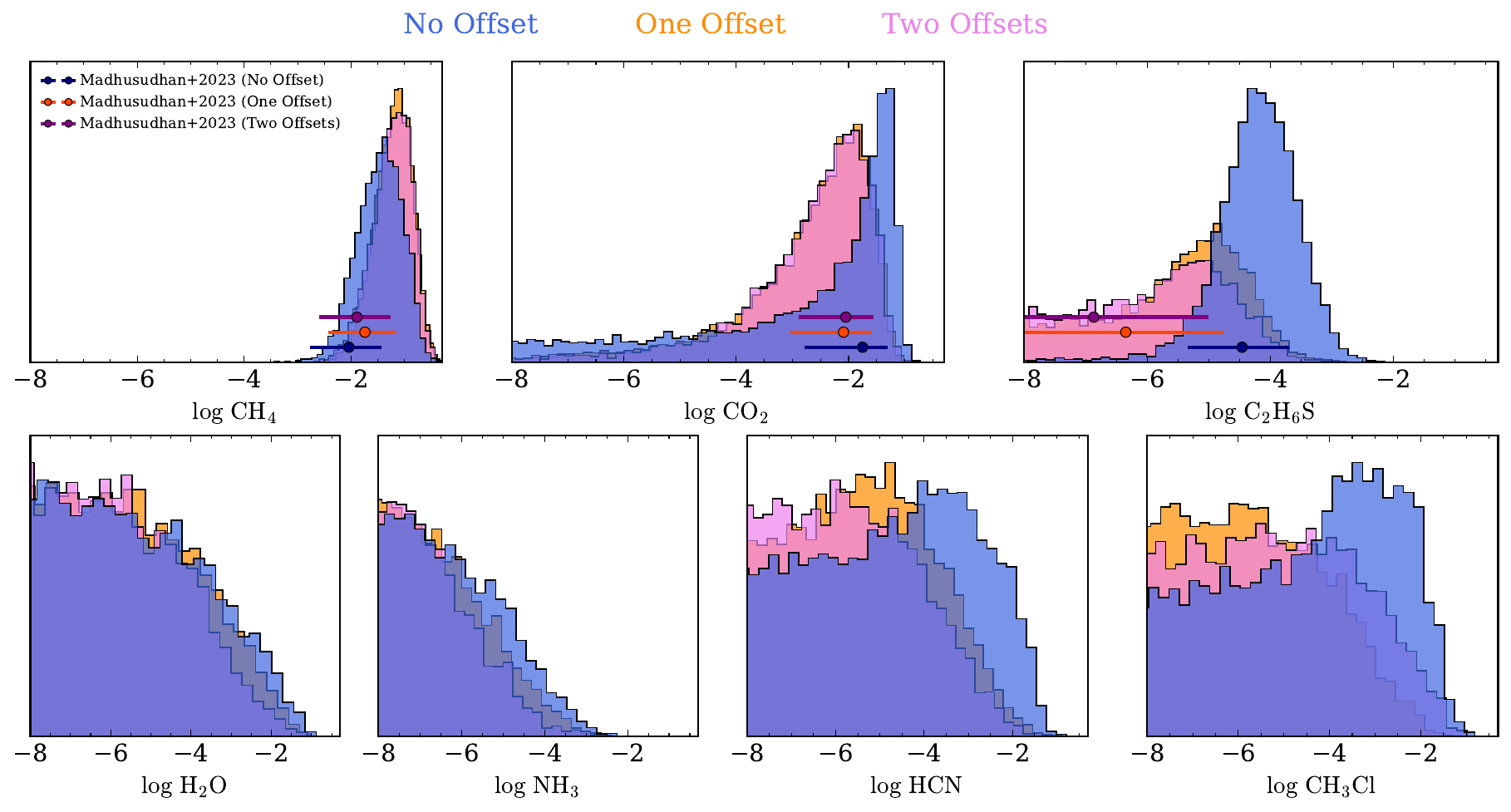}
    \caption{Reproduction of the results from \citet{Mad23} using their full-resolution NIRISS + NIRSpec data. Retrieval results are overplotted for three retrieval models: (i) with no offset between NIRISS and NIRSpec (blue), (ii) an offset between NIRISS and NIRSpec (orange), and (iii) two offsets, one between NIRISS and NIRSpec G395H NRS1 and one between NIRISS and NIRSpec G395H NRS2 (violet). The statistically favoured retrieval model has a single offset. Our \texttt{POSEIDON} retrieval framework produces consistent results with  \citet{Mad23} when using the same priors and data. Any differences in detections for our other retrievals are therefore due to differences in data reduction rather than differences inherent to our retrieval approach.}
    \label{fig:Madhu_reproducton_histograms}
\end{figure*}

Figure~\ref{fig:Madhu_reproducton_histograms} provides a reproduction of \citet{Mad23}'s Figure~4. We find broadly consistent posterior distributions to \citet{Mad23} when using their data. Our CH$_4$ abundances are higher by $\approx$ 0.8\,dex, but consistent within 1$\sigma$. We note that our CO$_2$ posterior for the `no offset' case has a more prominent tail to low abundances than Figure~4 in \citet{Mad23}. This difference arises because \citet{Mad23} renormalised the probability densities for each retrieval in their Figure~4 to have the same peak probability density, while our Figure~\ref{fig:Madhu_reproducton_histograms} preserves the relative probability density differences between the retrievals. We highlight that our visualisation method has the advantage of ensuring the same integrated area for each distribution (i.e. total probability), which \citet{Mad23}'s Figure~4 does not respect. We stress this point because it artificially suppresses the CO$_2$ tail for the `no offset' case in \citet{Mad23}'s Figure~4, which consequently oversells the confidence of their claimed CO$_2$ detection.

\begin{figure*}[htb!]
    \centering
    \includegraphics[width=\linewidth]{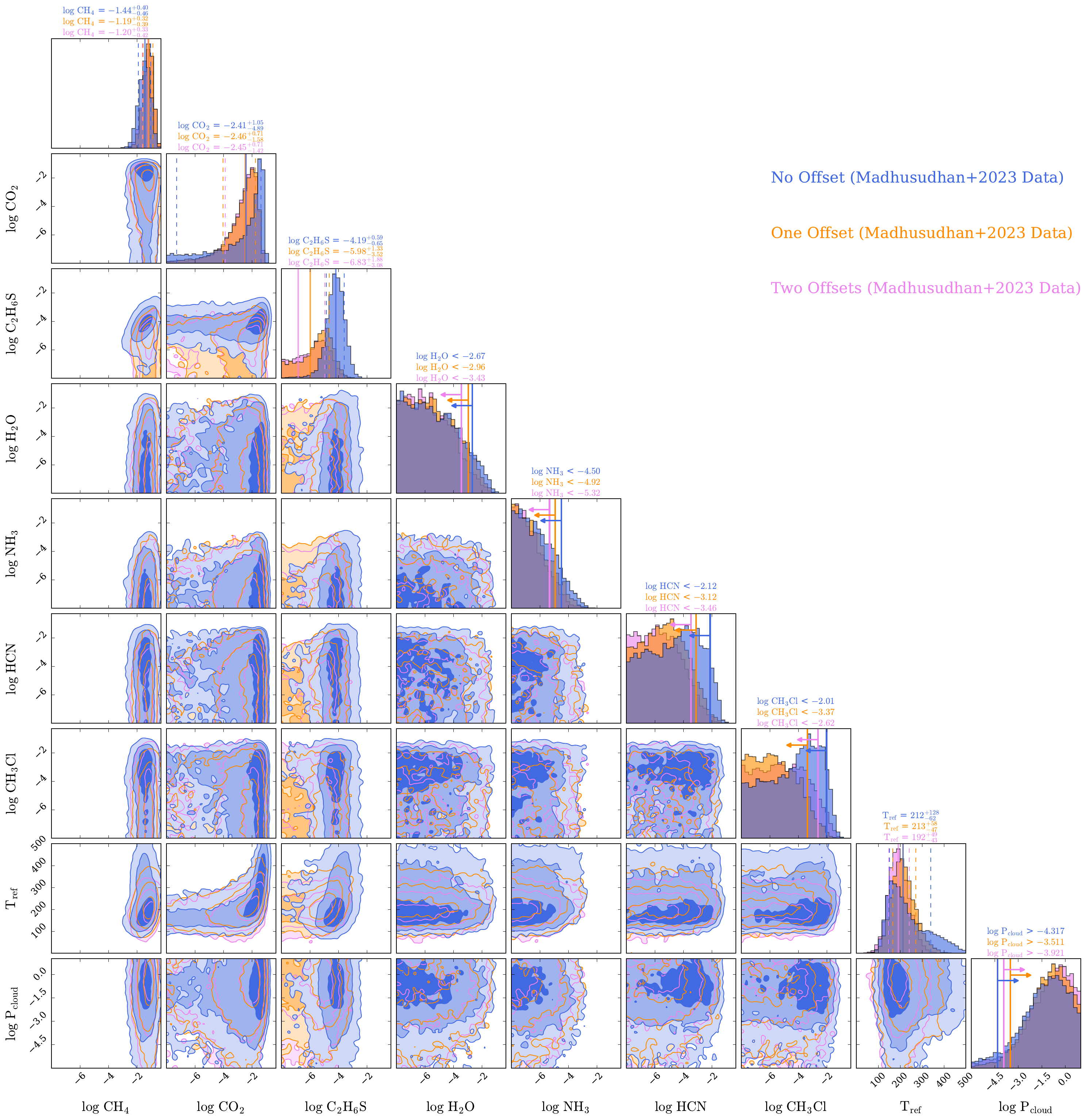}
    \caption{Corner plots corresponding to Figure~\ref{fig:Madhu_reproducton_histograms}, showing correlations between key atmospheric parameters. Parameters with only an upper or lower limit are marked by a vertical line and arrow at the 95\% credible interval. We include 1$\sigma$ credible regions for CO$_2$ and DMS, even though they are not robustly detected, for comparison with \citet{Mad23}.}
    \label{fig:Madhu_reproducton_corner}
\end{figure*}

Finally, Figure~\ref{fig:Madhu_reproducton_corner} shows correlations between notable atmospheric properties from these retrievals on the \citet{Mad23} data. We highlight a strong correlation between the CO$_2$ abundance and the temperature at the top of the atmosphere ($T_{\mathrm{ref}}$), with most of the parameter space that favours a high CO$_2$ abundance corresponding to temperatures far in excess of the planetary equilibrium temperature ($\approx$ 250\,K). Thus, even if one assumes CO$_2$ is present in K2-18\,b's atmosphere, the planet would be far too warm to support habitable liquid water surface conditions.

\begin{deluxetable}{lccccc}[ht!]
\tablecaption{Bayesian model comparison using the \citet{Mad23} data}
\tablewidth{\textwidth}
\tablehead{
\colhead{Retrieval Model} & \colhead{Bayesian Evidence ($\ln \mathcal{Z}$)} & \colhead{Bayes Factor ($\mathcal{B}_{\mathrm{Ref}, \, i}$)} & \colhead{Detection Significance} & \colhead{Classification}
}
\startdata
    \multicolumn{5}{c}{\textbf{No Offset}} \\
    Ref. & 28826.8 & --- & --- & --- \\
    No CH$_4$ & 28818.2 & 5541 & 4.5$\sigma$ & Strong Evidence \\
    No CO$_2$ & 28825.7 & 3.1 & 2.1$\sigma$ & Weak Evidence \\
    No DMS & 28825.6 & 3.2 & 2.1$\sigma$ & Weak Evidence \\
    \hline
    \multicolumn{5}{c}{\textbf{One Offset}} \\
    Ref. & 28830.5 & --- & --- & --- \\
    No CH$_4$ & 28820.4 & 25848 & 4.9$\sigma$ & Strong Evidence \\
    No CO$_2$ & 28828.9 & 5.2 & 2.4$\sigma$ & Weak Evidence \\
    No DMS & 28830.2 & 1.3 & N/A & No Evidence \\
    \hline
    \multicolumn{5}{c}{\textbf{Two Offsets}} \\
    Ref. & 28829.9 & --- & --- & --- \\
    No CH$_4$ & 28819.6 & 30638 & 4.9$\sigma$ & Strong Evidence \\
    No CO$_2$ & 28828.0 & 6.4 & 2.5$\sigma$ & Weak Evidence \\
    No DMS & 28829.8 & 1.1 & N/A & No Evidence \\
\enddata
\tablecomments{$\mathcal{Z}$ is the Bayesian evidence of each retrieval, $\mathcal{B}_{\mathrm{Ref}, \, i}$ is the Bayes factor between the reference model and the nested model without molecule `$i$', and the `classification' follows the Jeffreys' scale \citep[e.g.][]{Trotta2008}.}
\label{tab:Madhu_reproducton_Bayesian_model_comparison}
\end{deluxetable}

\section{Spectrophotometric Light Curve Fits with \texttt{FIREFLy}}\label{appendix:firefly_spectroresiduals}
\begin{figure*}
    \centering
    \includegraphics[width=\linewidth]{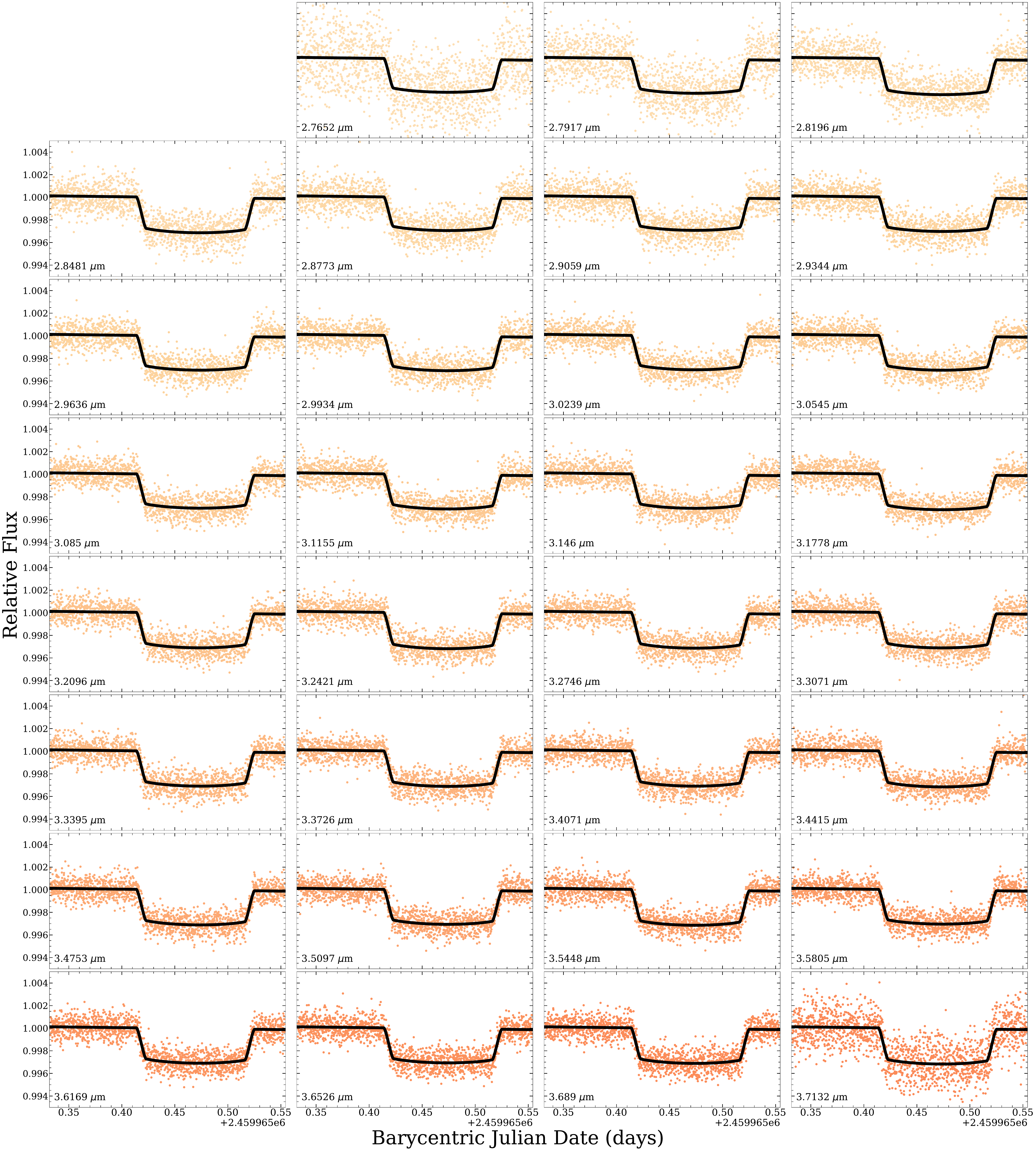}
    \caption{\texttt{FIREFLy} Spectroscopic light light curve fits for NRS1.}
    \label{fig:firefly_spectrophoto_nrs1}
\end{figure*}

\begin{figure*}
    \centering
    \includegraphics[width=\linewidth]{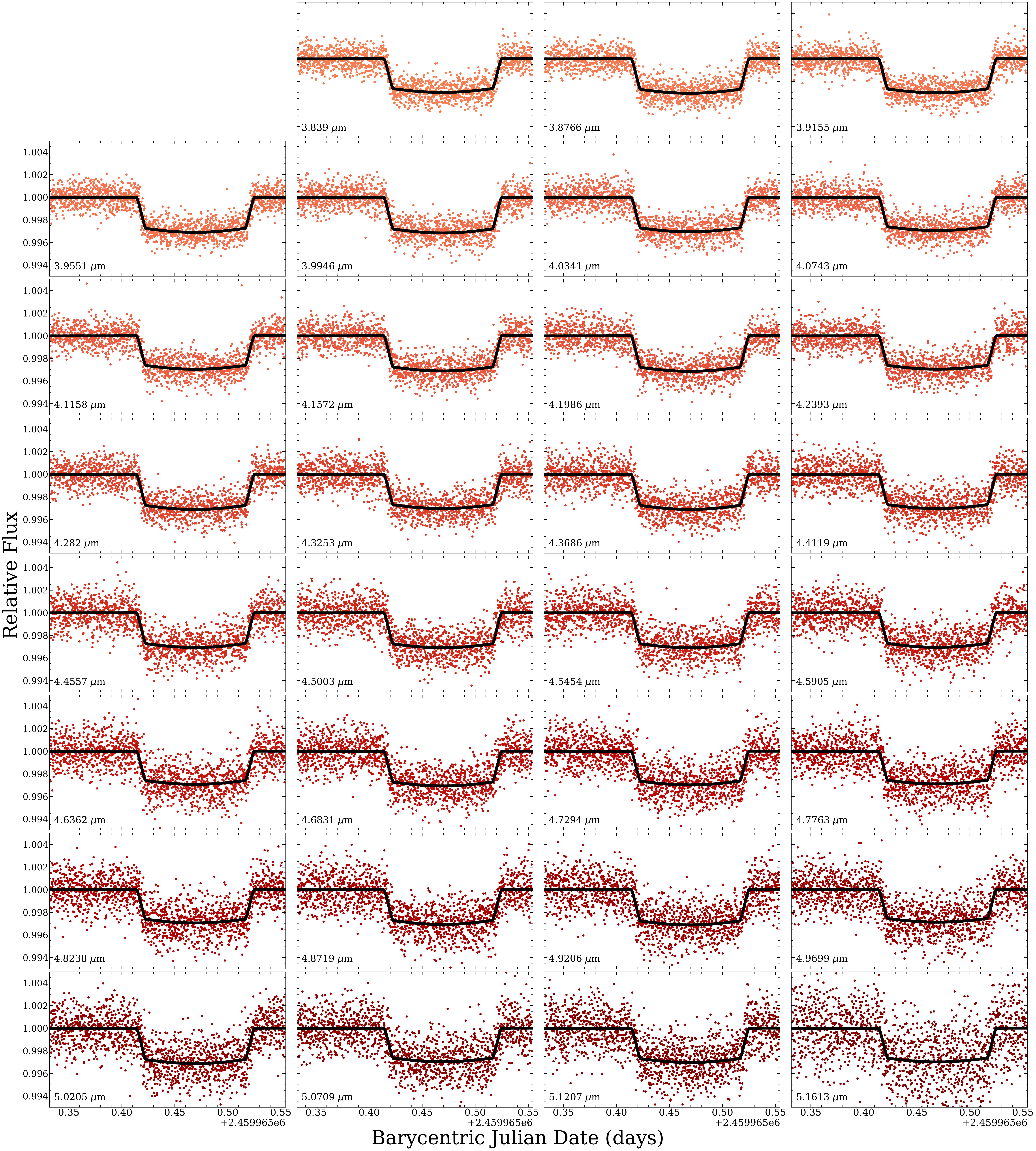}
    \caption{\texttt{FIREFLy} Spectroscopic light light curve fits for NRS2.}
    \label{fig:firefly_spectrophoto_nrs2}
\end{figure*}

\begin{figure*}
    \centering
    \includegraphics[width=\linewidth]{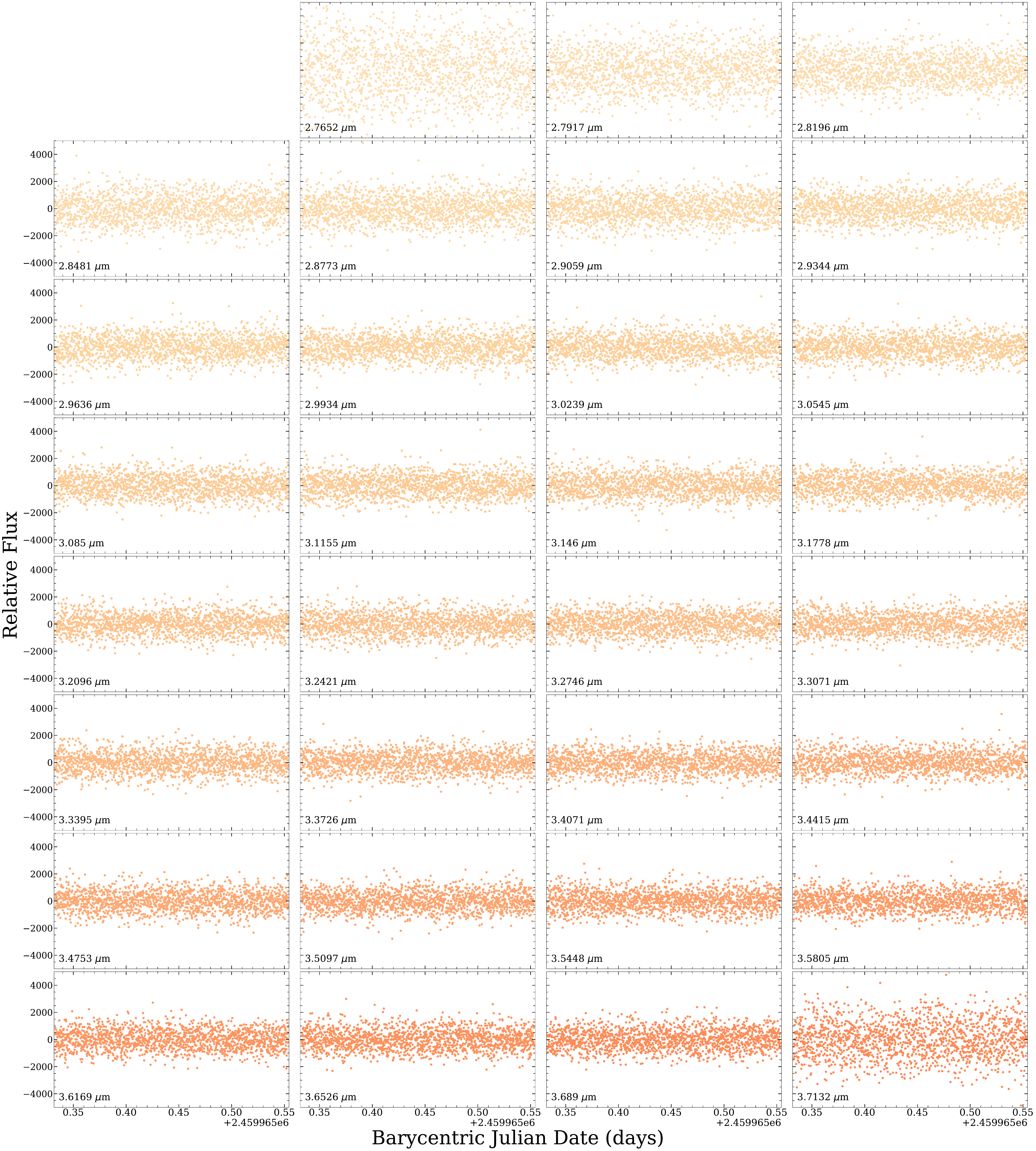}
    \caption{\texttt{FIREFLy} Spectroscopic light light curve residuals for NRS1.}
    \label{fig:firefly_spectrophoto_residual_nrs1}
\end{figure*}

\begin{figure*}
    \centering
    \includegraphics[width=\linewidth]{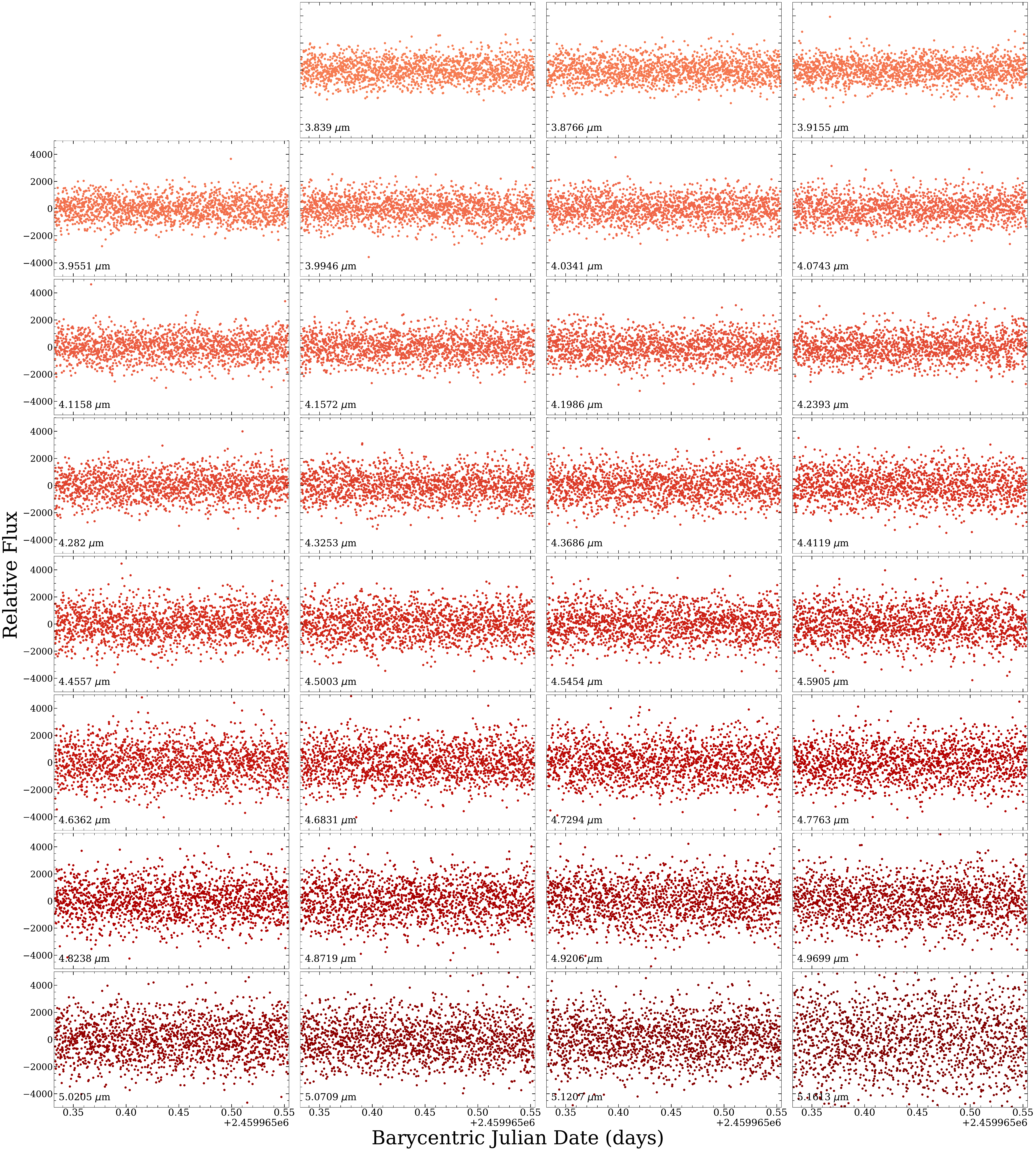}
    \caption{\texttt{FIREFLy} Spectroscopic light light curve residuals for NRS2.}
    \label{fig:firefly_spectrophoto_residual_nrs2}
\end{figure*}

Here we present the spectroscopic light curve fits and residuals for our $R\approx100$ \texttt{FIREFLy} reduction. 
We show the NRS1 fits in Figure \ref{fig:firefly_spectrophoto_nrs1} and the NRS2 fits in Figure \ref{fig:firefly_spectrophoto_nrs2}.
Additionally, we show the residuals of the NRS1 fits in Figure \ref{fig:firefly_spectrophoto_residual_nrs1} and the residuals of the NRS2 fits in Figure \ref{fig:firefly_spectrophoto_residual_nrs2}.
Each figure's panels all have the same x and y limits.
The lack of trends in each bin demonstrates the fidelity of our spectrophotometric fits.


\section{Significance Comparison Between NIRSpec and NIRISS Reductions}\label{appendix:comparisons}
\begin{figure*}[h!]
    \centering
    \vspace{3cm}
    \includegraphics[width=\linewidth]{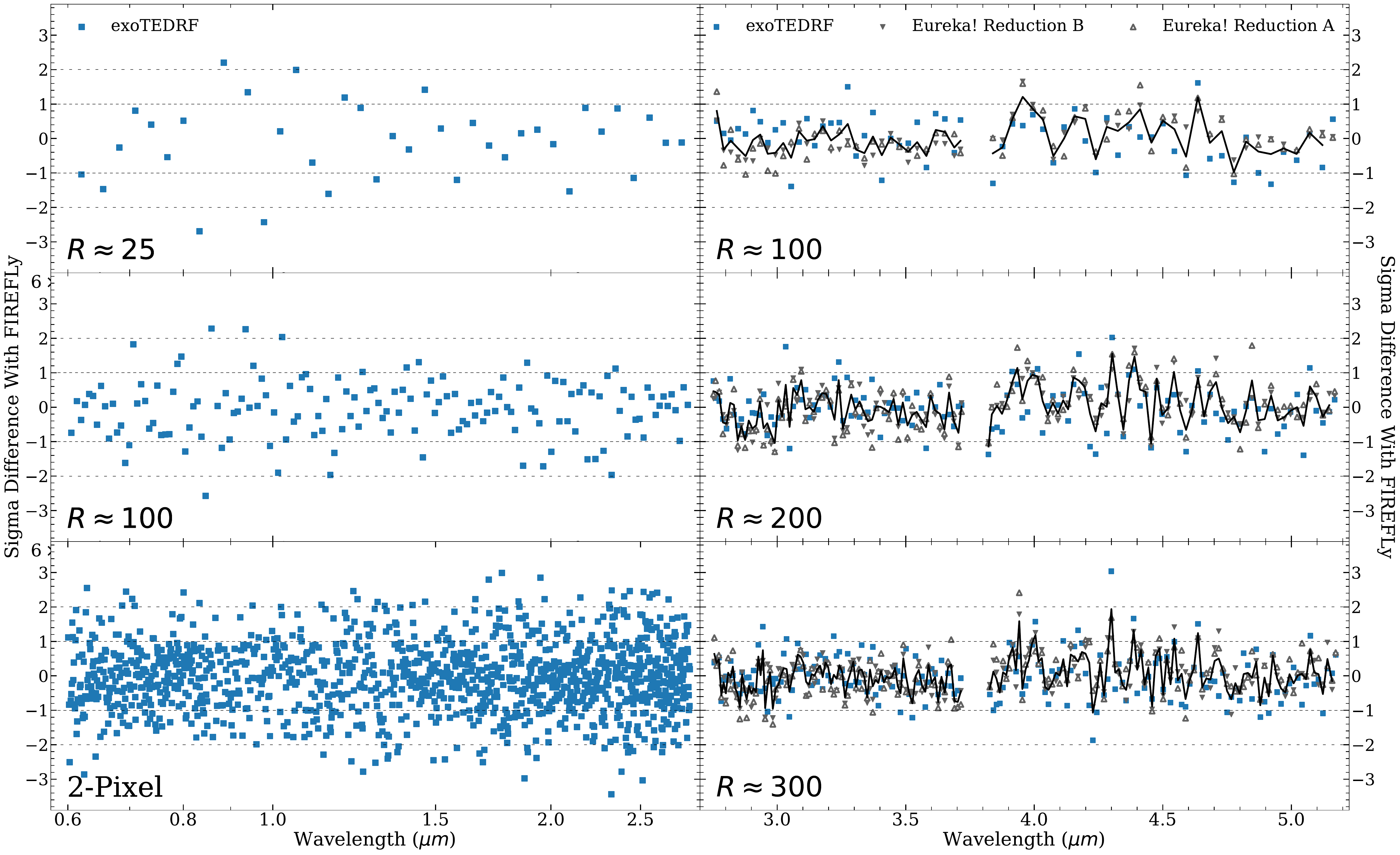}
    \caption{
    Significance of the discrepancy in transit depths between the \texttt{FIREFLy} reduction and the three other independent reductions. On the left three panels, we plot the significance between NIRISS reduction at $R\approx25$ (top), $R\approx100$ (middle), and at the 2-pixel level (bottom). On the right three panels, we plot the significance between NIRSpec reductions at $R\approx 100$ (top), $R\approx200$ (middle), and $R\approx300$ (bottom).
    Using the same color scheme as Figure \ref{fig:reductions}, we plot as blue squares the \texttt{exoTEDRF} reduction and as gray triangles the \texttt{Eureka!} reductions, distinguishing between the two with unfilled downward-pointing (A) and filled upward-pointing (B) triangles.
    We plot as a black solid line on the right three panels the average deviation between the three, and as gray horizontal lines the 1-$\sigma$ (densely-dashed) and 2-$\sigma$ (sparsely-dashed) cutoffs.
    Our reductions are consistent with one another, though the rate of $>2\sigma$ outliers increases with higher resolution.
    }
    \label{fig:significance}
\end{figure*}

For illustrative purposes, we perform an comparison between reductions in which we calculate the significance of the discrepancy in the transmission spectrum between the \texttt{FIREFLy} NIRSpec reduction and the other three reductions at $R\approx100$, $R\approx200$, and $R\approx300$.
We do the same for NIRISS, but at $R\approx25$, $R\approx100$, and 2-pixel binning.
We show the results of this in Figure \ref{fig:significance}.
We attribute the higher number of $>2\sigma$ outliers at shorter wavelengths between our $R\approx25$ NIRISS reductions to slight differences in our NIRISS binning scheme that are more pronounced there.
Our reductions are consistent with one another, typically within a 1 to 1.5-$\sigma$ range of the \texttt{FIREFLy} reduction.
The higher resolving power reductions have more outliers, but we expect this due to the larger number of data points.
This indicates that the differing choices we apply in our reductions each yield consistent results and are reasonable for the purposes of marginalizing over in our retrieval analysis.

\end{document}